\documentclass[10pt,aps,prb,twocolumn,showpacs,superscriptaddress,nobalancelastpage,notitlepage,longbibliography]{revtex4-2}

\usepackage{graphicx,psfrag,color,mathbbol,amsthm,amsmath,amsfonts,amssymb,bm,braket,dsfont,footmisc,mathrsfs,multirow,latexsym,times}

\usepackage[colorlinks]{hyperref}
\hypersetup{
 colorlinks=true,
 citecolor=magenta,
 linkcolor=blue,
 urlcolor=cyan}

\newcommand{\bfE}{\mathbf{E}}

\newcommand{\bfG}{\mathbf{G}}

\newcommand{\bfX}{\mathbf{X}}

\newcommand{\tr}{\text{tr}}
\newcommand{\norm}[1]{
\lVert#1
\rVert}

\newcommand{\id}{\mathds{1}}

\definecolor{gray}{rgb}{0.8,0.8,0.8}

\newcommand{\rev}[1]{\textcolor[rgb]{0,0,0}{#1}}
\definecolor{revred}{rgb}{0,0,0}

\newcommand{\normnoscl}[1]{\lVert#1\rVert}

\begin{document}

\title{Topological zero modes and edge symmetries of metastable Markovian bosonic systems}

\author{Vincent P. Flynn}
\thanks{vincent.p.flynn@dartmouth.edu}
\affiliation{\mbox{Department of Physics and Astronomy, Dartmouth College, 6127 Wilder Laboratory, Hanover, New Hampshire 03755, USA}} 

\author{Emilio Cobanera} 
\thanks{cobanee@sunyit.edu}
\affiliation{\mbox{Department of Mathematics and Physics, SUNY Polytechnic Institute, 100 Seymour Rd, Utica, NY 13502, USA}}
\affiliation{\mbox{Department of Physics and Astronomy, Dartmouth College, 6127 Wilder Laboratory, Hanover, New Hampshire 03755, USA}} 

\author{Lorenza Viola}
\thanks{lorenza.viola@dartmouth.edu}
\affiliation{\mbox{Department of Physics and Astronomy, Dartmouth College, 6127 Wilder Laboratory, Hanover, New Hampshire 03755, USA}} 

\begin{abstract} 
Tight bosonic analogs of free-fermionic symmetry-protected topological phases, and their associated edge-localized  excitations, have long evaded the grasp of condensed-matter and AMO physics. In this work, building on our initial exploration [Phys. Rev. Lett. {\bf 127}, 245701 (2021)], we identify a broad class of quadratic bosonic systems subject to Markovian dissipation that realize \rev{\textit{tight}} bosonic analogs of the Majorana and Dirac edge modes characteristic of topological superconductors and insulators, respectively. To this end, we establish a general framework for \textit{topological metastability} for these systems, by leveraging pseudospectral theory as the appropriate mathematical tool for capturing the non-normality of the Lindbladian generator. The resulting dynamical paradigm, which is characterized by both a sharp separation between transient and asymptotic dynamics and a non-trivial topological invariant, is shown to host edge-localized modes, which we dub Majorana and Dirac bosons. Generically, such modes consist of one conserved mode and a canonically conjugate generator of an approximate phase-space translation symmetry of the dynamics. The general theory is exemplified through several representative models exhibiting the full range of exotic boundary physics that topologically metastable systems can engender. In particular, we explore the extent to which Noether's theorem is violated in this dissipative setting and the way in which certain symmetries can non-trivially modify the edge modes. Notably, we also demonstrate the possibility of anomalous parity dynamics for a bosonic cat state prepared in a topologically metastable system, whereby an equal distribution between even and {\em odd} parity sectors is sustained over a long transient. For both Majorana and Dirac bosons, observable multitime signatures in the form of anomalously long-lived quantum correlations and divergent zero-frequency power spectral peaks are proposed and discussed in detail. Our results \rev{point to a new paradigm for} 
 symmetry-protected topological physics in free bosons, embedded deeply in the long-lived transient regimes of metastable dynamics.
\end{abstract}

\date{\today}

\maketitle

\section{Introduction}
\subsection{Context and motivation}

Indistinguishable quantum particles come in two flavors: fermions and bosons. While the distinction is kinematical and, as such, unrelated to any Hamiltonian specification, it can be explained most clearly when the particles are independent, or ``free''. Systems of free fermions (bosons) are described by Hamiltonians that are {\em quadratic} in their respective  canonical fermionic (bosonic) operators, and have long played a paradigmatic role as tractable -- either genuinely non-interacting or mean-field -- models for both equilibrium and non-equilibrium many-body physics \cite{Ripka}. For a quadratic fermionic Hamiltonian (QFH), there is always a state of lowest energy, the ground state, which captures the statistical behavior of the system in equilibrium at (and close to) zero temperature. A quantum phase transition is a phase transition at zero temperature that occurs as some parameter of the Hamiltonian is varied. Generically, the phases of free-fermion quantum matter are gapped, display no local order parameter, and the energy gap closes at a phase transition.  Since there is no local order parameter, there is no general Landau theory of the quantum phases of QFHs. Rather, the general theory of such phases is based on a different set of notions: that of protecting global symmetries, space dimension, and topological invariants. The phases of free fermions are examples of {\em symmetry-protected topological} (SPT) phases of quantum matter \cite{RyuClassification}. 

In the absence of local order parameters, how can one tell apart the different SPT phases of free fermions? A compelling answer is provided by the bulk-boundary correspondence. This powerful principle states that the topological invariant that characterizes an SPT phase also mandates the emergence of {\em robust zero-energy boundary-localized modes} \cite{ZirnbauerBBC,AlaseWH}. These zero modes (ZMs) are regarded as the main experimental manifestation of the underlying SPT phase. For example, the integer quantum Hall regimes in two (spatial) dimensions are SPT phases. In this case, the protecting symmetry is particle number and the topological invariant is the Chern number of the occupied single-particle energy bands. A measurement of the quantized Hall conductance probes directly the associated chiral surface modes. In one dimension, the Su-Schrieffer-Heeger model of polyacetylene also displays topologically-mandated edge modes \cite{SuSSH}. The protecting symmetries are particle number, spin rotations, (spinful) time reversal, and a many-body particle-hole symmetry that exchanges fermionic creation and annihilation operators. Likewise, superconductors can exist in a variety of SPT phases. While particle number cannot be one of the protecting symmetries, the superconducting classes are protected by a combination of spin symmetry, time reversal, and many-body particle-hole. The celebrated Majorana chain of Kitaev provides a paradigmatic example for $p$-wave topological superconductivity and can display edge ZMs \cite{KitaevMajorana}, which are protected against \rev{(weak)} perturbations that do not \rev{change the symmetry of the model \cite{AlaseWH}}. Altogether, SPT phases of free fermions are distinguished by the following key features: (i) The translationally invariant (bulk) system is gapped; (ii) The system displays certain combinations of protecting many-body symmetries; (iii) The ground state has an associated topological number, which can only change across a quantum phase transition as long the protecting symmetries are preserved; (iv) When the topological number is non-trivial and the system is terminated \rev{by imposing open (`hard-wall') boundary conditions}, the ground energy level is degenerate due to low-energy surface quasi-particles \footnote{Importantly, for a number-conserving system, its phase is determined by spectral properties of the Hamiltonian and the filling fraction. In addition, while in one dimension the topological surface modes are isolated in the gap, in higher dimensions the surface modes are organized into surface bands that either connect the lowest bulk band to zero energy (for superconductors), or cross the Fermi energy.}.

Having said that, if one understands the bulk-boundary correspondence as the relationship between a topological invariant and quasiparticle surface or edge modes, then there is nothing particularly ``fermionic,'' or symmetry-protected, about it. Rather, it is a general property of the Helmoltz wave equation in a structured medium. The point was forcefully made in Ref.\,\onlinecite{Haldane2008}, where a bulk-boundary correspondence for photonic crystals was identified and experimentally confirmed only one year later \cite{Wang2009}. These advances launched in earnest the new field of topological photonics, which has since witnessed a dramatic development \cite{OzawaPhotonics}. From a fundamental perspective, one may ask whether topological photonics truly represents the mirror image of topological electronics. If the answer is in the affirmative, then quantum statistics would appear to have very little to do with topological physics. More generally, to what extent can systems of free bosons exhibit SPT physics analogous to the above? The answer is complicated, and indeed depends on what concepts are emphasized. 

Consider, for example, the concept of a ``quantum phase.'' Unlike a QFH, a quadratic bosonic Hamiltonian (QBH) may not have a ground state: take a simple two-mode QBH like $H=\omega(a_1^\dag a_1 - a_2^\dag a_2)$, with frequency $\omega>0$. For such a QBH (which can arise, e.g., in cavity QED systems \cite{WiersigcQED}), the energy eigenvalues $E_{n_1,n_2} = \omega(n_1-n_2)$, with $n_{1,2}$ non-negative integers, are unbounded in both directions without external constraints placed on the total particle number. We say such Hamiltonians are \textit{thermodynamically unstable}, in the sense that no well-defined Gibbs state $\rho \sim e^{-\beta H}$ exists. Since the ground state plays such a crucial role in fermionic SPT physics, the most conservative extension into the bosonic realm is to consider only those QBHs that are thermodynamically stable. As it turns out, this subclass fails to exhibit \textit{any} of the characteristics of topological free-fermionic matter. The situation is neatly captured by three no-go theorems \cite{Squaring}, which we summarize as follows. If $H$ is a thermodynamically stable, gapped, translationally-invariant system, then: (i) $H$ can be adiabatically deformed into \textit{any} other QBH within the same class without closing the gap or breaking symmetries; (ii)  \textit{no} edge-localized ZMs emerge when the system is terminated at a boundary; and (iii) the ground state of $H$ has always {\em even bosonic parity}. \footnote{No-go (iii) does not require translation invariance.} In other words, there are no SPT phases of free-boson quantum matter. This is a puzzling conclusion, because the role of topology for QFHs is precisely to classify their gapped quantum phases. Fundamentally speaking, mean-field bosonic matter has very little in common with its fermionic counterparts from the point of view of quantum many-body physics. Nonetheless, thermodynamically stable bosonic systems, such as certain photonic, magnonic, and phononic crystals, may exhibit topologically mandated edge modes at higher, {\em non-zero} energies \cite{BaldesDisorder,PeanoSound,ShindouMagnon,FieteMagnon,OzawaPhotonics}. These topological features, however, are completely disconnected from the \rev{low-energy, low-temperature} physics. 

If one is willing to accept the loss of a many-body ground state and remove the constraint of thermodynamic stability, a gap condition may still be imposed by requiring that the quasiparticle energies (which are now necessarily not strictly positive) are gapped. That is, one can require that the quasiparticle energy bands are bounded away from zero.  \textit{Thermodynamically unstable}, ``gapped'' QBHs can then display genuinely topological ZMs under open boundary conditions. At face value, this (rather significant) compromise allows for the possibility of obtaining tight bosonic analogs of the ZMs characteristic of fermionic SPT phases. In Ref.\,\onlinecite{Decon}, we presented a QBH model hosting Hermitian edge modes that are canonically conjugate (once properly normalized), commute with the Hamiltonian in the infinite-size limit, are determined by a winding number, and exist in a well-defined (complex) spectral gap. Nonetheless, we are confronted with another major issue: due to the intrinsic non-Hermiticity of the dynamical matrix that govern the system's behavior, these bosonic ZMs are \textit{intrinsically unstable}, in a dynamical sense \cite{BarnettEdge,BaldesFlow,Decon}. Specifically, tools from Krein stability theory \cite{YakuKrein} reveal that any QBH hosting bosonic ZMs is either (i) dynamically unstable (characterized by unbounded evolution of observables), or (ii) able to be destabilized by \textit{arbitrarily small} perturbations. While the instability of these modes may have interesting implications in its own \cite{BarnettSelective,PeanoTopoSqueeze}, and the modes themselves can be thought to as ``shadows" of conventional Majorana ZMs \cite{Decon}, again we find ourselves far removed from the world of topological many-body physics. What (if anything) can be done to bring us closer to a bosonic analogue of the rich SPT physics enjoyed by free-fermionic matter? 

Our fundamental realization is the need to let go of a more subtle assumption of the fermionic paradigm: unitarity. Specifically, in this work we will provide extensive evidence that \textit{open} systems of free bosons {\em can} display SPT-like many-body physics that comes as close as possible to the SPT physics of QFHs \cite{Bosoranas}. In this sense, our work fully embraces the idea of ``topology by dissipation'' introduced in Ref.\,\onlinecite{DiehlTopByDiss, BardynTopByDiss}. However, there is a crucial difference. For fermions, as considered in these works, dissipation is a twist that one can add to a very well-developed theory for Hamiltonian systems: with dissipation or not, free fermions can be topological in the many-body sense. As we have emphasized, bosons do {\em not} seem amenable to that. Topology by dissipation could then very well be the only hope to bring about bosonic SPT physics without strong interactions. 

The program we undertake begins by appropriately adapting the key ingredients of free fermionic SPTs to the open bosonic setting. Focusing on the simplest case of Markovian dissipation in one spatial dimension, we retain the non-interacting property by restricting to the class of dynamics described by ``quasi-free,'' or {\em quadratic (Gaussian) Lindblad generators} \cite{TeretenkovQuadLindblad,BarthelQuadLindblad,PolettiGaussian}. Under certain stability assumptions, the notion of a ground state naturally maps to that of a steady state (SS), while the many-body gap condition maps to one placed on the Lindblad, or spectral gap. With these identifications made, we answer the question: to what extent can SPT-like physics emerge in quadratic bosonic Lindbladians (QBLs) possessing a unique SS and a finite spectral gap? Remarkably, signatures of SPT physics are found to emerge in a newly identified dynamical phase that we deem \textit{topologically metastable}. Topological metastability is, in turn, a specific instance of the more general \textit{dynamical metastabilty}, a phenomenon in which the dynamical stability of the system changes abruptly in the infinite-size limit. As we have pointed out in Ref.\,\onlinecite{Bosoranas}, dynamical metastability may or may not be topological in nature. Topological metastability arises precisely from requiring non-trivial bulk topology, on top of dynamical metastability. The key features of topologically metastable systems may be summarized as follows \cite{Bosoranas}: 
\begin{itemize}
\item A unique SS and a finite spectral gap are maintained for all finite system sizes. In particular, dynamical stability is present for all finite system sizes\vspace*{-1.5mm}.
\item Tight bosonic analogues of Majorana fermions, which we deem \textit{Majorana bosons} (MBs), emerge localized on opposite ends of the chain. They consist of an approximate ZM and \rev{an approximate-symmetry generator (SG)} 
and are canonically conjugate, despite macroscopic spatial separation\vspace*{-1.5mm}. 
\item A manifold of {\em degenerate quasi-steady states} manifest in the finite-size chains. Physically, they are displacements \rev{(generated by the SG)} of the unique steady state\vspace*{-1.5mm}. 
\item Both the \rev{ZM}, and the quasi-steady states persist in a transient dynamical regime whose duration diverges with system size. Further, their existence elicits divergent zero-frequency peaks in certain power spectra. 
\end{itemize}

\rev{Working within the aforementioned identification scheme for linking closed and dissipative many-body phenomena, the first three of these features (save for the split roles of SGs and ZMs) closely resemble the generic features of topologically non-trivial free fermionic systems subject to open boundary conditions. For instance, in addition to the mathematical similarities between the relevant edge modes, the joint presence of a steady state and a manifold of quasi-steady states of the third, are reminiscent of the (nearly) degenerate ground states found in, e.g., the Kitaev chain. 
A less-obvious similarity arises by analogizing the long, but \textit{finite} lifetimes of the ZM and the quasi-steady states with the (generically) exponentially small, but \textit{non-zero}, energies associated to Majorana fermions in \textit{finite} systems. Both phenomena are finite-size effects in nature, and may be related to one another by identifying the exponentially small energies in the fermionic case to the exponentially small decay rates in the bosonic case.}



\rev{Perhaps the most dramatic conceptual difference between topological fermions and topologically metastable bosons arises when one considers their bulk physics. As previously noted, for free fermions topological transitions are inextricably tied to a bulk quantum phase transition (without spontaneous symmetry-breaking). In particular, topological phases retain a unique (bulk) ground state. In sharp contrast, topologically metastable bosonic systems are necessarily bulk-unstable, despite remaining stable for all finite sizes. Consequently, they lack a bulk steady state altogether. This leads us to conclude that fair comparisons can only be made by considering \textit{finite} systems with \textit{open boundary conditions.} We summarize the main comparisons for systems lacking total number symmetry in Table \,\ref{CompTable}.}

\begin{table}[h!]
\begin{tabular}{|c|c|}
\hline\hline
\textbf{Topologically non-trivial QFH} & \textbf{Topologically metastable QBL} 
\\ \hline\hline
 \multicolumn{2}{|c|}{\em Edge modes (finite $N$)}
\\ \hline
Approximate ZM pairs & Approximate ZM and SG pairs
\\
Exponentially small energies & Exponentially small decay rates  
\\
Canonically conjugate & Canonically conjugate
\\
Hermitian & Hermitian
\\
Robust to weak perturbations & Robust to weak perturbations
\\ \hline
\multicolumn{2}{|c|}{\em ``Low-energy" boundary physics (finite $N$)}
\\ \hline
(Nearly) degenerate ground states & Manifold of quasi-steady states
\\ \hline
\multicolumn{2}{|c|}{\em ``Low-energy" boundary physics ($N\to\infty$)}
\\ \hline
Degenerate ground states & No steady states, unstable 
\\ \hline 
\multicolumn{2}{|c|}{\em Bulk physics ($N\to\infty$)}
\\ \hline
Non-zero bulk invariant & Non-zero bulk invariant 
\\
Unique ground state & No steady state, unstable
\\
\hline\hline
\end{tabular}
\caption{\rev{Comparisons between topologically non-trivial QFHs and topologically metastable QBLs, both in the absence of total number symmetry. The first five rows correspond to properties of the edge modes that arise in finite systems with open boundary conditions. The next two rows correspond to ``low-energy" (ground/steady state) features for a system with 2 ($N$ finite) or 1 ($N\to\infty$) boundaries. The last two rows correspond to bulk (i.e., boundary-less) properties.}} \label{CompTable}
\label{Realm}
\end{table}

\subsection{Outline and summary of main results}

This paper solidifies and expands greatly on the above core ideas. From a technical standpoint, a main tool for investigation is {\em pseudospectral theory} \cite{TrefethenPS,BottcherToe}, which provides the appropriate mathematical framework for convergence and stability analysis in the presence of {\em non-normal} dynamical generators. In terms of general results, we identify four major ones:

\begin{itemize}
\item We provide a self-contained proof that there exists a \textit{canonical} correspondence between ZMs and a class of linear symmetry generators in QBLs, despite the explicit breakdown of Noether's theorem in the dissipative Markovian setting. That is, a {\em partial restoration} of Noether's theorem is afforded by QBLs\vspace*{-1.5mm}. 

\item We introduce and apply two {\em design principles} for engineering QBLs with desired features. The first constitutes an explicit mapping from topologically non-trivial QFHs to QBLs that host MBs. The second provides a reservoir-engineering scheme for designing a QBL that relaxes to the quasiparticle vacuum of a given QBH\vspace*{-1.5mm}. 

\item We uncover the existence of tight bosonic analogues of the Dirac-fermion edge modes characteristic of topological insulators in QBLs that possess total-number symmetry. These bosonic modes, which we call \textit{Dirac bosons}, are tightly connected to two pairs of MBs and, as such, possess many of the same notable properties\vspace*{-1.5mm}. 

\item We demonstrate a strong connection between topological metastability and the existence of arbitrarily long-lived \textit{two-time quantum correlation functions}. Specifically, the macroscopically separated MBs show such long-lived correlations.
\end{itemize}

To illustrate the consequences of these general results, we provide several representative models -- beyond the dissipative bosonic Kitaev chain we introduced in Ref.\,\onlinecite{Bosoranas} and we also revisit here. The first new model is a {\em purely dissipative} ($H=0$) chain that descends from the fermionic Kitaev chain and most closely mirrors the purely dissipative setting considered for \rev{quadratic fermionic Lindbladians (QFLs)} \cite{DiehlTopByDiss,BardynTopByDiss}. This model demonstrates that a coherent, Hamiltonian contribution is {\em not} needed for topological metastability. It also possesses a new type of MBs, whereby both members of the pairs are approximate ZMs \textit{and} generate an approximate symmetry. We call these \textit{non-split} MBs. The second new model consists of the bosonic Kitaev chain Hamiltonian \cite{ClerkBKC,Decon} with a specially engineered dissipator that ensures relaxation to a {\em pure} SS. Remarkably, while the dissipator explicitly breaks translation invariance, a {\em restricted} translation invariance is retained, allowing for relevant spectral properties to be computed. Purity of the SS grants us analytical access to the exact dynamics of the quasi-steady states; in particular, these may exhibit highly non-trivial bosonic parity dynamics, with transient {\em odd bosonic parity}, in an appropriate sense. The third new model is a number-symmetric chain that possesses the aforementioned Dirac bosons. Being this the first example exhibiting such modes, we study their algebraic and dynamical properties in great detail. 

In more detail, the content is organized as follows. In Sec.\,\ref{BGSec}, we first establish relevant notation and second-quantization formalism in the closed-system setting of quadratic fermionic and bosonic Hamiltonians; we then move to Markovian quantum systems, and provide the necessary background about the Lindblad formalism and quadratic Lindbladians, with emphasis on symmetry properties; lastly, we summarize basic notions and results about pseudospectra. In Sec.\,\ref{QBLSec}, we present three main foundational results on QBLs: the correspondence between certain conserved quantities and symmetry generators, and two design protocols for generating QBLs with certain non-trivial features to be utilized in later examples, as mentioned above. In Sec.\,\ref{DynSec}, we introduce the notion of dynamical metastability in one-dimensional (1D) bulk-translationally invariant QBLs. We then focus on our main goal of exhibiting topologically nontrivial metastable dynamics, by synthesizing dynamical metastability and non-trivial bulk topology. We do so by separately discussing the case where number symmetry is broken due to the presence of ``bosonic pairing'' of either Hamiltonian or dissipative nature (Sec.\,\ref{TopoMS1}), and the case where number symmetry is unbroken (Sec.\,\ref{TopoMS2}). In particular, we present general results of, and several example models displaying Majorana and Dirac bosons, respectively. Observable signatures are explored in detail in Sec.\,\ref{MultiSec}, whereby multitime correlation functions and their associated power spectra are computed and analyzed in detail for several of our models. 

A number of additional results are included in separate appendixes. In particular, SPT phases of free fermions are further discussed in \rev{App.}\,\ref{QFHBG}, including a detailed analysis of the edge ZMs emerging in the topologically non-trivial regimes of the two paradigmatic Su-Schrieffer-Heeger and Kitaev chains.  In \rev{App.}\,\ref{Proofs} and \rev{App.}\,\ref{CatParApp}, we collect a number of proofs and explicit calculations supporting claims made in the main text. Finally, \rev{App.}\,\ref{DDW} presents a fourth new model which, unlike all models discussed in the main text, support {\em two} MBs of the same type (e.g., ZMs) on one edge in the topological regime and can thus be of independent interest.  

\subsection{Relation to existing work} 

With the program and our key results laid out, we wish to further place our contributions in the context of existing work. Firstly, concepts of metastability motivated by classical statistical physics have been extended to Markovian systems, and studied in great detail \cite{GarrahanMeta,MacieszczakMeta}. In essence, this form of metastability is characterized by multi-step relaxation, mandated by the presence of large gaps in the Lindblad spectra. While our {\em dynamically} metastable systems possess no such spectral gaps, it turns out they do exhibit \textit{pseudospectral gaps}. That is, the nontrivial pseudospectral (determined, as we will see, by the spectra under semi-infinite boundary conditions) remains gapped away from the exact spectrum in such a way to mandate anomalous transient dynamics (see Fig.\,Sec.\,\ref{DynSec} below). In fact, pseudospectra has been conjectured \cite{SatoAnomaly} to also play a role in the recent discoveries of anomalous relaxation dynamics in Markovian systems exhibiting a non-Hermitian skin-effect \cite{UedaSkin,MoriSlow}. They have further been successfully applied to study anomalous dynamics in random quantum circuits \cite{Marko1,Marko2,Marko3} and explicitly non-Hermitian Hamiltonians \cite{UedaTopoPhaseNH,SatoAnomaly}. 

A second branch of research which, while not motivated by the many-body physics of topological free fermions,  shares notable points of contact with our analysis, pertains to topological amplification \cite{PorrasTopoAmp,PorrasIO,NunnenkampTopoAmp,NunnenkampDisorder,NunnenkampRestore}. The standard approach to topological amplifiers employs input-output theory. Central to the input-output treatment is the {\em susceptibility matrix}, or Green's function, say, $\bm{\chi}(\omega)$, which connects incoming fields at frequency $\omega$ to outgoing fields at frequency $\omega$. For photonic systems \rev{with} $N$ modes, whose coherent and incoherent dynamics can be cast in the form of a Lindblad master equation, this susceptibility matrix takes the form $\bm{\chi}(\omega) = i(\omega\mathds{1}_{2N}-\mathbf{G})^{-1}$, with $\mathbf{G}$ being the dynamical matrix we will extensively discuss in Sec.\,\ref{QBLs}. It is then known \rev{from the above works} that topological amplification can \rev{{\em only}} take place if $\bm{\chi}^{-1}(\omega)$ winds around the origin, in a suitable sense. 

As it turns out, in our language this is, in the simplest case, equivalent to nontrivial winding of a certain spectral, ``rapidity band'' about the point $i\omega$. According to our pseudospectral approach, this implies that there exist pseudonormal modes with pseudoeigenvalues $i\omega+\lambda$, with $\lambda>0$. Such pseudonormal modes must necessarily amplify in the transient, and hence contribute to gain in the output signal. Moreover, we see that when these systems amplify zero-frequency ($\omega=0$) signals, they must possess MBs and all of the associated non-trivial transient dynamics they entail. To summarize, in our framework, general topological amplifiers are classified as dynamically metastable, while those that amplify zero-frequency input signals are classified as topologically metastable. We further observe that the introduction and application of the ``doubled matrix'' approach in Refs.\,\onlinecite{PorrasTopoAmp,PorrasIO} can be connected naturally to pseudospectral theory (see App.\,\ref{dblmatps}).

\section{Background}
\label{BGSec}

\subsection{Warm up: Quadratic bosonic and fermionic Hamiltonians}
\label{HamSec}

In the language of second quantization, a closed system of independent bosons (fermions) in equilibrium is described by a time-independent QBH (QFH). Despite the profound physical differences that exist between bosons and fermions, these Hamiltonians are diagonalized in essentially the same way. The key step is to solve a commutator equation involving the quadratic Hamiltonian of interest, to find the normal modes of the system. In this section, we summarize some machinery adapted to this task both for fermions and bosons, following exactly the same logic and comparing the mathematical structures that emerge. In doing so, we also introduce several concepts and notational conventions we will use in the more general setting of open quadratic dynamics in the next section and throughout this paper. 

Let \(c_i\), \(i=1,\dots,N\),  denote a set of fermionic annihilation operators satisfying the canonical anticommutation relations, $\{c_i,c_j^\dag\} = \delta_{ij}1_\mathcal{F}$ and $\{c_i,c_j\}=0$, with $1_\mathcal{F}$ the fermionic Fock space identity. A QFH is given by
\begin{equation*}
H_F = \frac{1}{2}\sum_{i,j=1}^N\left( \mathbf{K}_{ij}c_i^\dag c_j + \bm{\Delta}_{ij} c_i^\dag c_j^\dag + \text{H.c.}\right),
\end{equation*}
with $\mathbf{K}=\mathbf{K}^\dag$ the $N\times N$ \textit{hopping matrix} and $\bm{\Delta}^T = -\bm{\Delta}$ the $N\times N$ \textit{pairing matrix}. Such Hamiltonians are more compactly expressed in terms of the \textit{fermionic Nambu array},
\begin{equation*}
\Psi = [c_1,c_1^\dag,\ldots,c_N,c_N^\dag]^T.
\end{equation*}
It then follows that
\begin{equation*}
H_F = \frac{1}{2}\Psi^\dag \mathbf{H}_F\Psi,
\end{equation*}
where the $2N\times 2N$ \textit{Bogoliubov-de Gennes} (BdG) $\mathbf{H}_F$ is a block matrix with the $(ij)$-th block given by \cite{Ripka}
\begin{equation*}
[\mathbf{H}]_{ij} = \begin{bmatrix}
\mathbf{K}_{ij} & \bm{\Delta}_{ij} \\ -\bm{\Delta}_{ij}^* & -\mathbf{K}_{ij}^*
\end{bmatrix}.
\end{equation*} 
The stated conditions on $\mathbf{K}$ and $\bm{\Delta}$ imply that (i) the matrix $\mathbf{H}_F$ Hermitian; and (ii) $\mathbf{H}_F = -\bm{\tau}_1\mathbf{H}_F^T\bm{\tau}_1$ in terms of the matrices $\bm{\tau}_j \equiv \mathds{1}_N\otimes \bm{\sigma}_j$, with $\mathds{1}_N$ the $N\times N$ identity matrix and $\bm{\sigma}_j$, $j=1,2,3$, the usual Pauli matrices. 

Condition (ii) can be formalized in terms of a \textit{fermionic projector}, which we define according to
\begin{equation}
\mathcal{F}(\mathbf{M}) \equiv \frac{1}{2}\left( \mathbf{M} - \bm{\tau}_1\mathbf{M}^T\bm{\tau}_1\right),
\label{fermP}
\end{equation}
where $\mathbf{M}$ is any $2N\times 2N$ complex matrix. Property (ii) then says that fermionic BdG matrices are fixed points of this projection, i.e., $\mathcal{F}(\mathbf{H}_F) = \mathbf{H}_F$. While this explains the ``fermionic" moniker, the use of the word ``projector'' follows from the fact that $\mathcal{F}$, viewed as a linear map on the space of $2N\times 2N$ complex matrices, is idempotent: 
$\mathcal{F}^2=\mathcal{F}$. We may also define the complementary \textit{bosonic projector}, 
\begin{equation}
\mathcal{B}(\mathbf{M}) = \frac{1}{2}\left( \mathbf{M} + \bm{\tau}_1\mathbf{M}^T\bm{\tau}_1\right) = \mathbf{M} - \mathcal{F}(\mathbf{M}). 
\label{bosP}
\end{equation}
It is complementary in the sense that $\mathbf{M} = \mathcal{F}(\mathbf{M}) + \mathcal{B}(\mathbf{M})$. Moreover, these projectors are orthogonal in the sense that $\mathcal{B}(\mathcal{F}(\mathbf{M})) = 0=\mathcal{F}(\mathcal{B}(\mathbf{M}))$. Given a matrix $\mathbf{M}$, we call $\mathcal{F}(\mathbf{M})$ and $\mathcal{B}(\mathbf{M})$ its \textit{fermionic} and \textit{bosonic projections}. Fixed points of these projectors will be called \textit{fermionic} and \textit{bosonic matrices}, respectively. While the ``bosonic" moniker will be made clear later, it is natural to ask what happens if $\mathbf{H}_F$ has a non-zero bosonic projection. If this is the case, then 
\begin{align*}
H_F = \frac{1}{2}\Psi^\dag \mathbf{H}_F \Psi = \frac{1}{2}\Psi^\dagger \mathcal{F}(\mathbf{H}_F)\Phi - \frac{1}{2}\tr[\mathcal{B}(\mathbf{H}_F)]\,1_\mathcal{F}.
\end{align*}
Thus, each Hermitian fermionic matrix $\mathcal{F}(\mathbf{H}_F)$ defines an equivalence class of QFHs that differ by a constant shift. We eliminate this redundancy by requiring $\mathbf{H}_F$ be fermionic.

To diagonalize a QFH, one seeks a Bogoliubov transformation mapping the original degrees of freedom to a set of independent fermionic quasiparticles. That is, we search for a set of transformed fermionic operators $d_n$, $n=1,\ldots, N$ with
\begin{equation*}
[H_F, {\color{revred}d_n}] = -\epsilon_n d_n, 
\label{FermTr}
\end{equation*}
with $\epsilon_n\geq0$ the associated \textit{quasiparticle energies}. If $\ket{E}$ is an eigenstate of $H_F$ with energy $E$, then, if $d_n\ket{E}$ ($d_n^\dag \ket{E}$) is non-zero, it is a state with energy $E-\epsilon_n$ ($E+\epsilon_n$). That is, $d_n$ and $d_i^\dag$ annihilate and create a quasiparticle of energy $\epsilon_n$. It follows that the many-body ground state is the state annihilated by each $d_n$, and that eigenstates of $H_F$ are built up from it by populating quasiparticles states.  

A simple way to to find these quasiparticles is to introduce the \textit{fermionic hat map}. To each numerical vector $\vec{\alpha}\in\mathbb{C}^{2N}$, we associate a linear form
\begin{equation*}
\widehat{\vec{\alpha}} \equiv \vec{\alpha}^\dag \Psi = \alpha_1^*c_1 
+\alpha_{{2}}^{{\color{revred}*}} c_1^\dag + \cdots+\alpha_{{2N-1}}^{{\color{revred}*}}c_{N} +\alpha_{2N}^*c_N^\dag.
\end{equation*}
This map has two notable properties:
\begin{eqnarray}
{\widehat{\vec{\alpha}}}^\dag &=& \widehat{\bm{\tau}_1\vec{\alpha}^*},
\label{FermHatDag} \\
\{\widehat{\vec{\alpha}},\widehat{\vec{\beta}} \,\mbox{}^{\dag} \} &=& \vec{\alpha}^\dag \vec{\beta}\,1_{\mathcal{F}},\label{FermAntiComm}
\end{eqnarray}
with the right hand-side of the second equation being the linear form associated to a numerical vector $\bm{\tau}_1\vec{\alpha}^*$. With some algebra, one may further verify that
\begin{equation}
\label{HFcomm}
[H_F,\widehat{\vec{\alpha}}] = -\widehat{\mathbf{H}_F\vec{\alpha}}.
\end{equation}
If we then take $\vec{\alpha}=\vec{d}_n$, with $\vec{d}_n$ an eigenvector of $\mathbf{H}_F$ with eigenvalue $\epsilon_n$, it follows that 
\begin{equation*}
[H_F,\widehat{\vec{d}_n}] = -\widehat{\mathbf{H}_F\vec{d}_n} = -\epsilon_n\widehat{\vec{d}_n}.
\end{equation*}
The eigenvalue equation $\mathbf{H}_F\vec{d}_n = \epsilon _n \vec{d}_n$ provides the \textit{fermionic BdG equation} and resembles the time-independent Schr\"{o}dinger equation on a $2N$-dimensional Hilbert space.  

Now, since $\mathbf{H}_F$ is fermionic, it follows that if $\vec{d}_n$ is an eigenvector with eigenvalue $\epsilon_n$, then ${\vec{d}}'_n \equiv \bm{\tau}_1\vec{d}_n^*$ is an eigenvector with eigenvalue $-\epsilon_n$. We can then construct the quasiparticles as $d_n = \widehat{\vec{d}_n}$, with $\vec{d}_n$ being the $N$ orthonormal eigenvectors of $\mathbf{H}_F$ with eigenvalues $\epsilon_n\geq 0$. From Eq.\,\eqref{FermHatDag}, it follows that $\widehat{\vec{d}_n'} = d_n^\dag$, whereby orthonormality and Eq.\,\eqref{FermAntiComm} ensure the quasiparticles satisfy the canonical anti-commutation relations. 

Since the dynamics generated by $H_F$ is unitary, determining the evolution of $\Psi$ suffices to determine the evolution of any observable built up from products and sums of fermionic creation and annihilation operators. The Heisenberg equation of motion (in units where $\hbar=1$), 
\begin{eqnarray}
\frac{d}{dt}  \Psi (t) = i [H_F, \Psi(t)] = -i \mathbf{H}_F \Psi(t) , 
\label{eomf}
\end{eqnarray}
can be mapped to a linear time-invariant (LTI) dynamical system on \(\mathds{C}^{2N}\) by using the above hat map. Consider a general linear form $\widehat{\vec{\alpha}}$ in the Heisenberg picture and the Ansatz \cite{Decon}
\begin{equation*}
\widehat{\vec{\alpha}}(t) = \vec{\alpha}^\dag \Psi(t) = \vec{\alpha}^\dag(t)\Psi(0) \equiv \widehat{\vec{\alpha}(t)},
\end{equation*}
where the right hand-side is the linear form associated to the now \textit{time-dependent} coefficient vector $\vec{\alpha}(t)$. From Eq.\,\eqref{HFcomm}, it follows that $\widehat{\vec{\alpha}}(t)$ is a solution to Eq.\,\eqref{eomf} if and only if $\vec{\alpha}(t)$ is a solution to the LTI matrix equation
\begin{equation*}
\dot{\vec{\alpha}}(t) = i\mathbf{H}_F \vec{\alpha}(t).
\end{equation*}
The quasiparticles are then interpreted dynamically as \textit{normal modes} of this LTI system, i.e., $\vec{d}_n(t) = e^{i\epsilon_n t}\vec{d}_n(0)$. Since $\epsilon_n\ge 0$, this corresponds to {\em bounded motion} for all times. 

Let us now run through the exactly same ideas for bosons. Let \(a_i\), \(i=1,\dots,N\),  denote a set of bosonic annihilation operators satisfying the canonical commutation relations (CCRs), $[a_i,a_j^\dag] = \delta_{ij}1_\mathcal{F}$ and $[a_i,a_j]=0$, with $1_\mathcal{F}$ now being the bosonic Fock-space identity. A QBH is given by
\begin{equation}
H_B = \frac{1}{2}\sum_{i,j=1}^N\left( \mathbf{K}_{ij}a_i^\dag a_j + \bm{\Delta}_{ij} \rev{a_i^\dag a_j^\dag} + \text{H.c.}\right),
\end{equation}
with $\mathbf{K}=\mathbf{K}^\dag$ the $N\times N$ \textit{hopping matrix} and $\bm{\Delta}^T = \bm{\Delta}$ the $N\times N$ \textit{pairing matrix} (note that bosonic pairing matrices are symmetric, unlike fermionic ones which are antisymmetric). As with fermions, we define the \textit{bosonic Nambu array},
\begin{equation}
\Phi \equiv [a_1,a_1^\dag,\ldots,a_N,a_N^\dag]^T.
\label{bNa}
\end{equation}
It then follows that
\begin{equation}
H_B = \frac{1}{2}\Phi^\dag \mathbf{H}_B\Phi,
\end{equation}
where $\mathbf{H}_B$ is the $2N\times 2N$ block matrix with the $(ij)$-th block given by 
\begin{equation}
[\mathbf{H}]_{ij} = \begin{bmatrix}
\mathbf{K}_{ij} & \bm{\Delta}_{ij} \\ \bm{\Delta}_{ij}^* & \mathbf{K}_{ij}^*
\end{bmatrix}.
\end{equation} 
The stated conditions on $\mathbf{K}$ and $\bm{\Delta}$ imply that (i) the matrix $\mathbf{H}_B$ is Hermitian; and (ii) bosonic, i.e., $\mathbf{H}_B = \mathcal{B}(\mathbf{H}_B)$. Analogously to the fermionic case, any non-zero fermionic projection of $\mathbf{H}_B$ simply shifts the Hamiltonian according to
\begin{equation}
H_B = \frac{1}{2}\Phi^\dag \mathbf{H}_B \Psi = \frac{1}{2}\Phi^\dagger \mathcal{B}(\mathbf{H}_B)\Phi - \frac{1}{2}\tr[\bm{\tau}_3\mathcal{F}(\mathbf{H}_B)]\,1_\mathcal{F}.
\end{equation}
An equivalence class of QBHs is thus defined uniquely by a Hermitian, bosonic matrix $\mathbf{H}_B$. 

Diagonalization proceeds by searching for a Bogoliubov transformation to a set of bosonic creation and annihilation operators $b_n$, $n=1,\ldots,N$ satisfying the CCRs and
\begin{equation}
\label{BoseQuasi}
[H_B,b_n] = -\epsilon_n b_n, 
\end{equation}
for some $\epsilon_n\in\mathbb{R}$. In sharp contrast to their fermionic counterparts and Eq.\eqref{FermTr}, this is not always possible, however. To understand why, we define the \textit{bosonic hat map}. To each numerical vector $\vec{\alpha}\in\mathbb{C}^{2N}$, we associate a linear form
\begin{equation*}
\widehat{\vec{\alpha}} \equiv  \vec{\alpha}^\dag \bm{\tau}_3\Psi = \alpha_1^*a_1 
-\alpha_{{2}}^{{\color{revred}*}} a_1^\dag + \cdots+\alpha_{{2N-1}}^{{\color{revred}*}}a_{N} - \alpha_{2N}^*a_N^\dag, 
\end{equation*}
where $\bm{\tau}_3$ is included for later convenience. This map has two notable properties (cf. Eqs.\,\eqref{FermHatDag} and \eqref{FermAntiComm}):
\begin{eqnarray}
\widehat{\vec{\alpha}}\,\mbox{}^\dag &=& -\widehat{\bm{\tau}_1\vec{\alpha}^*},
\label{BoseHatDag}
\\
{[}\widehat{\vec{\alpha}},\widehat{\vec{\beta}} \, \mbox{}^\dag{]} &=& \vec{\alpha}^\dag\bm{\tau}_3 \vec{\beta}\,1_{\mathcal{F}},\label{BoseComm}
\end{eqnarray}
With some algebra, one may further verify that
\begin{equation}
\label{HBcomm}
[H_B,\widehat{\vec{\alpha}}] = -\widehat{\mathbf{G}\vec{\alpha}},\quad \mathbf{G} \equiv \bm{\tau}_3\mathbf{H}_B, 
\end{equation}
where we have introduced the bosonic BdG Hamiltonian $\mathbf{G}$. The hunt for quasiparticle excitations then requires solving the \textit{bosonic BdG equation}, $\mathbf{G}\vec{b}_n = \epsilon_n\vec{b}_n$. However, unlike the fermionic BdG Hamiltonian, $\mathbf{G}$ is non-Hermitian whenever bosonic pairing is present, $\bm{\Delta}\neq 0$. Instead, it is generally \textit{pseudo-Hermitian} \cite{MostafazadehIthruIII}, that is, $\mathbf{G}^\dag = \bm{\tau}_3\mathbf{G}\bm{\tau}_3$. Non-Hermiticity eliminates the guarantee that $\mathbf{G}$ has real eigenvalues, or even that it is diagonalizable at all. 

It turns out that the desired Bogoliubov transformation exists if and only if $\mathbf{G}$ is diagonalizable with an entirely real spectrum \cite{Decon}. In this case, there are a set of $N$ eigenvectors $\vec{b}_n$ of $\mathbf{G}$ satisfying (i) $\mathbf{G}\vec{b}_n = \epsilon_n \vec{b}_n$; and (ii) $\vec{b}_n^\dag \bm{\tau}_3\vec{b}_m = \delta_{nm}$. The linear forms $b_n \equiv \widehat{\vec{b}_n}$ then satisfy the CCRs and Eq.\,\eqref{BoseQuasi}. The remaining $N$ eigenvectors $\vec{b}_n'=\bm{\tau}_1\vec{b}_n^*$ correspond to the eigenvalues $-\epsilon_n$ and $\widehat{\vec{b}_n'} = -b_n^\dag$. Constructing the eigenstates of $H_B$ proceeds exactly as in the fermionic case with one notable exception. If there exists both a positive and a negative quasiparticle energy, then then $H_B$ is thermodynamically unstable (unbounded in both directions). Thus, there is no well-defined ground state (nor a well-defined Gibbs state, even if allowing for negative temperatures). Instead, the state annihilated by each $b_n$ is simply a {\em quasiparticle vacuum}.

As a consequence of the fact that $\mathbf{G}$ is, generically, no longer normal, the dynamical perspective for bosons is much richer than it is for fermions. While the dynamics generated by $H_B$ is still unitary at the many-body level (allowing us again to build up the dynamics of any observable from that of $\Phi$), the dynamics of the Nambu array is {\em effectively non-Hermitian}, that is,
\begin{eqnarray}
\frac{d}{dt}  \Phi (t) = i [ \rev{H_B}, \Phi(t)] = -i \mathbf{G} \Phi(t). 
\label{eomb}
\end{eqnarray}
Adopting the Ansatz $\widehat{\vec{\alpha}}(t) = \vec{\alpha}^\dag(t)\Phi(0)$, as we did in the fermionic case, now leads to the  LTI dynamical system
\begin{equation}
\dot{\vec{\alpha}}(t) = i\mathbf{G} \vec{\alpha}(t).
\label{LTI}
\end{equation}
If $\mathbf{G}$ is diagonalizable with entirely real spectrum, then the eigenvectors associated to the quasiparticles are normal modes of this system with $\vec{b}_n(t) = e^{i\epsilon_n t}\vec{b}_n(0)$. However, if $\mathbf{G}$ fails to meet these requirements, there will be (possibly generalized, in the sense of the Jordan canonical form) eigenvectors with possibly non-real eigenvalues.  In this case, $H_B$ is {\em dynamically unstable}, i.e., there exists observable expectation values that exhibit unbounded motion (see Ref.\,\onlinecite{Decon} for a detailed account of dynamical instabilities in QBHs). The normal mode picture for bosons is thus more general than the quasiparticle picture, as it persists even when the system is dynamically unstable. For this reason, it is more appropriate to refer to $\mathbf{G}$ as the \textit{dynamical matrix} (as opposed to the BdG Hamiltonian) of the QBH in the general case. 

\subsection{Open Markovian bosonic dynamics}

\subsubsection{Basics of Lindblad formalism}

Our focus in this paper will be on Markovian open quantum systems, whose dynamics are governed by a semigroup {\em Lindblad master equation} (LME) of the form $\dot{\rho}(t) = \mathcal{L}(\rho(t))$, with $\rho(t)$ a density operator describing the state of the system at time $t>t_0\equiv 0$ and the (time-independent) superoperator $\mathcal{L}$ -- the {\em Lindbladian} -- being the generator of the dynamics. Physically, a LME provides the most general form of continuous-time quantum dynamics that obeys complete positivity and trace preservation. In its canonical (or diagonal) form, the Markovian generator may be written as \cite{LindbladGenerators, Alicki-Lendi}
\begin{align}
\mathcal{L}(\rho) &= -i[H,\rho] +\sum_{\alpha}\Big( L_\alpha \rho L_\alpha^\dag-\frac{1}{2}\{L_\alpha^\dag L_\alpha,\rho\}\Big) \nonumber \\
&\equiv -i[H,\rho]+\sum_\alpha\mathcal{D}[L_\alpha](\rho),
\label{SLind}
\end{align}
where in second line we have defined the {\em dissipator} $\mathcal{D}[A]$. Here, $H=H^\dagger$ represents the Hamiltonian, coherent contribution to the dynamics, whereas $\{L_\alpha\}$, $\alpha=1,\ldots, d$, are the Lindblad (or ``jump'') operators, with each $\mathcal{D}[L_\alpha]$ phenomenologically accounting for a distinct, irreversible ``noise channel''. It is well known that the representation ${\cal L}(H, \{L_\alpha\})$ of the generator in terms of Hamiltonian and Lindblad operators is not unique, nor is the separation between a Hamiltonian and dissipative component \cite{Alicki-Lendi,TicozziMarkov,GoughNetwork}. In applications, the choice of a given representation is typically dictated by physical requirements; for instance, Lindblad operators are often naturally traceless. In a similar venue, it may be preferable to work in a non-diagonal representation, whereby both the Hamiltonian and the Lindblad operators are expressed in terms of a fixed set of operators, which may enjoy special mathematical (e.g., completeness) or physical (e.g., locality) properties. Specifically, let $\{A_j \}$, $j=1,\ldots,d'$, such a set of operators, with $L_\alpha \equiv \sum_{j} \ell_{j \alpha} A_j$,  $\ell_{j \alpha} \in {\mathbb C}$. The Gorini-Kossakowski-Lindblad-Sudarshan (GKLS) representation for the Markovian generator $\mathcal{L}$ then reads \cite{Gorini,Alicki-Lendi}
\begin{align}
\mathcal{L}(\rho) & = -i[H,\rho] +\sum_{j k} \mathbf{M}_{j k}\Big( A_k \rho A_j^\dag-\frac{1}{2}\{A_j^\dag A_k,\rho\}\Big)
\nonumber \\
&\equiv -i[H, \rho]+\sum_{j k} \mathbf{M}_{j k}\,\mathcal{D}[A_k,A_j^\dagger](\rho), 
\label{SGKLS}
\end{align}
where the square, positive-semidefinite (relaxation or GKLS) matrix $\mathbf{M}_{jk} = \sum_{\alpha=1}^d \ell_{j \alpha}^* \ell_{k \alpha}$ accounts for the non-unitary contribution to the dynamics, and we have introduced the shorthand notation $\mathcal{D}[A,B](\rho) \equiv A\rho B - \{BA,\rho\}/2$. Clearly, $\mathcal{D}[A,A^\dagger] = \mathcal{D}[A]$, as previously defined. 
 
In addition to the above Schr\"{o}dinger dynamics for the (not necessarily pure) state of the system, one can define an equivalent Heisenberg picture. As usual, all states are stationary in this picture and it is the operators associated to physical observables or auxiliary quantities (e.g., the electromagnetic potential) which evolve in time. An observable $B=B^\dagger$ then obeys the equation of motion $\dot{B}(t) = \mathcal{L}^\star(B(t))$, with the dual (Heisenberg) Markovian generator being given by 
\begin{eqnarray*}
\mathcal{L}^\star(B) &\equiv & i[H,B] + \sum_{\alpha}\Big(L_\alpha^\dag B L_\alpha - \frac{1}{2}\{L_\alpha^\dag L_\alpha,B\}\Big) \nonumber \\
&=& i[H,B] + \sum_{jk} \mathbf{M}_{j k} \Big(A_k^\dag B A_j- \frac{1}{2}\{A_k^\dag A_j,B\}\Big) ,
\end{eqnarray*}
in its canonical and GKLS form, respectively. As for unitary dynamics, the mathematical relationship between $\mathcal{L}$ and $\mathcal{L}^\star$ follows from demanding that expectation values agree in the two pictures at all times, that is, \(\tr(B(t) \rho) = \tr(B\rho(t))\), for all observables and density operators. Unlike the unitary case, however, Markovian evolution is {\em not} multiplicative in general: that is, generically, $(B_1 B_2)(t) \neq B_1(t) B_2(t)$. 

If the underlying Hilbert space $\mathcal{H}$ is finite-dimensional, say, of dimension $D$, the LME describes an LTI dynamical system whose generator $\mathcal{L}$ may be thought of as a (generally) non-Hermitian linear operator acting on a $D^2$-dimensional space. A number of general conclusions can then be made about the spectral properties of $\mathcal{L}$ and, in turn, the ensuing transient and asymptotic relaxation dynamics into the steady state manifold \cite{SchirmerMarkov,BaumGeneral}. In particular, the SS manifold always comprises at least one SS, say, $\rho_\text{ss}$, satisfying $\mathcal{L}(\rho_\text{ss})=0$. Furthermore, the spectrum of the Lindbladian, $\sigma(\mathcal{L})$, is bound to the closed left-half complex-plane, with any non-real eigenvalues existing in complex conjugate pairs. In the important situation where the only element in $\sigma(\mathcal{L})$ with vanishing real part is zero, the SS is unique, and arbitrary initial states $\rho_0$ asymptotically undergo exponential relaxation to $\rho_\text{ss}$, that is, $\rho_\text{ss}$ is globally asymptotically stable \cite{TicozziMarkov}.  Let the {\em spectral gap} (or dissipative gap) of $\mathcal{L}$ be defined by 
\begin{equation}
\Delta_\mathcal{L} \equiv |\sup\text{Re}[\sigma(\mathcal{L})\setminus\{0\}]| \geq 0,
\label{sgap}
\end{equation} 
that is, by the closest distance between the imaginary axis and the set of non-zero eigenvalues of $\mathcal{L}$. Then, $\Delta_\mathcal{L}$ sets the asymptotic decay rate through the inequality 
\[d_\text{max}(t) \equiv \sup_{\rho_0} \norm{e^{t\mathcal{L}}(\rho_0) - \rho_\text{ss}}_\text{tr}\leq K e^{-\Delta_\mathcal{L}t},  \quad K>0, \] 
where the trace distance $\norm{A}_\text{tr} \equiv \tr[\sqrt{A^\dag A}]$. The minimum time it takes for the above $d_\text{max}(t)$ to fall below a predetermined accuracy $\epsilon>0$ defines the {\em mixing time} of the Markovian semigroup, in direct analogy to classical Markov processes. Since the prefactor $K$ may be {\em a priori} very large, the mixing time plays a central role in characterizing the convergence behavior, by also accounting for the non-exponential pre-relaxation regime. Notably, dissipative ``quasi-free'' fermionic dynamics, described by \rev{QFLs}, are known to exhibit rapid convergence to stationarity \cite{TemmeHyper}, including via the occurrence of cutoff phenomena \cite{VernierCutoff}.

If $\mathcal{H}$ is infinite-dimensional, as it is necessarily the case for bosonic systems, a rigorous mathematical description becomes substantially more challenging and major departures from the above picture may arise. On the one hand, proving the existence of a generator of the form in Eq.\,\eqref{SLind} requires appropriate boundedness assumptions to hold; on the other hand, even when a Markovian generator can be shown or assumed to exist, the resulting Lindbladian may lack a SS, and relating spectral properties to dynamical evolution is far more involved in general \cite{Klausen}. Leaving mathematical rigor aside, LMEs have been remarkably successful in describing a wide class of open quantum systems, notably, across quantum optics and photonics \cite{GardinerNoise,OzawaPhotonics,NoriEPs}. Infinite dimensionality opens up the possibility for the Hamiltonian and other observables to be unbounded. As a consequence, ${\cal L^\star}$ may now obtain eigenvalues with {\em strictly positive} real part, e.g., in the case of optical pumping with $H=\omega a^\dag a$ and $L=\sqrt{\kappa}a^\dag$. With these caveats in mind, we now narrow our focus to the class of quadratic dissipative bosonic systems we will be specifically interested in throughout this work.

\subsubsection{Quadratic bosonic Lindbladians}
\label{QBLs}

As before, let $a_j^\dag$ and $a_j$ denote canonical creation and annihilation operators for bosons, arranged in a bosonic Nambu array $\Phi^\dagger$ as in Eq.\,\eqref{bNa}. The bosons can be regarded as quasi-free if the Lindblad generator is of the quadratic form
\begin{equation}
\mathcal{L}(\rho) = -i[H,\rho] + \sum_{i,j=1}^{2N}\mathbf{M}_{ji} \mathcal{D}(\Phi_i,\Phi_j^\dagger)(\rho),
\label{qLME}
\end{equation}
with $H=\frac{1}{2}\Phi^\dag \mathbf{H}\Phi $ being a QBH. We will call $\mathcal{L}$ (and its adjoint $\mathcal{L}^\star$) {\em quadratic bosonic Lindbladians} (QBLs). We note that a QBL so defined is not the most general instance of quasi-free bosonic dynamics, as it is also possible for $H$ to include a term linear in the creation and annihilation operators \cite{BarthelQuadLindblad}. In the context of continuous-variable quantum information, such quasi-free Lindblad generators are also often referred to as {\em Gaussian} \cite{GenoniGaussian}. Notably, Gaussian Markovian generators may be equivalently defined in terms of their preservation of the set of Gaussian states \cite{PolettiGaussian}. Owing to the non-uniqueness in representing the generator, linear Hamiltonian contributions can often be removed by way of an appropriate constant shift of the Lindblad operators \cite{TicozziMarkov}, hence the creation and annihilation operators. In any case, our primary focus here will be {\em purely quadratic} Gaussian generators, of the form given in Eq.\,\eqref{qLME}.

Gaussian LMEs are exactly solvable thanks to the fact that a quadratic Lindblad generator maps any polynomial in the creation and annihilation operators to another such polynomial of the same or lesser degree. As a rule, the dynamics of a form of odd (even) order depends on all the forms of odd (even) order of lesser degree. Hence, the linear forms satisfy a closed equation of motion and so do quadratic forms, up to a contribution proportional to the identity operator (the form of degree zero). To obtain a closed system of equations for cubic forms, one needs to consider cubic and and linear forms together, and so on. In short, one can map Gaussian LME to a hierarchy of linear ordinary differential equations. In this sense, a Gaussian LME is exactly solvable. However, fully taking advantage of these facts requires some machinery. 

Our discussion in Sec.\,\ref{HamSec} showed that fermionic projections $\mathcal{F}$ play no role in the dynamics of closed bosonic systems. Remarkably, dissipation brings them right back into the picture to play a very distinct role. Let $\mathcal{D}^\star$ denote the dissipator of a general QBL. Then, by writing $\mathbf{M}=\mathcal{B}(\mathbf{M})+\mathcal{F}(\mathbf{M})$ in Eq.\,\eqref{qLME}, and leveraging the CCRs, one can show that  
\begin{align*}
\mathcal{D}^\star(A) = \frac{1}{2}\sum_{i,j}\mathcal{B}(\mathbf{M})_{ij}\!\left(\left[[\Phi_i^\dag,A],\Phi_j \right]+ \left[\Phi_i^\dag,[A,\Phi_j]\right]\right)&
\\
+\,\mathcal{F}(\mathbf{M})_{ij}\!\left(\left\{[\Phi_i^\dag,A],\Phi_j \right\}+ \left\{\Phi_i^\dag,[A,\Phi_j]\right\}\right)&,
\end{align*}
with complete generality. Thinking of $A$ as a form of degree $d$, we see that the first term reduces the degree of $A$ by 2, while the second preserves the degree. Thus, the bosonic projection of $\mathbf{M}$ is responsible for connecting the degree-$d$ and degree-$(d-2)$ sectors of the system's operator algebra, while the fermionic projections leaves the degree-$d$ sector invariant. 

Finally, we are ready to present the equations of motion for linear and quadratic forms. Let \(Q\equiv \Phi\Phi^\dagger\) denote the square array of elementary quadratic forms \(a_ia_j\) and so on. Then, the Heisenberg equations of motion for arbitrary linear and quadratic form follow from the compact formulas 
\begin{eqnarray}
\left\{  \begin{array}{l}
\!\dot{\Phi}(t) = \mathbf{G}\Phi(t)\label{lineom} ,\\
\!\dot{Q}(t) = -i(\mathbf{G}Q(t) - Q(t)\mathbf{G}^\dag) + \bm{\tau}_3\mathbf{M}\bm{\tau}_3\,1_F , 
\label{quadeom}
\end{array}\right.
\end{eqnarray}
where 
\begin{equation}
\mathbf{G}\equiv \bm{\tau}_3\mathbf{H}-i\bm{\tau}_3\mathcal{F}(\mathbf{M})
\label{dynM}
\end{equation}
is the dynamical matrix associated to the QBL. Remarkably, the dynamics of a linear bosonic form is entirely controlled by a fermionic matrix in the case of pure dissipation ($\mathbf{H}=0)$. For future reference, we note the relationships 
\begin{equation}
\label{QBLSP}
\mathcal{L}^\star(\widehat{\vec{\alpha}}) = \widehat{i\widetilde{\mathbf{G}}\vec{\alpha}},\quad \widetilde{\mathbf{G}} \equiv \bm{\tau}_3 \mathbf{G}^\dag \bm{\tau}_3.
\end{equation}
It follows that the dynamics generated by $\mathcal{L}^\star$, restricted to the subalgebra of linear forms, are encoded by an associated LTI system within the BdG space $\mathbb{C}^{2N}$ generated by $i\widetilde{\mathbf{G}}$. That is, 
\begin{equation}
\widehat{\vec{\alpha}}(t) = \widehat{\vec{\alpha}(t)} \Longleftrightarrow\dot{\vec{\alpha}}(t) = i\widetilde{\mathbf{G}}\vec{\alpha}(t).
\end{equation}
Notice that $\widetilde{\mathbf{G}} = \mathbf{G}$ if and only if $\mathcal{D}=0$ (cf. Eq.\,\eqref{LTI}). 

Since the bosonic Fock space is infinite-dimensional, a QBL can become dynamically unstable. Then, by definition, there exists an observable $A$ and a state $\rho$ such that $\braket{A\,}\!(t)$ is unbounded in time. As a rule, dynamical instability can be diagnosed through the dynamical matrix by way of the {\em rapidity spectrum}, that is, $\sigma(-i\mathbf{G})$. In addition to the spectral gap $\Delta_{\cal L}$ of Eq.\,\eqref{sgap}, the \textit{stability gap} is defined by  
\begin{equation}
\Delta_S \equiv \max\text{Re}\,\sigma(-i\mathbf{G}).
\label{StGap}
\end{equation}
If $\Delta_S<0$ ($\Delta_S>0$, then the QBL is dynamically stable (unstable). The crossover regime $\Delta_S=0$ can be either stable or unstable contingent on the existence of a steady state \cite{BarthelQuadLindblad}. If the QBL is dynamically stable, then the stability gap determines the spectral gap of $\mathcal{L}$ by way of the simple formula $\Delta_\mathcal{L} = -\Delta_S$ \cite{Prosen3QBoson}.
Moreover, there exists a unique, globally attractive, and Gaussian $\rho_{ss}$. Its first and second moments are stationary solutions to Eqs.\,\eqref{lineom}, that is, 
\begin{eqnarray}
\left\{ 
\begin{array}{l}
\mathbf{G} \braket{\Phi}_{ss}=0,\\
 -i\mathbf{G}\!\braket{Q}_{ss}+i\braket{Q}_{ss}\!\mathbf{G}^\dagger + \bm{\tau}_3\bf{M}\bm{\tau}_3=0.
 \end{array} \right.
 \end{eqnarray}
The condition of a strictly negative stability gap identifies $-i\mathbf{G}$ as a {\em Hurwitz matrix} in the language of matrix analysis. Then, uniqueness of the SS follows from the uniqueness of the solutions of the above two equations when $-i\mathbf{G}$ is Hurwitz \cite{Prosen3QBoson}. In particular, $\braket{\Phi}=0_{ss}$ necessarily. If $\Delta_S\geq 0$, there can be either zero or infinitely many SSs \cite{BarthelQuadLindblad}. Since \(\braket{\mathcal{F}(Q)} = \bm{\tau}_3/2\) automatically in any state, one can focus on $\braket{B(Q)}_{ss}$ which, after an appropriate basis transformation, is unitarily equivalent to the SS covariance matrix \cite{KrugerGaussian}.

\vfill

\subsubsection{Symmetries, conserved quantities, 
and 
\\
bulk-translationally invariant QBLs} 

\label{BTIQBLs}

Just as in the closed-system case, symmetries and conserved quantities provide crucial information about the system. While the notion of a \rev{\em conserved quantity}, i.e., an observable $Q$ such that $\dot{Q}(t)=0$, has a straightforward characterization ($\mathcal{L}^\star(Q)=0$), two distinct notions of symmetry emerge in the Markovian setting \cite{ProsenSpin,VictorSymCQ,BaumGeneral}. A unitary or anti-unitary operator $S$ is a \textit{weak symmetry} of the dynamics if, for all $\rho$, $(S\rho S^{-1})(t) = S\rho(t)S^{-1}$ or, equivalently, $[\mathcal{L},\mathcal{S}]=0$, with $\mathcal{S}(\rho)\equiv S \rho S^{-1}$. Further, $S$ is a \textit{strong symmetry} if $[H,S]=0$ and $[L_{\color{revred}\alpha},S]=0$, for all ${\color{revred}\alpha}$. \rev{For simplicity, unless otherwise noted, in this work we will generically refer to a weak symmetry as just a symmetry. }

If the dynamics possesses a {\em continuous} unitary family of strong symmetries $S(\theta)$, the existence of a corresponding conserved quantity is guaranteed. In fact, the conserved quantity $Q$ coincides with the generator of the one-parameter group $S(\theta) = e^{i\theta Q}$. This fact generalizes naturally to higher-dimensional Lie groups and is reminiscent of Noether's theorem, which establishes the existence of conserved quantities given a continuous symmetry. However, in sharp contrast to the Hamiltonian case, (i) the existence of a continuous weak symmetry does not imply the existence of a conserved quantity; and (ii) the existence of a conserved quantity does not imply the existence of either a weak \textit{or} strong continuous symmetry. In this sense, we say the quantum Hamiltonian form of Noether's theorem breaks down in the presence of Markovian dissipation. We further note that $S(\theta) = e^{i\theta Q}$ is a weak symmetry of the dynamics if and only if
\begin{equation}
\label{GeneralSG}
\mathcal{L}^\star([Q,A]) - [Q,\mathcal{L}^\star(A)] = 0, \qquad \forall A, 
\end{equation} 
that is, $\mathcal{L}^\star$ commutes with the adjoint action of $Q$.

The class of QBLs we consider are further characterized by (weak) discrete {\em bulk-translation symmetry} in 1D. Here, bulk-translation invariance refers to the presence of translation invariance when suitably far from a well-defined boundary. Specifically, bulk-translationally invariant QBLs may correspond to one of four main configurations\vspace*{0mm}: 
\begin{itemize}
\item A finite number of lattice sites on a chain (open boundary conditions, OBCs)\vspace*{-1.5mm}; 
\item A finite number of lattice sites on a ring (periodic boundary conditions, PBCs)\vspace*{-1.5mm};
\item An infinite number of lattice sites extending only in one direction (semi-infinite boundary conditions, SIBCs)\vspace*{-1.5mm};
\item An infinite number of lattice sites extending in two directions (bi-infinite boundary conditions, BIBCs).
\end{itemize} 
All the Hamiltonian and dissipative couplings are assumed to be independent of lattice site, modulo BCs, and of a {\em finite range} $R$, with $R \ll N$ \cite{GBTJPA}. In this way, the equation of motion for operators at site $j$ only involve operators at most a distance $R$ away, e.g., $R=1$ for  nearest-neighbor (NN) hopping. To each lattice site $j$, let us associate $d_{\text{int}}$ bosonic degrees of freedom arranged in a site-local Nambu array, $\phi_j \equiv  [a_{j,1},a_{j,1}^\dag,\ldots,a_{j,d_{\text{int}}},a_{j,d_{ \text{int}} }^\dag]^T$. The Hamiltonian contribution is then constrained to a particular simple form, 
\begin{equation*}
H = \frac{1}{2}\sum_{j}\sum_{r=-R}^R \phi_j^\dag \mathbf{h}_r \phi_{j+r},
\end{equation*}
with $\mathbf{h}_r$ a $2d_{\text{int}} \times 2d_{\text{int}}$ matrix that encodes the coherent hopping and pairing mechanisms between sites $j$ and $j+r$. Necessarily, $\mathbf{h}_r^\dag=\mathbf{h}_{-r}$ and $\mathbf{h}_r^* = \bm{\tau}_1\mathbf{h}_r\bm{\tau}_1$. In the above expression, the sum over $j$ encodes the BCs. Likewise, the Heisenberg-picture dissipator must take the form
\begin{equation*}
\mathcal{D}^\star(A) = \sum_j \sum_{r=-R}^R \Big( \phi_j^\dag A \mathbf{m}_r \phi_{j+r} - \frac{1}{2}\Big\{\phi_j^\dag \mathbf{m}_r \phi_{j+r},A\Big\} \Big), 
\end{equation*}
with $\mathbf{m}_r$ a $2d_{\text{int}} \times 2d_{\text{int}}$ matrix that encodes the incoherent/dissipative damping, pumping, and pairing mechanisms between sites $j$ and $j+r$. Necessarily, $\mathbf{m}_r^\dag = \mathbf{m}_{-r}$, and again, the $j$-sum encodes the BCs.  

Working with a fixed number of lattice sites $j=1,\ldots,N$, the dynamical matrix and the GKLS matrix are then given by
\begin{eqnarray*}
\mathbf{G}_N &=& \mathds{1}_N\otimes \mathbf{g}_0 +\sum_{r=1}^R \mathbf{S}_N^r\otimes \mathbf{g}_r+\mathbf{S}_N^\dag{}^r\otimes \mathbf{g}_{-r},
\\
\mathbf{M}_N &=& \mathds{1}_N\otimes \mathbf{m}_0+\sum_{r=1}^R \mathbf{S}_N^r\otimes \mathbf{m}_r+\mathbf{S}_N^\dag{}^r\otimes \mathbf{m}_{-r}, 
\end{eqnarray*}
in terms of the BC-dependent left-shift operator
\begin{equation*}
\mathbf{S}_N \equiv \sum_{j=1}^{N-1}\vec{e}_j\vec{e}_{j+1}^{\,\dag}+ \begin{cases}
0,& \text{OBCs},
\\
\vec{e}_N\vec{e}_{1}^{\,\dag},&\text{PBCs},
\end{cases}
\end{equation*}
with $\vec{e}_j$ the $j$'th canonical basis vector of $\mathbb{C}^N$. When focusing exclusively on OBCs (PBCs), we will usually denote $\mathbf{S}_N=\mathbf{T}_N$ ($\mathbf{V}_N$). By using Eq.\,\eqref{dynM}, the internal coupling matrices of the dynamical matrix are identified by letting
\begin{equation*}
\mathbf{g}_r \equiv \bm{\tau}_3\mathbf{h}_r -\frac{i}{2}\bm{\tau}_3(\mathbf{m}_r - \bm{\tau}_1\mathbf{m}^*_{r}\bm{\tau}_1),
\end{equation*}
with $\bm{\tau}_j=\mathds{1}_{d_{\text{int}}}\otimes \bm{\sigma}_j$ in this context. Notably, we have $\mathbf{g}_r^* = -\bm{\tau}_1 \mathbf{g}_r \bm{\tau}_1$. Under OBCs (PBCs), the matrices $\mathbf{h}_r, \mathbf{m}_r$ belong to the class of banded block-Toeplitz (banded block-circulant) matrices \cite{GBTJPA}. We label semi-infinite and bi-infinite lattices according to $j=1,2,\ldots,\infty$ and $j=0,\pm 1,\pm2,\ldots,\pm\infty$, respectively. In the case of SIBCs (BIBCs), the dynamical matrices belong to the classes of banded block-Toeplitz (banded block-Laurent) operators. For OBCs, PBCs, SIBCs, and BIBCs, we denote the matrices $\mathbf{X}=\mathbf{G},\mathbf{M}$ by $\mathbf{X}^\text{OBC}_N$, $\mathbf{X}^{\text{PBC}}_N$, $\mathbf{X}^\text{SIBC}$, and $\mathbf{X}^{\text{BIBC}}$, respectively (see Table\,\ref{BCTable}). When the particular BC is unimportant, or when we wish to refer to the full family of configurations, we simply write $\mathbf{X}$. 
\begin{table}
\begin{tabular}{|c|c|}
\hline\hline
{\bf Boundary conditions} & {\bf Dynamical/GKLS matrix type}
\\
\hline\hline
Open & Banded block-Toeplitz matrix
\\
\hline
Periodic & Banded block-circulant matrix
\\
\hline
Semi-infinite & Banded block-Toeplitz operator
\\
\hline 
Bi-infinite & Banded block-Laurent operator
\\
\hline
\end{tabular}
\caption{The correspondence between BCs and the structure of the dynamical and GKLS matrices 
$\mathbf{G}$ and $\mathbf{M}$ in bulk-translationally invariant QBLs.}
\label{BCTable}
\end{table}

In the translationally invariant cases (PBCs and BIBCs), we may introduce the conserved crystal momenta $k$ through Fourier modes, $b_{k,m} \equiv \sum_{j} e^{-ijk}a_{j,m}$. Here, $k$ takes on either discrete (PBCs) or continuous (BIBCs) values in the Brillouin zone $[-\pi,\pi]$. The dynamics of the Fourier modes, organized again in the form of a Nambu array, $\widetilde{\phi}_k \equiv [b_{k,1},b_{-k,1}^\dag,\ldots,b_{k,d_{\text{int}}},b_{-k,d_{\text{int}}}^\dag]^T$, are then governed by a $2d_{\text{int}}\times 2d_{\text{int}}$ $k$-dependent dynamical matrix $\mathbf{g}(k)$, that is, 
\begin{equation}
 \dot{ {\widetilde{\phi}}}_k (t) = -i\mathbf{g}(k) \widetilde{\phi}_k (t)
,\quad \mathbf{g}(k) \equiv  \sum_{r=-R}^R \mathbf{g}_r e^{ikr} .
\label{transl}
\end{equation}
We call $\mathbf{g}(k)$ the {\em Bloch dynamical matrix} to draw analogy with the Bloch Hamiltonian of condensed-matter systems. Similarly, we define the \textit{rapidity bands}, which are the dissipative generalization of energy bands, to be the eigenvalues of $-i\mathbf{g}(k)$. These provide the rapidities of the translationally invariant configurations PBCs and BIBCs. Consistent with the known symmetries of Lindbladian eigenvalues, the property $\mathbf{G}^* = -\bm{\tau}_1\mathbf{G}\bm{\tau}_1$ (or, equivalently, $\mathbf{g}(k)^* =-\bm{\tau}_1\mathbf{g}(-k)\bm{\tau}_1$ with $\bm{\tau}_1$ understood here to be $2d\times 2d$) ensures that if $\lambda(k)$ is a rapidity band, then so is $\lambda(k)^*$.

Thus, by exploiting translational invariance, a simple description of the rapidities for PBCs and BIBCs is possible, and solving for the dynamics of a large class of operators of interest amounts to solving the finite-dimensional LTI system in Eq.\,\eqref{transl}. The dynamics of QBLs for which translational symmetry is broken by OBCs or SIBCs are considerably more difficult to describe. While the spectral theory of block-Toeplitz matrices and operators is well-established in the mathematical-physics literature \cite{GBTJPA, TrefethenPS, BottcherToe}, it turns out that spectral properties are not, in general, the appropriate tool to consider, due to the fact that the relevant dynamical matrices need not be Hermitian or even normal. 

\subsection{The pseudospectrum}
\label{SPSToe}

Pseudospectral theory is an extension of spectral theory adapted to non-normal matrices and operators. As our work will illustrate, it is a powerful alternative to the traditional spectral analysis in terms of eigenvalues and invariant subspaces, because it can predict approximate dynamical {\em transient} behavior that is hard to identify directly from the exact normal modes of the dynamical system. In addition, the notion of the pseudospectrum can accommodate  the idea of an ``approximate mode'' in a mathematically sharp framework. 

Let $\bfX$ be an $n\times n$ complex matrix. A complex number $\lambda$ is in the spectrum of $\bfX$, $\lambda \in \sigma(\bfX)$, if $\bfX-\lambda \id_n$ is not invertible. To motivate the notion of the pseudospectrum, let us recast this definition in an equivalent way: if $\{\lambda_i\}$ is a a sequence converging to $\lambda$, and $\bfX-\lambda_i$ is invertible for all $i$, then $\lambda\in\sigma(\bfX)$ if $\norm{(\bfX-\lambda_i)^{-1}}\to\infty$ for some matrix norm. The choice of norm is of no consequence if ${\cal H}$ is finite-dimensional, because all norms are equivalent (induce the same topology). This definition of the spectrum suggests a natural generalization. Given some fixed matrix norm $\norm{\cdot}$ and $\epsilon>0$, the $(\epsilon,\norm{\cdot})$-{\em pseudospectrum} of \(\bfX\) is defined as \begin{equation}
\label{PSnorm}
\sigma_{\epsilon,\norm{\cdot}}(\bfX) \equiv \set{\lambda\in\mathbb{C} : \norm{(\bfX-\lambda\id_n)^{-1}}>1/\epsilon},
\end{equation}
with the understanding that $\norm{(\bfX-\lambda)^{-1}} = \infty$ for $\lambda\in\sigma(\bfX)$; that is, the spectrum is always a subset of the pseudospectrum. The elements of the pseudospectrum are the $(\epsilon,\norm{\cdot})$-eigenvalues, or {\em pseudoeigenvalues} of \(\bfX\). If $\norm{\cdot}$ is induced by a vector norm, that is, $\norm{\bfX}\equiv \text{sup}_{\norm{\vec{v}}=1} \norm{\bfX \vec{v}}$, we have the equivalent (and more useful for our applications) definition:
\begin{equation}
\label{PSvec}
\sigma_{\epsilon,\norm{\cdot}}(\bfX) = \set{\!\lambda\in\mathbb{C} : \exists\vec{v}\in\mathbb{C}^n,\,\norm{\vec{v}}=1,\,\norm{(\bfX-\lambda)\vec{v}}<\epsilon}.
\end{equation}
The normalized vectors $\vec{v}$, with $\norm{(\bfX-\lambda)\vec{v}}<\epsilon$, are the $(\epsilon,\norm{\cdot})$-{\em pseudoeigenvectors} associated to the $(\epsilon,\norm{\cdot})$-eigenvalue $\lambda$. When  $\epsilon$, $\norm{\cdot}$ are either understood or inconsequential, we will drop one or both of them in our notations.

The pseudospectrum becomes increasingly relevant as the matrix $\mathbf{X}$ becomes highly nonnormal. One reason is the way it relates to perturbations. It follows from the definition that if $\lambda\in\sigma_{\epsilon,\norm{\cdot}}(\bfX)$, then there exists a perturbation $\bfE$ of size $\norm{\bfE}<\epsilon$, such that $\lambda\in\sigma(\bfX+\bfE)$ \cite{TrefethenPS}. Hence,
\begin{equation}
\label{PSpert}
\sigma_{\epsilon,\norm{\cdot}}(\bfX) = \bigcup_{\bfE:\,\norm{\bfE}<\epsilon} \sigma(\bfX+\bfE) .
\end{equation}
Another reason can be seen simply by considering the case where $\norm{\cdot}=\norm{\cdot}_2$ is the matrix norm induced by the standard (Euclidean) vector $2$-norm. From Eq.\,\eqref{PSpert}, a perturbation of size $\epsilon$ can only shift the spectrum of a normal matrix by at most $\epsilon$. However, if $\bfX$ is nonnormal, small perturbations can drastically modify the spectrum -- one manifestation of this being the non-Hermitian skin-effect (NHSE) \cite{YaoSkin,SlagerTopoSkin,SatoTopoSkin}. In this case, the pseudospectra can dramatically influence the transient dynamics of an LTI system generated by $\bfX$, e.g., $\dot{\vec{v}} = \bfX \vec{v}$. In particular, the pseudospectra bounds the maximal dilation of norm in the sense of 
\begin{align}
\label{pab}
\sup_{t\geq 0}\norm{e^{t\bfX}} \geq \frac{\alpha_\epsilon(\bfX)}{\epsilon},\quad 
\alpha_\epsilon(\bfX) \equiv \sup\text{Re}\,\sigma_\epsilon(\bfX).
\end{align}
The quantity $\alpha_\epsilon$ is called the {\em pseudospectral abscissa} and measures the extent to which the $\epsilon$-pseudospectrum of $\bfX$ extends towards, or into, the right-half complex plane \cite{TrefethenPS}. 

The bound of Eq.\,\eqref{pab} is particularly relevant for Hurwitz matrices, that is, matrices with spectrum bounded in the left-half complex plane. If $\alpha_\epsilon(\bfX)$ is positive and large compared to $\epsilon$, $\norm{e^{t\bfX}}$ will experience {\em transient growth} before asymptotically decaying to zero. That is, highly non-normal, but asymptotically stable, dynamical systems can appear unstable during a transient period. In fact, given a particular $\epsilon$-pseudoeigenvector \rev{$\vec{v}$} with $\epsilon$-\rev{pseudoeigenvalue} $\lambda$, we have
\begin{equation}
\label{pseudomode}
\!\!\norm{e^{t\bfX}\vec{v}-e^{\lambda t}\vec{v}} \leq \norm{(\bfX-\lambda)\vec{v}}t + \mathcal{O}(t^2) < \epsilon t +\mathcal{O}(t^2),
\end{equation}
so that $\vec{v}$ evolves like a normal mode with eigenfrequency $\lambda$ for sufficiently small (set by $\epsilon$) timescales. We call such modes \textit{pseudonormal modes}. We note that $2$-norm pseudospectra can be directly related to singular values, as we briefly discuss in App.\,\ref{dblmatps}.

The theory of pseudospectra is especially well-developed and useful for the Toeplitz matrices and operators which, as we have seen, describe bulk-translationally invariant QBLs. While this theory applies generally to block-Toeplitz matrices \cite{TrefethenPS,BottcherToe}, we recount here only the non-block case for simplicity. Let $\mathbf{X}_N^\text{OBC}$, denote a banded, $N\times N$ Toeplitz matrix, $\mathbf{X}^\text{SIBC}$ its associated Toeplitz operator, and $\mathbf{X}^\text{BIBC}$ its associated Laurent operator, respectively. Explicitly,
\begin{equation*}
\mathbf{X}_N^\text{OBC} = x_0 \mathds{1}_N + \sum_{r=1}^R (x_r \mathbf{T}_N^r + x_{-r} \mathbf{T}_N^\dag{}^r), \quad x_0,x_{\pm r} \in {\mathbb C} ,
\end{equation*}
with $R$ the range. With this, we may define the complex-valued {\em symbol} $x(z) \equiv \sum_{r=-R}^R x_r z^r$, with $z\in\mathbb{C}$. Upon comparison, it follows that $x(e^{ik})$ is precisely the non-block analogue of the Bloch dynamical matrix we introduced for translationally invariant QBLs. As in that case \footnote{To spell out the analogy further, one arrives at the Bloch dynamical matrix following a Fourier transform. Similarly, the action of $\mathbf{X}^\text{BIBC}$ on plane wave modes $\vec{f_k}=\sum_{j}e^{ijk}\vec{e}_j$ is characterized by $x(e^{ik})$ according to $\mathbf{X}^\text{BIBC}\vec{f}_k = x(e^{ik})\vec{f}_k$.}, one finds that $x(e^{ik})$ determines the spectrum of the Laurent operator $\mathbf{X}^\text{BIBC}$. The spectrum of $\mathbf{X}^\text{SIBC}$ may also be characterized as the spectrum of $\mathbf{X}^\text{BIBC}$, together with all complex numbers about which the symbol winds. Explicitly, let the integer-valued winding number of $x(e^{ik})$ about \rev{a complex number $\lambda$ not in the image of the unit circle under $x$, i.e.,}
$\lambda \not\in \{x(e^{ik}): k\in[-\pi,\pi]\}$, be given by
\begin{eqnarray*}
\nu(\lambda) & \equiv & \frac{1}{2\pi i}\oint_{|z|=1}\!\frac{x'(z)}{x(z)-\lambda}dz \\
& =& \frac{1}{2\pi i}\int_{-\pi}^\pi \!\frac{d}{dk}\ln[x(e^{ik}) - \lambda]\,dk.
\end{eqnarray*}
Then, if $\nu(\lambda)\neq 0$, $\lambda$ is in the spectrum of $\mathbf{X}^\text{SIBC}$. 

\begin{figure}[t]
\includegraphics[width=\columnwidth]{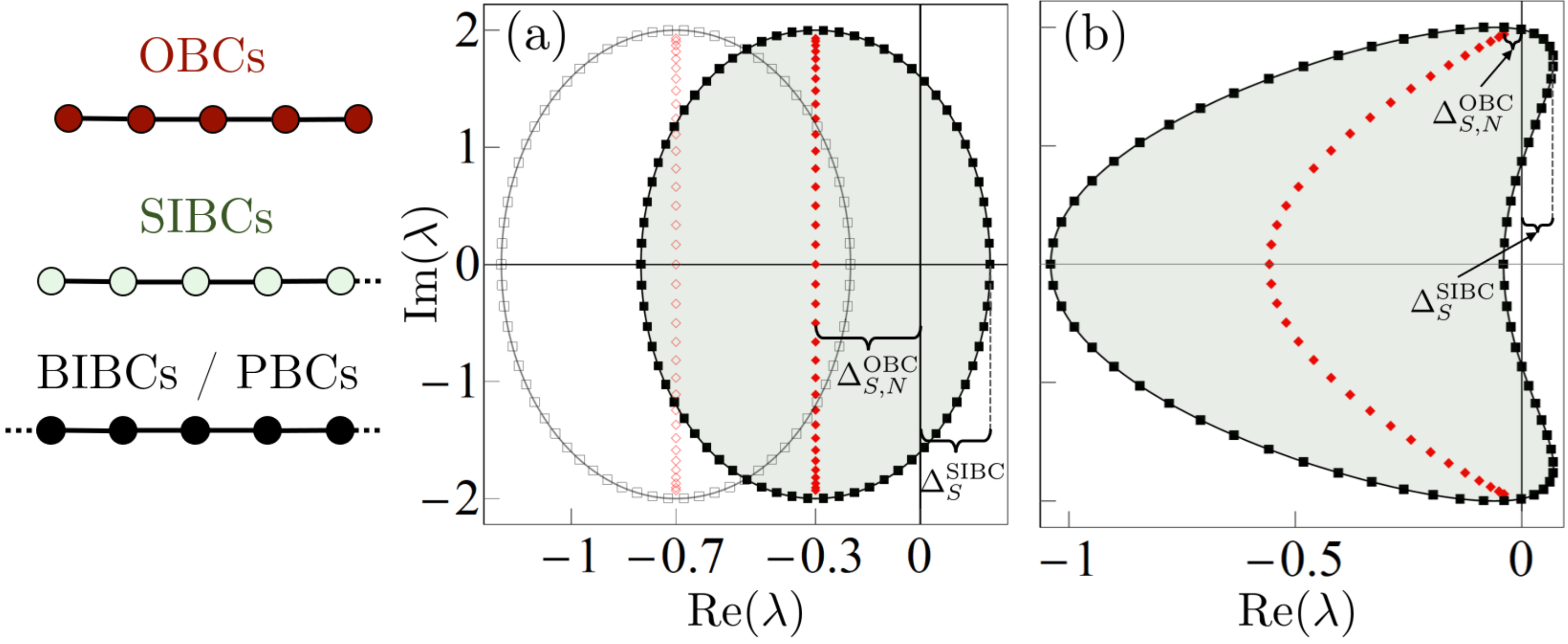}
\caption{Representative behavior of rapidity spectra under different BCs, for a QBL describing the dissipative bosonic Kitaev chain model of Sec.\,\ref{DBKCSec}. The solid curves give the bulk BIBC spectra, whereas the points along the curves are the rapidities for PBCs. The points on the vertical lines are the rapidities for OBCs. The shaded regions denote the SIBC spectra. 
{\bf (a)} The doubly-degenerate rapidity spectrum when $\Gamma=0$. The filled (open) markers represent the topologically metastable (anomalously relaxing) regime, with $\kappa/\Delta = 0.6$ $(1.4)$. 
{\bf (b)} The doubly-degenerate rapidity spectrum when $\Gamma=0.12$, so that the winding around $\lambda = 0$ is zero. This represents non-topological dynamical metastability. In all cases, $J=2$, $\Delta=0.5$, $\mu=0$, and $N=25$. Adapted from Ref.\,\cite{Bosoranas}.
}
\label{DBKCrapidities}
\end{figure}

Finding a general and sharp characterization of the spectrum of $\mathbf{X}_N^\text{OBC}$ remains an open problem. At best, it is known that the eigenvalues fall on curves contained in the SIBC spectrum as $N\to\infty$. Shockingly, however, {\em the finite-$N$ spectra need not converge to the semi-infinite spectra}, as illustrated in the Fig.\,\ref{DBKCrapidities}. The spectra shown are those of Toeplitz matrices associated to the dynamical matrix of a particular QBL model to be considered in Sec.\,\ref{DBKCSec}. In this example, the Toeplitz (OBC) spectra of the system with $N$ sites fall on a vertical line contained with in the 2D shaded area corresponding to the SIBC spectrum. This suggests that the spectral properties of the infinite-size limit are generically distinct from those of the \rev{corresponding} finite-size systems. In fact, a precise mathematical relationship does exist when one zooms out to the pseudospectra. Namely, we have \cite{TrefethenPS}
\begin{equation}
\lim_{\epsilon\to 0}\lim_{N\to\infty}\sigma_\epsilon(\mathbf{X}_N^{\text{OBC}}) = \sigma(\mathbf{X}^\text{SIBC}) \neq \lim_{N\to\infty}\lim_{\epsilon\to 0}\sigma_\epsilon(\mathbf{X}_N^{\text{OBC}}) .
\label{nonComm}
\end{equation}
That is, the two limits do \emph{not} commute in general: While the right hand-side reflects the generic discontinuity in finite-size spectra as $N\to\infty$, the left hand-side of the above equation can be understood as follows. Given $\lambda$ in the SIBC spectrum and an arbitrary $\epsilon>0$, then $\lambda$ is in the $\epsilon$-pseudospectra of the finite-size OBC system for all suitably large $N$. \rev{This can be understood intuitively by considering an edge-localized (see Sec.\,\ref{DynSec} for a brief discussion on localization) eigenvector of the semi-infinite system corresponding to an eigenvalue $\lambda$ not in the bulk bands, and then projecting this eigenvector onto a finite chain of length $N$. Generically, this will not provide an exact eigenvector of the finite chain but, rather, an approximate one with pseudoeigenvalue $\lambda$ and accuracy $\epsilon$ set by the localization length of the original eigenvector.} In this sense, we can say that the SIBC spectrum ``imprints itself'' into the $\epsilon$-pseudospectra of the finite-size OBC system, for arbitrary $\epsilon$.

\section{Some fundamental results about QBL{s}}
\label{QBLSec}

\subsection{A correspondence between conserved quantities and symmetry generators} 
\label{ZMWSG}

The edge-localized Majorana ZMs of a free-fermion Hamiltonian are an indication that the system is in an SPT phase. A mode is, by definition, a linear \rev{combination} of creation and annihilation operators and a ZM, in particular, commutes with the Hamiltonian. There are two other important classes of operators that commute with the Hamiltonian. They are the generators of continuous symmetry groups and the observables associated to conserved quantities. Hence, for Hamiltonian systems, three conceptually quite different objects arise as the solutions of one and the same equation: \([H, A]=0\).

The situation is different for Markovian systems. The symmetry generators and conserved quantities of a LME are characterized by two very different equations which need not share any solutions \cite{VictorSymCQ}. On the one hand, the adjoint action of a symmetry generator commutes with \(\mathcal{L}^\star\), Eq.\,\eqref{GeneralSG}; on the other hand, a conserved quantity is represented by an observable in the kernel of \(\mathcal{L}^\star\). In preparation for searching for SPT-like physics in QBLs, we will develop in this section a theory of modes that are ZMs when regarded as symmetries or when regarded as conserved quantities, but not necessarily both. In this sense, dissipation ``splits'' the set of ZMs, and a ZM that is both a conserved quantity and a symmetry generator occurs only in special circumstances. This suggests that if one \rev{is} going to rely on boundary physics as an indicator of SPT-like behavior in QBLs, then one should consider both kinds of ZMs since there is no clear reason as yet to privilege edge-localized conserved quantities over symmetry generators, or the other way around. 

Let $\mathcal{L}^\star$ be a Heisenberg-picture QBL. Referring to Eq.\,\eqref{QBLSP}, it follows that $\widehat{\vec{\alpha}}$ is a ZM if and only if $\vec{\alpha}\in\ker\widetilde{\mathbf{G}}$. Note that we can, without loss of generality, take $\widehat{\vec{\alpha}}$ to be \rev{Hermitian \footnote{This follows from the property ${\mathcal L}^\star(A^\dagger) = [{\mathcal L}^\star(A)]^\dagger$, for all $A$. Therefore, if 
$\hat {\vec{\alpha}}$ is not Hermitian to begin with, we may form the linear combination $\hat {\vec{\alpha}}+ \hat {\vec{\alpha}}^\dagger$, 
which is Hermitian and still belongs to ker ${\mathcal{L}^\star}$, as desired.}.} 
We denote the real vector space of Hermitian ZMs by $\mathcal{Z}$. Likewise, by virtue of Eq.\,\eqref{QBLSP}, we have 
\begin{equation*}
\mathcal{L}^\star([\widehat{\vec{\alpha}},A]) - [\widehat{\vec{\alpha}},\mathcal{L}^\star(A)] = [\widehat{-i\mathbf{G}\vec{\alpha}},A], \quad \forall A. 
\end{equation*}
Referring to Eq.\,\eqref{GeneralSG}, $\widehat{\vec{\alpha}} (=\widehat{\vec{\alpha}}^\dag)$ is a SG if and only if $\widehat{-i\mathbf{G}\vec{\alpha}}$ is proportional to the identity. Since it is linear in $a$ and $a^\dag$, it must be zero. It follows that $\widehat{\vec{\alpha}}$ generates a symmetry if and only if $\vec{\alpha}\in\ker\mathbf{G}$. Since the operator $e^{i\theta\widehat{\vec{\alpha}}}$ has the form of Weyl displacement operator, we refer to this class of symmetries as \textit{Weyl symmetries} and their corresponding generators as \textit{Weyl SGs}. The corresponding real vector space of Weyl SGs will be denoted by $\mathcal{W}$. We refer to the both ZMs and Weyl SGs as \textit{Noether modes}.

Note that for closed-system evolution ($\mathcal{D}=0$), the dynamical matrix is $\bm{\tau}_3$-pseudo-Hermitian, i.e., $\mathbf{G}=\widetilde{\mathbf{G}}$ \footnote{In fact, $\mathbf{G}$ is $\bm{\tau}_3$-pseudo-Hermitian whenever $\mathcal{F}(\mathbf{M})=0$, which ensures {\em unitality} of the dynamics, $\mathcal{L}(1_{\mathcal{F}})=0$.}. Hence, as expected,  $\mathcal{Z}=\mathcal{W}$. Interestingly, in the purely dissipative case \mbox{($H=0)$}, the dynamical matrix is skew $\bm{\tau}_3$-pseudo-Hermitian, i.e., $\mathbf{G}=-\widetilde{\mathbf{G}}$. Again, $\mathcal{Z}=\mathcal{W}$. In such cases, we say the Noether modes are \textit{non-split}.

Generically, Noether-modes are \textit{split}, in the sense that a given mode is either a ZM or an SG, and not both. Remarkably, however, for QBLs we are able to establish a direct one-to-one correspondence between ZMs and Weyl SGs. One may show that, given a fixed ZM, there always exists a corresponding \textit{canonically conjugate} Weyl SG. This canonical isomorphism, which we anticipated in Ref.\,\onlinecite{Bosoranas}, is rather surprising given the general decoupling of ZMs and SGs in open quantum systems \cite{VictorSymCQ}. Formally, we have the following: 

\medskip

\noindent \textbf{Theorem 1.} {\em For \rev{an arbitrary} QBL, we have $\dim_\mathbb{R}\mathcal{Z} = \dim_\mathbb{R}\mathcal{W}$. If, in addition, the zero rapidity hosts only length-one Jordan chains, then for each ZM there is a canonically conjugate Weyl SG, and vice-versa.}

\begin{proof}
Consider the antilinear operator $\mathcal{C}$ defined by $\mathcal{C}\vec{\alpha} = \bm{\tau}_1\vec{\alpha}^*$ so that $\widehat{\vec{\alpha}}\,\mbox{}^\dag = -\widehat{\mathcal{C}\vec{\alpha}}$. In particular, $\widehat{\vec{\alpha}}$ is Hermitian if and only if $\mathcal{C}\vec{\alpha} = -\vec{\alpha}$. Since $-\mathbf{G} = \bm{\tau}_1 \mathbf{G}^{\color{revred}*}\bm{\tau}_1 = \mathcal{C}\mathbf{G}\mathcal{C}$, it follows that $\ker\mathbf{G}$ is invariant under $\mathcal{C}$. Since $\mathcal{C}^2=\mathds{1}_{2N}$, we have a real structure on $\ker\mathbf{G}$. That is, $\ker{\mathbf{G}} \cong \mathcal{C}_- \!\oplus i\mathcal{C}_-$, with $\mathcal{C}_-$ the real vector space of kernel vectors $\vec{\alpha}$, with $\mathcal{C}\vec{\alpha}=-\vec{\alpha}$. ``Quantizing" these vectors with the bosonic hat map yields the set of Weyl SGs. Moreover, $\dim_\mathbb{R}\mathcal{W}=\dim_\mathbb{R}\mathcal{C}_-  = \dim_\mathbb{C}\ker\mathbf{G}$. An identical analysis yields $\dim_\mathbb{R}\mathcal{Z}= \dim_\mathbb{C}\ker\widetilde{\mathbf{G}}$. Elementary linear algebra yields $\dim_\mathbb{C}\ker \mathbf{G}=\dim_\mathbb{C}\ker\widetilde{\mathbf{G}}$, establishing the first claim.

For the second statement, let us take a biorthogonal basis $\{\vec{\gamma}^z_j, \vec{\eta}_j\}$ for $\ker\widetilde{\mathbf{G}}$. By definition, $\{\vec{\gamma}^z_j\}$ spans $\ker\widetilde{\mathbf{G}}$, $\{\vec{\eta}_j\}$ spans $\ker\widetilde{\mathbf{G}}^\dag$, and $\vec{\eta}_j^\dag \vec{\gamma}^z_k = \delta_{jk}$. The existence of such a basis hinges upon the length-one Jordan chain assumption. Per the invariance under $\mathcal{C}$, we can take $\mathcal{C}\vec{\gamma}_j^z = -\vec{\gamma}_j^z$ and $\mathcal{C}\vec{\eta}_j = -\vec{\eta}_j$. Now, let $\vec{\gamma}_j^s = -i\bm{\tau}_3 \vec{\eta}_j$. These vectors span $\ker{\mathbf{G}}$ and are odd under $\mathcal{C}$. Finally,
\begin{equation*}
[\widehat{\vec{\gamma}}_j^s,\widehat{\vec{\gamma}}_k^z] = \vec{\gamma}_j^s{}^\dag\bm{\tau}_3 \vec{\gamma}_k^z\,1_{\mathcal{F}} = i \vec{\eta}_j^\dag\vec{\gamma}^z_k\,1_{\mathcal{F}} = i\delta_{jk}\,1_{\mathcal{F}}. 
\end{equation*}
Thus, we have constructed canonically conjugate bases of $\mathcal{Z}$ and $\mathcal{W}$, as desired. 
\end{proof}

A relevant generalization of this result establishes a correspondence between the set of approximate ZMs and the generators of approximate symmetries. We say that a Hermitian operator $\widehat{\vec{\gamma}}^z$ is an \textit{approximate ZM} if $\norm{\widetilde{\mathbf{G}}\vec{\gamma}^z}<\epsilon$ for a prescribed accuracy $\epsilon>0$. These arise from $\epsilon$-pseudoeigenvectors of $\widetilde{\mathbf{G}}$ corresponding to the $\epsilon$-pseudoeigenvalue $0$. Since, in our work, the most useful norm is the $2$-norm, $\vec{\gamma}^z$ can equivalently characterized as a singular vector of $\widetilde{\mathbf{G}}$ corresponding to a singular value $s<\epsilon$ (see also App.\,\ref{dblmatps}). The second-quantized operator  then satisfies $\mathcal{L}^\star(\widehat{\vec{\gamma}}{}^z) = K\alpha$, with $|K|<\epsilon$ and $\alpha$ some linear form with bounded coefficients.  Generators of approximate Weyl symmetries are defined analogously. See App.\,\ref{ApproxThm} for a detailed account of this generalization, including an extension of Theorem 1 to the approximate setting.

\subsection{Constructive procedures for QBL design}

If they are robust against an appropriate set of perturbations, the edge-localized ZMs of a free-fermion Hamiltonian are likely to be an indication that the system is in a SPT phase. An SPT phase is a gapped phase of matter characterized by a degenerate ground state  and the absence of a local order parameter. The ZMs connect the various ground states essentially by changing the number of the quasiparticles they create or annihilate, with no associated energy cost. In order to search for this kind of physics in QBLs, one would greatly benefit from two key capabilities: 
\begin{enumerate}
\item The ability to synthesize QBLs with desirable physical properties (e.g., locality) and a rich array of possibilities in terms of their ZMs\vspace*{-1.5mm}.
\item The ability to synthesize QBLs with pure SSs.
\end{enumerate}
There is a systematic way to meet these needs and we describe them in this section. 

\subsubsection{A zero-mode-preserving map of Hamiltonian free fermions\\  to Markovian free bosons}
\label{MFtoMB}

Suppose we have a QFH $H_F$ with a midgap state. In the BdG formalism, this means there is a vector $\vec{\gamma}_+$ such that $\mathbf{H}_F\vec{\gamma}_+ = \epsilon_N \vec{\gamma}_+$, with $\epsilon_N\sim \mathcal{O}(e^{-N})$ exponentially small in system size and $\mathbf{H}_F$ the BdG Hamiltonian corresponding to $H_F$. Because $\mathcal{F}(\mathbf{H}_F) = \mathbf{H}_F$, there is also a vector $\vec{\gamma}_-$ satisfying $\mathbf{H}_F\vec{\gamma}_- = -\epsilon_N \vec{\gamma}_-$. Moreover, we can ensure that $\vec{\gamma}_- = \mathcal{C}\vec{\gamma}_+$ and $\vec{\gamma}_+^\dag \vec{\gamma}_-=0$. Upon defining the vector $\vec{\gamma}\equiv (\vec{\gamma}_+-\vec{\gamma}_-)/\sqrt{2}$, note that
\begin{equation}
\norm{\mathbf{H}_F \vec{\gamma}} = \epsilon_N \frac{\norm{\vec{\gamma}_++\vec{\gamma}_-}}{\sqrt{2}} = \epsilon_N \sim \mathcal{O}(e^{-N})
\end{equation}
and, by construction, $\mathcal{C}\vec{\gamma}=-\vec{\gamma}$. If $\Psi$ is the Nambu array of fermionic operators, then $i\vec{\gamma}{}^\dag\Psi$ is a Majorana fermion whose commutator with $H_F$ is exponentially small in system size. 

Now consider the QBL defined by $$\mathbf{H}=0, \qquad \mathbf{M} =\mathbf{H}_F+\mathbf{B},$$ with $\mathbf{B}=\mathbf{B}^\dag$ any bosonic matrix such that $\mathbf{H}_F + \mathbf{B} \geq 0$. This QBL is purely dissipative (has zero Hamiltonian) and its dynamical matrix is $\bfG = -i\bm{\tau}_3 \mathbf{H}_F$. The main virtue of this ``transmutation'' of fermions into bosons is that it preserves ZMs. If $\mathbf{H}_F$ has a midgap state, then we can construct a Hermitian operator $\gamma \equiv \vec{\gamma}{}^\dag \tau_3 \Phi$ that (i) is approximately conserved; and (ii) generates an approximate symmetry. Since Noether modes are non-split in purely dissipative systems, (i) and (ii) follow simply from
\begin{equation*}
\norm{\bfG\vec{\gamma}} = \norm{\bm{\tau}_3 \mathbf{H}_F \vec{\gamma}} \leq \norm{\mathbf{H}_F\vec{\gamma}} \sim \mathcal{O}(e^{-N}).
\end{equation*}
Importantly, this operator has the same spatial profile as the Majorana fermion, and hence is localized on one edge of the chain. The existence of a second, canonically conjugate Hermitian mode localized on the opposite edge follows from Theorem 2 in App.\,\,\ref{ApproxThm}. 

A potential drawback to our procedure is that the constructed QBL may fail to be dynamically stable. The reason is that there is no simple relationship between the spectrum of $\mathbf{H}_F$ and $\bm{\tau}_3 \mathbf{H}_F$, in general. If there is one because \(\bm{\tau}_3\) commutes with \(\mathbf{H}_F\), it means that the QFH commutes with the fermion number operator. A closely related but nonetheless surprising fact is that a pair of unitarily equivalent fermionic matrices may lead to QBLs with different stability properties. Nonetheless, we will also see how useful \rev{our} map turns out to be in practice.

\subsubsection{Designing QBLs with pure steady states via dualities}
\label{PureSSApp} 

We now provide a constructive procedure for engineering a QBL that is guaranteed to relax to the quasiparticle vacuum (QPV) of an arbitrary, but fixed, dynamically stable QBH. This procedure hinges on the existence of a particular duality transformation that maps any number-non-conserving QBH to a number-conserving one \cite{Dualities}.

Specifically, consider a dynamically stable QBH $H$, defined by its dynamical matrix $\mathbf{G}_0$. Dynamical stability ensures $\mathbf{G}_0$ is diagonalizable and has an entirely real spectrum. In turn, these features ensure that there exists a positive-definite matrix $\mathbf{S}$, such that $\mathbf{G}_0^\dag = \mathbf{S}\mathbf{G}_0\mathbf{S}^{-1}$. Moreover, $\mathbf{S}$ and, more importantly, its unique positive-definite square root $\mathbf{R}=\mathbf{S}^{1/2}$ provide a Bogoliubov transformation $\Phi\mapsto \mathbf{R}^{-1}\Phi$ that maps $H$ to a number-conserving Hamiltonian $H^D$. The dynamical matrix of $H^D$ is $\mathbf{G}^D = \mathbf{R}\mathbf{G}_0\mathbf{R}^{-1}$. This transformation can be understood physically by noting that the bosonic covariance matrix of the QPV $\ket{\widetilde{0}}$ of $H$ is
\begin{align*}
\mathcal{B}(\mathbf{Q}_\text{QPV}) = \braket{\widetilde{0}|\mathcal{B}(Q)|\widetilde{0}} =  \frac{1}{2}\mathbf{S}^{-1} = \frac{1}{2}\mathbf{R}^{-2}.
\end{align*}

We claim that the QBL $\mathcal{L}$ defined via the Hamiltonian $H$ and a GKLS matrix $\mathbf{M} \equiv \kappa(\mathbf{S} + \bm{\tau}_3)$, with $\kappa>0$ is (i) dynamically stable; and (ii) relaxes to the unique SS $\rho_\text{ss} = \ket{\widetilde{0}}\bra{\widetilde{0}}$. Firstly, one may show that this QBL is well-defined, in that $\mathbf{M} \geq 0$. This follows from the explicit form of $\mathbf{S}$ given in Ref.\,\onlinecite{Dualities}. Dynamical stability follows from the observation that the dynamical matrix of $\mathcal{L}$ is $\mathbf{G}_\kappa = \mathbf{G}_0 -i\kappa \mathds{1}_{2N}$. Hence, the rapidites are simply $-\kappa \pm i \omega_n$, with $\omega_n$ the quasiparticle energies of $H$. Dynamically stability ensures that the SS is unique. Moreover, Gaussianity allows us to characterize it entirely by its covariance matrix $\mathcal{B}(\mathbf{Q}_\text{ss})$. We compute this by solving the Lyapunov equation Eq.\,\eqref{quadeom} directly. Thus, 
\begin{eqnarray*}
\mathcal{B}(\mathbf{Q}_\text{ss}) &=& \int_0^\infty e^{-i\mathbf{G}_\kappa t}\bm{\tau}_3\mathcal{B}(\mathbf{M}) \bm{\tau}_3 e^{i\mathbf{G}_\kappa^\dag t} \,\rev{dt}.
\end{eqnarray*}
By construction, $\mathcal{B}(\mathbf{M}) = \kappa \mathbf{S}$. Recalling that $\mathbf{S}$ provides a Bogoliubov transformation, we have $\bm{\tau}_3 \mathbf{S}\bm{\tau_3} = \mathbf{S}^{-1}$. Also,
\begin{eqnarray*}
e^{-i\mathbf{G}_\kappa t} = e^{-\kappa t}e^{-i\mathbf{G}_0 t}.
\end{eqnarray*}
With this, we can simplify the integrand by computing
\begin{eqnarray*}
e^{-i\mathbf{G}_0 t}\mathbf{S}^{-1}e^{i\mathbf{G}_0^\dag t} &=& \mathbf{S}^{-1} e^{-i\mathbf{S}\mathbf{G}_0\mathbf{S}^{-1} t}e^{i\mathbf{G}_0^\dag t}
\\
&=&  \mathbf{S}^{-1} e^{-i\mathbf{G}_0^\dag t}e^{i\mathbf{G}_0^\dag t} = \mathbf{S}^{-1}.
\end{eqnarray*}
Altogether, this yields the solution 
\begin{eqnarray*}
\mathcal{B}(\mathbf{Q}_\text{ss}) &=& \kappa \mathbf{S}^{-1}\int_0^\infty e^{-2\kappa t}\,dt = \frac{1}{2}\mathbf{S}^{-1}.
\end{eqnarray*}
Since both the QPV and the SS are Gaussian, have the same mean vector (zero), and the same covariance matrix, they are the same state. Alternatively, this may be shown by noting that the dissipator becomes diagonal in a particular Hamiltonian \rev{normal-mode basis} (as we will see in the concrete example of Sec.\,\ref{PureSSSec}).

\section{Dynamical metastability}
\label{DynSec}

We have seen thus far that (i) the dynamical matrices of 1D bulk-translationally invariant QBLs under OBCs are block-Toeplitz matrices; and (ii) such matrices have non-trivial pseudospectra influenced by the bulk topology. What are, then, the physical consequences of (ii) given (i)? 

The spectrum of a translationally invariant QBL can be described in terms of curves in the complex plane labeled by the crystal momentum \(k\), namely, the rapidity bands of the system. The pseudospectrum of a bulk-translationally invariant QBL is, instead, best described in terms of the band winding number. Consider a QBL with distinguishable rapidity bands, that is, the bands are described by \(M\) complex-valued
functions $\{\lambda_m(k),\lambda_m(k)^*\}_{M=1}^d$, and let $\lambda \in {\mathbb C}$ not belong to any of the rapidity bands. In the language of Ref.\,\onlinecite{KawabataNHSymTopo}, $\lambda$ is a {\em point-gap} of the complex spectrum. For each band $m=1,\ldots, M$, the winding number of the band about $\lambda$ is given by 
\begin{equation*}
\nu_{m}(\lambda) = \frac{1}{2\pi i}\int_{-\pi}^\pi \frac{d}{dk}\ln[\lambda_m(k) - \lambda]\,dk.
\end{equation*}
As we mentioned in Sec.\,\ref{SPSToe}, the SIBC rapidity spectrum consists of the rapidity bands and all the point-gaps for which $\nu_{m}(\lambda) \ne 0$ for at least one $m$. \rev{Due to the Bloch-like structure of the dynamical matrix ${\mathbf{g}}(k)$ [Eq.\,\eqref{transl}], and following the same logic that leads one to conclude that mid-gap modes are localized in Hermitian systems, the modes associated to a point-gap are necessarily {\em boundary-localized} \cite{SlagerTopoSkin,GBTJPA}}. Generically, a positive winding number elicits left-localized modes, while a negative one elicits right-localized modes \cite{TrefethenPS}. Since we adopt the convention where the semi-infinite chain retains its left boundary, the right localized modes are lost \footnote{One could, alternatively, consider a chain extending infinitely to the left, which will retain all right-localized modes and lose the left-localized ones.}. As we also mentioned, the OBC rapidities are more complicated to describe and less predictable. At a minimum, they are a discrete subset of the SIBC spectrum and lie on continuous arcs within the SIBC spectrum. This has especially notable consequences when the OBC spectra consists entirely of point-gapped rapidities. Such systems turn out to exhibit the NHSE \cite{YaoSkin,SlagerTopoSkin,SatoTopoSkin,SatoAnomaly}, whereby {\em all} finite-size normal modes are edge-localized. 

Strikingly, it is even possible for the OBC chain to be dynamically stable for all finite $N$, despite having an unstable semi-infinite limit. Physical intuition stipulates that the dynamics of increasingly large-$N$ truncations should well-approximate those of the semi-infinite limit. How can we possibly meld this with the spontaneous loss of dynamical stability, or the spectral discontinuity in general? The pseudospectrum is precisely the tool for answering this question. Mathematically, we have seen that the SIBC spectrum imprints itself into the pseudospectrum of the finite OBC chain, in the precise sense of Eq.\,\eqref{nonComm}. To understand this intuitively, consider a localized normal mode of the semi-infinite chain, corresponding to a rapidity far separated from the OBC rapidities for any finite $N$. Truncating this mode to fit within the finite chain will not yield a normal mode, but rather a pseudonormal modes with pseudorapidity (pseudoeigenvalue of $-i\mathbf{G}$) equal to the semi-infinite rapidity. Edge localization implies that early-time dynamics of this truncated mode cannot sense the existence of the two boundaries and hence will behave as though the system were semi-infinite in extent. As $N$ grows, the pseudospectral accuracy parameter $\epsilon$ goes to zero, and so these pseudonormal modes will be indistinguishable from exact normal modes over appreciably long timescales.
 
Making the discussion more concrete, let $\Delta_{S,N}^\text{OBC}$ and $\Delta_S^\text{SIBC}$ denote the stability gaps of a length-$N$ OBC chain and the corresponding semi-infinite chain, respectively, as defined in Eq.\,\eqref{StGap}. Note that $\Delta_S^\text{SIBC}$ is also the stability gap for BIBCs. In general, we always have
\begin{equation}
\Delta_{S,\infty}^\text{OBC} \equiv \lim_{N\to\infty} \Delta_{S,N}^\text{OBC}\leq \Delta_S^\text{SIBC}.
\end{equation}
However, if the OBC and SIBC spectra differ dramatically, a {\em strict} inequality $\Delta_{S,\infty}^\text{OBC} < \Delta_S^\text{SIBC}$ is possible. We distinguish two notable cases:

\begin{itemize}
\item[(i)] If both $\Delta_{S,\infty}^\text{OBC}$ and $\Delta^\text{SIBC}_S$ are negative, we say that the OBC chain is in an \textit{anomalously relaxing \rev{dynamical} phase}. This phase, which is dynamically stable {\em both} in the finite case and the infinite-size limit, is characterized by an increasingly long transient time whereby the system possesses {\em pseudonormal} modes decaying at rates much slower than the rate set by the finite-size stability gap. \rev{This} transient is followed by asymptotic decay, whose rate is set by the finite-size stability gap. We conjecture that a similar mechanism is responsible for the anomalously long relaxation dynamics found in dissipative systems exhibiting the so-called {\em Liouvillian skin effect} \cite{UedaSkin, MoriSlow,Mori}. \rev{This dynamical behavior} also bears some resemblance to metastability associated with spectral separation \cite{GarrahanMeta}, with the spectral separation here being between the finite-size stability gap, and the infinite-size (or pseudospectral) gap \vspace*{-1.5mm}. 

\item[(ii)] If $\Delta_{S,\infty}^\text{OBC} < 0 < \Delta^\text{SIBC}_S$, we say that the OBC chain is \textit{dynamically metastable} (or just \textit{metastable}, in our context, for short). This phase is characterized by an increasingly long transient time whereby the pseudo-normal modes whose pseudoeigenvalues are in the right half plane {\em amplify exponentially}. Once this transient concludes, all modes decay asymptotically with rate set by the finite-size stability gap. The terminology ``dynamical metastability"  refers to the metastable amplifying transient dynamical phase that eventually gives way to the necessarily stable asymptotic dynamics. \rev{Note that, since our notion aims to capture metastable behavior of the structural (stability) properties of the dynamical generator, it is, at the current stage of understanding, a distinct notion than metastability for states of open quantum systems that is known to stem from} large intra-spectral gaps \cite{GarrahanMeta,GarrahanCorrelation,MacieszczakMeta}. However, both cases are characterized by delayed relaxation to the true SS manifold.
\end{itemize}

\begin{figure}
\includegraphics[width=.75\columnwidth]{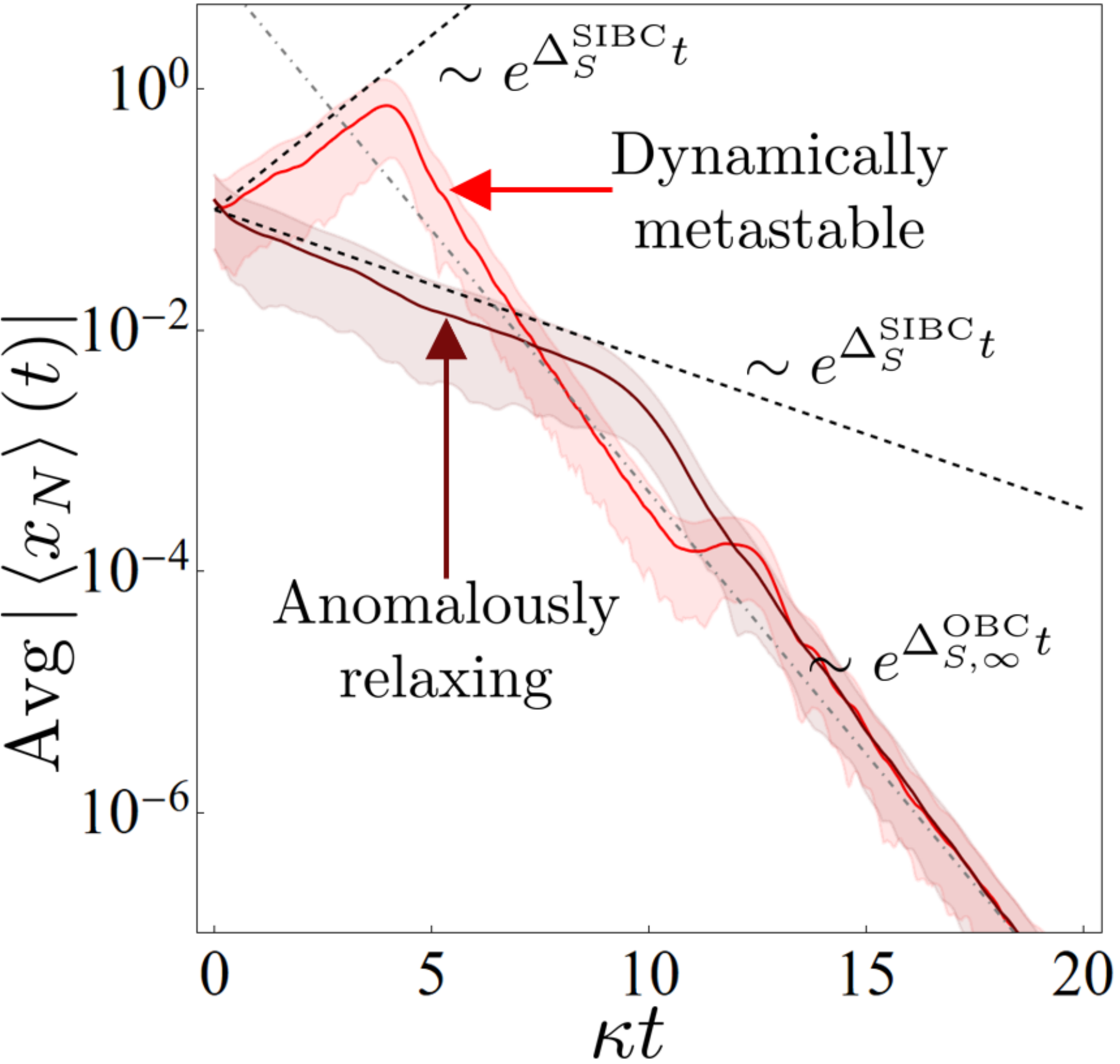}
\vspace*{-2mm}
\caption{Anomalously relaxing vs. dynamically metastable dynamics in the DBKC (Sec.\,\ref{DBKCSec}). The solid lines are the trajectory of $|\braket{x_N}(t)|$ averaged over 250 random initial conditions. The filled regions are $\pm$ one standard deviation from the mean. The black dashed lines are the dynamics predicted from the SIBC stability $\Delta_S^\text{SIBC}$ gap, while the gray dashed lines are the dynamics expected from the finite-size stability gaps, $\Delta_{S,\infty}^\text{OBC} = \lim_{N\to\infty}\Delta_{S,N}^\text{OBC}$. For this example, $\Delta_{S,N}^\text{OBC} = \Delta_{S,\infty}^\text{OBC} = -\kappa$. In all cases, $J=2$, $\Delta=0.5$, $\mu=\Gamma=0$, $N=25$. The dynamically metastable (anomalously relaxing) curve corresponds to $\kappa/\Delta = 0.6$ (1.4).}  
\label{BKCMeanDecay}
\end{figure} 

Examples of the above dynamical behavior are found in the dissipative bosonic Kitaev chain (DBKC) model whose rapidity spectra are shown in Fig.\,\ref{DBKCrapidities} (see Sec.\,\ref{DBKCSec} for a detailed discussion). In particular, the rapidities corresponding to open marker in Fig.\,\ref{DBKCrapidities}(a) show an anomalously relaxing phase when the dissipation rate $\kappa$ exceeds the two-photon pumping, $\Delta$, while the filled (darker) rapidities show a metastable phase. To explicitly illustrate the two-step nature of the relaxation dynamics that classes (i) and (ii) above support, in Fig.\,\ref{BKCMeanDecay} we plot the trajectory of $|\braket{x_N}(t)|$ averaged over an ensemble of random initial conditions. In the anomalously relaxing phase, exponential relaxation indeed proceeds in two steps. The transient decay rate is set by $\Delta_{S}^\text{SIBC}<0$, while the asymptotic decay rate is the true one, i.e., $\Delta_{S,N}^\text{OBC}<0$. The dynamically metastable phase shows a drastically different transient behavior, with amplification engendered by $\Delta_{S}^\text{SIBC}>0$. Dynamical stability nonetheless guarantees that the asymptotic dynamics coincide with those of the anomalously relaxing phase.

It is natural to ask how long these anomalous transient regimes persist. For this, consider the normal mode of the SIBC system corresponding to the rapidity with the largest real part. In the anomalously relaxing case, this is the slowest relaxing mode, while in the dynamically metastable case, it is the one that amplifies the fastest. Truncating this mode to fit into a chain of length $N$ will provide a pseudonormal mode of the OBC system of accuracy $\epsilon(N)$. As before, the well-behaved nature of the pseudospectrum infinite-size limit implies that $\epsilon(N)\to 0$ as $N\to\infty$. Referring to Eq.\,\eqref{pseudomode}, the lifetime of this pseudonormal mode will necessarily increase as $\epsilon(N)\to 0$. Thus, the transient timescale diverges as $N\to\infty$. One consequence of this fact is that, in the dynamically metastable case, the \textit{linear} mixing time of the QBL (roughly, the time it takes the first moments of arbitrary states to come suitably close to their SS value) will diverge \cite{Bosoranas}.

In what follows, we shall focus on showing how dynamically metastable phases can be further categorized into those that are topological in an appropriate sense (as in Fig.\,\ref{DBKCrapidities}(a)), and those that are not (as in Fig.\,\ref{DBKCrapidities}(b)). \rev{We further remark that a more comprehensive analysis of dynamical metastability will be presented in Ref.\,\onlinecite{MariamPaper}.}

\section{Topological dynamical metastability with broken number symmetry}
\label{TopoMS1}

\begin{table*}[t!]
\begin{tabular}{|c|c|c|c|c|c|}
\hline\hline
{\bf Model} & {\bf Hamiltonian} & {\bf Lindblad dissipator} & {\bf U(1) symmetry} & {\bf Noether modes} & {\bf Steady state}   \\
\hline\hline
$\quad$DBKC$\quad$ & BKC + DPA & Uniform onsite plus next-NN damping & No & Split & Mixed \\
Pure-SS & BKC & Uniform damping in BKC normal-mode basis; & No & Split & Pure\vspace*{-1mm} \\
DBKC & & site-local; restricted translation symmetry  &  &  &  \\
PDMC & $H=0$ & Uniform onsite and NN damping, pumping, pairing & No & Non-split & Mixed  \\
{DDWC} & {DPA+ NDPA}  & {Uniform onsite  damping and  pumping, NN pairing} & {No} & {Split, doubled}  & {Mixed} \\ \hline
DNSC & (i)TB & Uniform onsite and NN damping, onsite pumping   & Yes & Split & Mixed   \\
\hline\hline
\end{tabular}
\caption{Summary of the key features of the five topologically metastable models analyzed in this work. Abbreviations: 
DBKC = dissipative bosonic Kitaev chain; 
PDMC = purely dissipative Majorana chain; 
DDWC = dissipative double-winding chain; 
DNSC = dissipative number-symmetric chain;  
DPA = degenerate parametric amplifier; 
NDPA = non-degenerate parametric amplifier; 
NN = nearest-neighbor; 
(i)TB = (imaginary) tight-binding. 
In all models (except the pure SS DBKC), the purity of the SS is assessed numerically throughout the relevant regions of parameter space (data not shown). The DDWC, which is found to support {\em two} ZMs (SGs) on each edge, is discussed in App.\,\ref{DDW}, while the rest are in the main text.
}  
\label{Realm}
\end{table*}

Let us now restrict our focus on dynamically metastable QBLs, and further assume that they (i) are point-gapped at zero (i.e., their Bloch dynamical matrix $\mathbf{g}(k)$ is invertible for all $k$); and (ii) have at least one rapidity band winding about zero. We shall call QBLs belonging to this class \textit{topologically dynamically metastable} (or just \textit{topologically metastable}, for short). 
Physically, an additional important restriction stems from the possible presence of a (weak) U(1) number symmetry, generated by the total bosonic number operator. While we defer the discussion of number-symmetric QBLs to Sec.\ref{TopoMS2}, we examine first QBLs which, in analogy to fermionic topological superconducting phases, will feature some {\em bosonic pairing} mechanisms, at either the Hamiltonian or dissipative level.  For reference, a summary of all the illustrative models we will analyze in this work is provided in Table\,\ref{Realm}.

\subsection{Majorana edge bosons: General properties}
\label{MBGen}

From the definition of topological metastability\rev{, in addition to the spectral and pseudospectral properties of Toeplitz and Laurent matrices and operators discussed in Sec.\,\ref{SPSToe},} we may immediately conclude that $\mathbf{G}_N^\text{OBC}$ possesses at least one pseudonormal mode with zero pseudorapidity. Similarly, we can conclude that $\widetilde{\mathbf{G}}_N^\text{OBC}$ {\em also} possesses a pseudonormal mode with zero pseudorapidity. Following the conventions of Sec.\,\ref{ZMWSG}, we call these two pseudomodes $\vec{\gamma}^s$ and $\vec{\gamma}^z$, respectively, and ensure that the associated linear forms $\gamma^s \equiv \widehat{\vec{\gamma}}{}^s$ and $\gamma{}^z\equiv \widehat{\vec{\gamma}}^z$ are Hermitian, and have commutator\footnote{Note that the canonical correspondence assumes the technical assumptions of Theorem 2 in App.\,\ref{ApproxThm} are met.} equal to $i$.  We deem the pair $(\gamma^z,\gamma^s)$ \textit{Majorana bosons}. MBs enjoy a number of remarkable properties derived from their pseudospectral and topological origins:
 
\begin{itemize}
\item[(i)] An MB pair consists of 
one approximate ZM $\gamma^z$ and one  generator of an approximate Weyl symmetry $\gamma^s$. Both are necessarily {\em Hermitian}. That is, MBs are a particular instance of Noether modes\vspace*{-1.5mm}. 

\item[(ii)] The pair can always be normalized to satisfy {\em canonical} commutation relations, while the roles of $\gamma^z$ and $\gamma^s$ as approximate ZM and SG are maintained\vspace*{-1.5mm}.

\item[(iii)] One member of the pair is {\em exponentially localized} on the left half of the chain, while the other is localized on the right. This follows because, due to the adjoint relationship between $\widetilde{\mathbf{G}}$ and $\mathbf{G}$, the winding associated to $\gamma^z$ and $\gamma^s$ necessarily have opposite sign. This engenders the stated localization properties \cite{TrefethenPS,BottcherToe,MathFollowup}\vspace*{-1.5mm}. 

\item[(iv)] Combining (i)-(iii) allows us to construct a {\em spatially split bosonic degree of freedom} $\gamma^z + i \gamma^s$ whose quadrature components are the MBs. In the case where the MBs are non-split (in the sense of Sec.\,\ref{ZMWSG}), this creates a long-lived bosonic excitation in the system. 
\end{itemize}

It is worth to zoom in specifically on point (ii). While a more general and rigorous discussion will be provided in a mathematics-focused companion paper \cite{MathFollowup}, a typical instance of MBs takes on the form:
\begin{equation}
\gamma^z \equiv \mathcal{M}_z(N) \sum_{j=1}^N \delta^j x_j
,\quad \gamma^s \equiv \mathcal{M}_s(N) \sum_{j=1}^N \delta^{N-j}p_j, 
\end{equation}
with $\mathcal{M}_{z,s}(N)$ size-dependent normalization constants to be determined and $\delta$ real, with $|\delta|<1$. The above operators satisfy 
\begin{eqnarray}
\mathcal{L}^\star(\gamma^z) &=& \mathcal{M}_z(N) \delta^{N-1} \chi ,
\label{zMBtyp}
\\
\mathcal{L}^\star([\gamma^s,A]) - [\gamma^s,\mathcal{L}^\star(A)] &=&    \mathcal{M}_s(N)\delta^{N-1}[\xi,A] ,\quad
\label{sMBtyp}
\end{eqnarray}
along with the algebraic relationship
\begin{equation}
\label{algtyp}
[\gamma^z,\gamma^s] = i \mathcal{M}_z(N)\mathcal{M}_s(N)N\delta^{N-1}.
\end{equation}
Here, $\chi$ and $\xi$ represent (typically localized) Hermitian linear forms whose coefficients in the Nambu basis are {\em system-size independent}. The goal of normalization is to choose $\mathcal{M}_z(N)$ and $\mathcal{M}_s(N)$ such that: (i) the right hand-sides of Eqs.\,\eqref{zMBtyp} and \eqref{sMBtyp} go to zero as $N\to\infty$; and (ii) the right hand-side of Eq.\,\eqref{algtyp} is $i$. That is, we want {\em canonically conjugate modes that provide asymptotically exact ZMs and SGs.} Mathematically, these conditions are
\begin{eqnarray*}
&\text{(i) }&\lim_{N\to\infty} \mathcal{M}_z(N) \delta^{N-1} = \lim_{N\to\infty} \mathcal{M}_s(N) \delta^{N-1} = 0 ,
\\
&\text{(ii) }&  \mathcal{M}_z(N)\mathcal{M}_s(N)N\delta^{N-1} = 1.
\end{eqnarray*}
Remarkably, a scheme satisfying both (i) and (ii) always exists \cite{MathFollowup}. The simplest, and most natural choice is
\begin{equation*}
\mathcal{M}_z(N) = \mathcal{M}_s(N) = \frac{\delta^{-(N-1)/2}}{\sqrt{N}}.
\end{equation*}
With this, condition (ii) is clearly satisfied. As for (i),
\begin{equation*}
\lim_{N\to\infty} \mathcal{M}_z(N) \delta^{N-1} = \lim_{N\to\infty} \frac{\delta^{(N-1)/2}}{\sqrt{N}} = 0,
\end{equation*}
and similarly for $\mathcal{M}_s(N)$. This scheme, which we call {\em symmetric normalization}, will be employed throughout the text. Clearly, this choice is non-unique, and another may be adopted if certain features are desired over others. For example, if we wish for $\gamma^z$ to have bounded coefficients in the $x_j$ basis, then we may take $\mathcal{M}_z(N) = 1$ and $\mathcal{M}_s(N) = \delta^{-(N-1)}/N$. Crucially, the exponential profile of the MBs is {\em independent} of the choice of normalization. We also remark that this approach to MB normalization shares many features with the fermionic case, as detailed in App.\,\ref{QFHBG}.

While the dynamical features of the approximate ZM MB $\gamma^z$ is relatively easy to describe, those of $\gamma^s$ are initially more opaque. We illuminate one particular implication of the existence of this approximate symmetry by leveraging the uniqueness of the SS under OBCs. Denoting said SS by $\rho_\text{ss}$, we introduce the family of Weyl-displaced Gaussian states
\begin{equation*}
\rho_\theta = e^{i\theta\gamma^s}\rho_\text{ss}e^{-i\theta\gamma^s},\quad \theta\in\mathbb{R}.
\end{equation*}
Because $\gamma^s$ generates an approximate symmetry, the states $\rho(\theta)$ are \textit{quasi-steady}, in the sense that $\dot{\rho}_\theta(0) \sim 0 + \mathcal{O}(\theta\epsilon)$, with $\epsilon$ being an {\em exponentially small} accuracy parameter such that $0$ is in the $\epsilon$-pseudospectrum of $\mathbf{G}$.  Unlike the SSs of a QBL, these quasi-SSs possess {\em non-zero} mean vectors:
\begin{equation*}
\vec{m}_\theta(0) \equiv \tr[\Phi \rho_\theta(0)]  = i\theta\vec{\gamma}^s.
\end{equation*}
Since $\vec{\gamma}^s$ is edge-localized and a pseudonormal mode of the dynamical matrix, the quasi-steady mean vectors are exponentially localized on one edge of the chain and are long-lived. More explicitly, we may write 
\begin{equation}
\label{mbound}
\frac{\norm{\vec{m}_\theta(t) - \vec{m}_\theta(0)}}{\norm{\vec{m}_\theta(0)}} \leq \epsilon \, t \,e^{\Omega t},
\end{equation}
with $\Omega>0$ a system-size independent constant and $\epsilon \rightarrow 0$ as $N\rightarrow\infty$ (see Ref.\,\onlinecite{Bosoranas} for an explicit derivation). 
 On the other hand, being only a Weyl displacement of the SS, it follows that $\rho_\theta$ shares a covariance matrix (and hence all even moments) with $\rho_\text{ss}$. Interestingly, these states can be used to construct long-lived classical non-Gaussian states. Namely, any convex linear combination of the $\rho_\theta$'s will be long-lived and exhibiting a non-Gaussian, albeit still non-negative, Wigner function.

\subsection{A purely dissipative bosonic Majorana chain} 
\label{DBKC}

As the first concrete example of topological metastability, we focus on a purely dissipative ($H=0$) setting. The motivation for this is twofold. Firstly, the existence of a purely dissipative model displaying topological metastability demonstrates that the phenomena does not require any Hamiltonian contribution. In particular, this establishes that topological metastability is not a property of the Hamiltonian that is stabilized through dissipation but, rather, a fundamentally dissipative phenomenon. In fact, a similar philosophy underpins early analysis of topological physics in quasi-free open fermionic systems \cite{BardynTopByDiss}. Secondly, as detailed in Sec.\,\ref{ZMWSG}, the MBs of purely dissipative chain are necessarily non-split: Each MB is both a ZM and an SG. Thus, they provide the most natural bosonic generalization of the Majorana fermions of topological superconductivity.  

\begin{figure}[b!]
\includegraphics[width=\columnwidth]{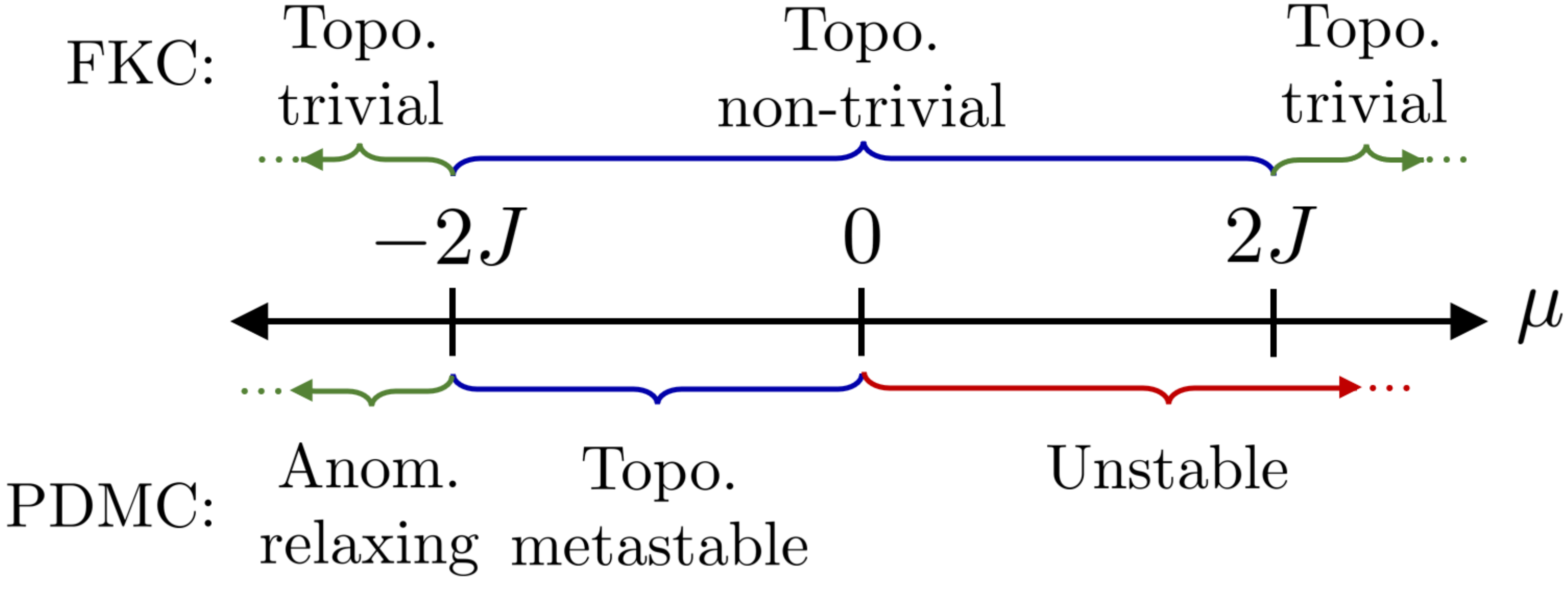}
\vspace*{-3mm}
\caption{Comparison between the topological phase diagram of the FKC and the topological stability phase diagram of the corresponding PDMC.}
\label{FKCPD}
\end{figure}

To accomplish this, let us apply the construction of Sec.\,\ref{MFtoMB} to the paradigmatic fermionic Kitaev chain (FKC) (see also App.\,\ref{QFHBG}), whose phase diagram is shown in Fig.\,\ref{FKCPD}. Within the topological phase and under OBCs, $H_{\text{FKC}}$ has two approximate Majorana ZMs (in the special case $\mu=0$, these approximate ZMs are  exact for odd $N$). The corresponding BdG Hamiltonian is $\mathbf{H}_{\text{FKC}} = \mathds{1}_N\otimes \mathbf{h}_0 + \mathbf{S}\otimes \mathbf{h}_1 + \mathbf{S}^\dag \otimes \mathbf{h}_1^\dag$, with the internal matrices given by
\begin{align*}
\mathbf{h}_0 = \begin{bmatrix}
-\mu & 0 \\ 0 & \mu
\end{bmatrix},\quad \mathbf{h}_1 = \begin{bmatrix}
-J & \Delta \\ -\Delta & J
\end{bmatrix}.
\end{align*}
Moving to the corresponding QBL, we set $\mathbf{M} \equiv \mathbf{H}_{\text{FKC}} + \alpha \mathds{1}_{2N}$, with $\alpha\geq |\min\sigma(\mathbf{H}_{\text{FKC}})|$, so that we ensure $\mathbf{M}\geq 0$. Physically, the Lindblad dissipator describing this purely dissipative Majorana chain (PDMC) has five distinct incoherent contributions: $\mathcal{D} \equiv \mathcal{D}_{-,0} + \mathcal{D}_{+,0} + \mathcal{D}_{-,1} + \mathcal{D}_{+,1} + \mathcal{D}_{p,1}$, with
\begin{eqnarray*}
\mathcal{D}_{-,0} &=& (\alpha-\mu) \sum_{j=1}^N\mathcal{D}[a_j,a_j^\dag] ,
\\
\mathcal{D}_{+,0} &=& (\alpha+\mu) \sum_{j=1}^N\mathcal{D}[a_j^\dag,a_j] ,
\\
\mathcal{D}_{-,1} &=& -J\sum_{j=1}^N\mathcal{D}[a_j,a_{j+1}^\dag]  + \mathcal{D}[a_{j+1},a_{j}^\dag] ,
\\
\mathcal{D}_{+,1} &=& J\sum_{j=1}^N\mathcal{D}[a_j^\dag,a_{j+1}]  + \mathcal{D}[a_{j+1}^\dag,a_{j}] ,
\\
\mathcal{D}_{p,1} &=& \frac{\Delta}{2} \sum_{j=1}^N \mathcal{D}[a_j,a_{j+1}] - \mathcal{D}[a_{j+1},a_{j}]-
(a\leftrightarrow a^\dag).
\end{eqnarray*}
From top to bottom, these mechanisms can be described as uniform onsite damping of strength $\alpha-\mu$, uniform onsite pumping of strength $\alpha+\mu$, uniform NN damping  of strength $J$, uniform NN pumping of strength $J$, and incoherent pairing of strength $\Delta$, respectively. 

Diagonalizing the GKLS matrix leads to a particularly interesting set of Lindblad operators.  Let $\{\vec{\psi}_{n,\pm}\}$ be the eigenvectors of $\mathbf{H}_{\text{FKC}}$ (and hence, $\mathbf{M}$) with eigenvalues $\{\pm\epsilon_n\}$. These represent the FKC quasiparticles ($+$) and quasiholes ($-$). The Lindblad operators are then $L_n^\pm = \sqrt{\alpha\pm\epsilon_n}\,\vec{\psi}^{\,\dag}_{n,\pm}\Phi$. Under PBCs, these correspond to delocalized plane-waves with $\pm\epsilon_n$ in the FKC bands. The OBC case depends on the topological phase. In the trivial phase, the $L_n^\pm$'s are delocalized standing waves with $\pm\epsilon_n$ in the bands. In the non-trivial phase, there are $2N-2$ standing waves with $\pm\epsilon_n$ confined to the bands and 2 edge modes with $\pm\epsilon_n\sim \pm e^{-N}$ in the gap. These two modes provide boundary dissipation and pumping for the corresponding bosonic chain.

The Bloch dynamical matrix of the PDMC is founds as 
$$-i \mathbf{g}(k) = (\mu + 2J \cos(k))\mathds{1}_2 - 2i\Delta \sin(k)\bm{\sigma}_1.$$ 
Thus, the rapidity bands for this model are given by
\begin{equation*}
\lambda_\pm(k) = \mu + 2J \cos(k) \pm 2i\Delta\sin(k) .
\end{equation*}
Both dynamical instability and non-zero winding are guaranteed whenever $|\mu/2J|<1$, which corresponds {\em exactly} to the non-trivial phase of the FKC. However, the OBC chain is unstable whenever $\mu>0$, see Fig.\,\ref{FKCPD}.  Explicitly, the OBC rapidities for the PDMC are given by
\begin{equation}
\label{PDCRap}
\lambda_m = \mu + 2i\sqrt{\Delta^2-J^2}\cos\left(\frac{m\pi}{N+1}\right),
\end{equation}
with $m=0,\ldots,2N-1$ (see App.\,\ref{PDCdetails} for a \rev{more} detailed spectral analysis). For simplicity, we take $\Delta\geq J>0$ and restrict $\mu<0$. The relevant stability gaps are then
\begin{equation*}
\Delta_{S,N}^\text{OBC} = \mu,\quad \Delta_{S}^\text{SIBC} = 2J+\mu .
\end{equation*}
Thus, we conclude that the PDMC (under OBCs) is topologically metastable whenever $\mu<0$ and $-\mu/ (2|J| ) <1$. 

For general parameter values, the calculation of MBs is performed numerically, \rev{as described in more detail in App.\,\ref{NumMBs}}. However, we may construct the (unnormalized) MBs analytically in the special case $J=\Delta$:
\begin{equation}
\label{PDCMBs}
\gamma_L = \sum_{j=1}^N \delta^{j-1} x_j,\quad \gamma_R = \sum_{j=1}^N \delta^{N-j} p_j,\quad \delta\equiv-\frac{\mu}{2J}.
\end{equation}
We immediately see that (i) both $\gamma_L$ and $\gamma_R$ are approximate ZMs and generate approximate symmetries;  and (ii) they have the exact same localization lengths as the MF edge modes of the FKC. While (ii) can be inferred directly from Eq.\,(\ref{PDCMBs}), (i) is verified by the relations
\begin{eqnarray*}
\mathcal{L}^\star(\gamma_L) &=& -2J \delta^N x_N,
\\
\mathcal{L}^\star(\gamma_R) &=& -2J \delta^N p_1,
\\
\mathcal{L}^\star([\gamma_L,A]) - [\gamma_L,\mathcal{L}^\star(A)] &=& -2iJ \delta^N[x_N,A], \,\,\forall A,
\\
\mathcal{L}^\star([\gamma_R,A]) - [\gamma_R,\mathcal{L}^\star(A)] &=& -2iJ \delta^N[p_1,A], \,\,\forall A.
\end{eqnarray*}
Moreover, the unnormalized MB algebra is $[\gamma_L,\gamma_R]=2iN\delta^{N-1}$. Normalization proceeds exactly as detailed in Sec.\,\ref{MBGen}. Finally, as anticipated, the MBs are non-split and have the same spatial profile as the Majorana fermions of the FKC \cite{KitaevMajorana}. In Sec.\,\ref{MultiSec}, we will show how non-split MBs may be distinguished from the more typical split MBs via two-time correlation functions.

\subsection{A dissipative bosonic Kitaev chain}
\label{DBKCSec}

\begin{figure*}
\includegraphics[width=2\columnwidth]{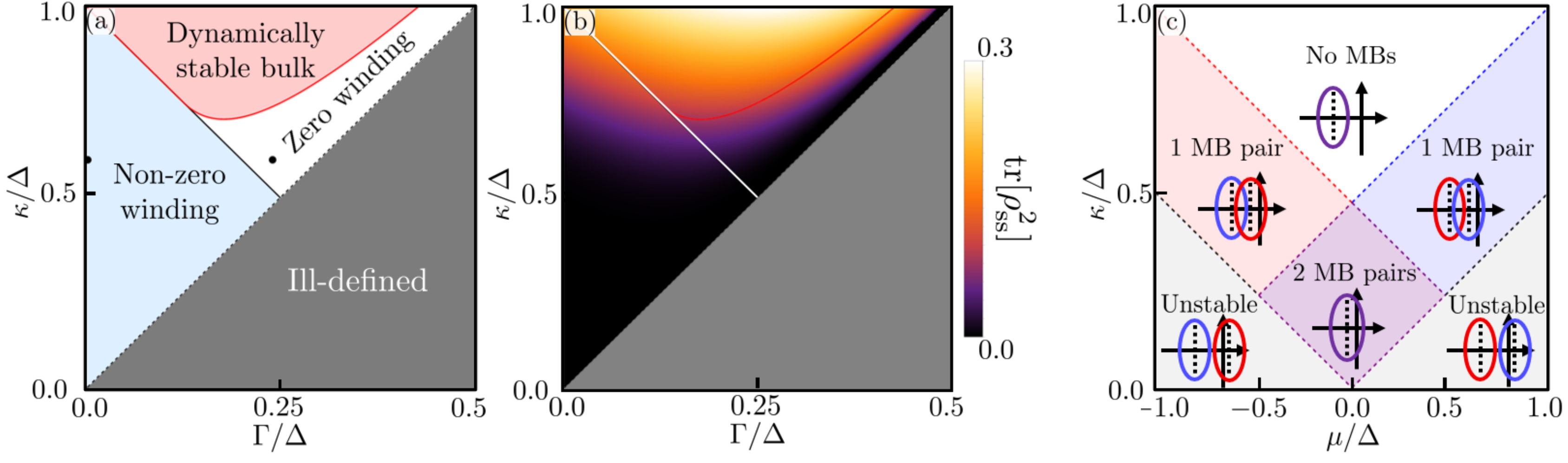}
\vspace*{-2mm}
\caption{{\bf (a)} The topological phase diagram of the DBKC with $J\geq \Delta>0$ and $\mu=0$. The ``Non-zero winding" region indicates the parameter regime where the rapidity bands wind around the origin. The ``Zero winding" region indicates the parameter regime where neither band winds around the origin. If the OBC chain is dynamically stable in either of these two regions, then it is dynamically metastable, additionally being topologically metastable in the former case. The ``Dynamically stable bulk" region occurs when the rapidity bands lie in the left-half plane. If the OBC chain is dynamically stable here, then it is anomalously relaxing.  In the ``Ill-defined" region, $\mathbf{M}$ is no longer positive-semidefinite.   The black dots at $(\Gamma/\Delta , \kappa /\Delta) = (0,0.6)$ and $(0.24,0.6)$ indicate the position of two representative parameter choices used in later figures. {\bf (b)} For later reference, the SS purity under OBCs in the same phase space as {\bf (a)} with $J=2$, $\Delta=0.5$, and $N=25$. 
{\bf (c)} The topological stability phase diagram of the DBKC under OBCs with $\Gamma=0$. The inset figures show representative rapidity band structure in each region. The number of bands winding around the origin determines the number of MB pairs. (a) and (c) adapted from Ref.\,\cite{Bosoranas}. 
}
\label{dbkcpds}
\end{figure*}

The MBs of the previous model represent the simplest bosonic extension of Majorana fermions. However, the model itself is atypical in the sense that QBLs generically possess both coherent (Hamiltonian) and incoherent processes.  As a quintessential example of topological metastability furnished by the interplay between these two dynamical contributions, we consider the  dissipative BKC  (DBKC) we introduced in Ref.\,\onlinecite{Bosoranas}. The Hamiltonian component now consists of the BKC Hamiltonian \cite{ClerkBKC}, modified with an isotropic degenerate parametric amplifier (DPA) term at each site. Explicitly,
\begin{equation}
\label{BKCHam}
H_\text{BKC} = \frac{i}{2}\sum_{j=1}^{N} \!\left(J a_{j+1}^\dag a_j + \Delta a_{j+1}^\dag a_{j}^\dag + \mu a_{j}^\dag{}^2\right) +\text{H.c.},
\end{equation}
with $J\geq \Delta\geq 0 $ being the hopping and pairing amplitudes, $\mu\in\mathbb{R}$ the uniform DPA strength, and $a_{N+j} \equiv a_j$ for PBCs, or $a_{N+j}\equiv 0$ for OBCs. By making the system open, we introduce two damping mechanisms:
\begin{equation*}
\mathcal{D} = \sum_{j=1}^N\bigg(\kappa - \frac{(2\Gamma)^2}{\kappa}\bigg)\mathcal{D}[a_j] + \mathcal{D}\Big[\sqrt{\kappa} a_j + \frac{2\Gamma}{\sqrt{\kappa}}a_{j+2}\Big],
\end{equation*}
where $\kappa\geq 0$ is a uniform onsite damping rate and $\Gamma \geq 0$ is a next-nearest neighbor (NNN) damping rate. BCs are enforced in the usual way, whereas we ensure the GKLS matrix $\mathbf{M}$ is positive-semidefinite by requiring $\kappa \geq 2\Gamma$. 

The Bloch dynamical matrix is given by
\begin{equation*}
\mathbf{g}(k) \!=\! -i [\kappa -iJ\sin(k)+2\Gamma \cos(2k)]\mathds{1}_2 + i [\mu+\Delta \cos(k)]\bm{\sigma}_1,
\end{equation*}
from which the two rapidity bands are determined to be
\begin{equation}
\lambda_\pm(k) = -(\kappa \pm \mu) \mp \Delta \cos(k) - i J \sin(k) - 2\Gamma \cos(2k) ,
\label{eq:rapB}
\end{equation}
where $k\in[-\pi,\pi]$ is the crystal momentum. The stability phase diagram for PBCs/BIBCs is obtained by determining when the bands cross the imaginary axis. In the unstable phase, two possibilities arise: (i) at least one band winds around the origin or (ii) neither band wind around the origin. A representative stability phase diagram, including the partition of the unstable phase into topological and non-topological regions, is shown in Fig.\,\ref{dbkcpds}(a).

\subsubsection{The parameter regime $\Gamma=0$}

The parameter regime where the NNN damping vanishes is especially interesting from the perspective of topological dynamical metastability. In this case, the OBC rapidities can be computed as 
\begin{equation*}
\lambda_{\pm,m} = -\kappa \pm \mu + i \sqrt{J^2-\Delta^2}\cos\left(\frac{m\pi}{N+1}\right),
\end{equation*}
with $m=0,\ldots,2N-1$. In the notation of Sec.\,\ref{DynSec}, the relevant stability gaps in this limit are
\begin{equation*}
\Delta_{S,N}^\text{OBC} = -\kappa +|\mu| ,\quad \Delta_{S}^\text{SIBC} = \Delta -\kappa +|\mu|.
\end{equation*} 
Example rapidities for $\Gamma=\mu=0$ are shown in Fig.\,\ref{DBKCrapidities}(a). Stability is guaranteed whenever $\kappa\geq |\mu|$, independent of $N$. Under this constraint, the winding numbers $\nu_\pm$ of the bands $\lambda_\pm(k)$ can be tuned to be either $(\nu_+,\nu_-) = (1,-1)$, $(1,0)$, $(0,-1)$, or $(0,0)$, see Fig.\,\ref{dbkcpds}(c). The first three regimes exhibit topological metastability while the fourth is anomalously relaxing. Each non-zero winding number corresponds to an MB pair. Thus, in a topologically metastable regime, the number of MB pairs can either be 1 or 2: in the former case, a particular edge hosts either a boundary-localized ZM or a SG, whereas in the latter case both a boundary-localized ZM and a SG are present on each edge. We remark that the existence of phases where $\sum\nu \neq 0$ {\em explicitly requires dissipation}, and is interesting from the perspective of Fredholm theory \cite{MathFollowup}.  

For the special case $J=\Delta$, we compute the MBs exactly:
\begin{align*}
\gamma_L^z &= \sum_{j=1}^N \delta_-^{j-1} x_j, & \gamma_R^s &= \sum_{j=1}^N \delta_-^{N-j} p_j ,
\\
\gamma_L^s &= \sum_{j=1}^N \delta_+^{j-1} x_j, & \gamma_R^z &= \sum_{j=1}^N \delta_+^{N-j} p_j,
\end{align*}
with $\delta_\pm = -(\mu\pm \kappa)/J$. The relevant non-vanishing commutators are $[\gamma_L^z,\gamma_R^s] = iN \delta_-^{N-1}$ and $[\gamma_L^s,\gamma_R^z]=iN\delta_+^{N-1}$. The pseudospectral origin of each mode and its role as approximate ZM / SG are  evident in the following identities:
\begin{align*}
\mathcal{L}^\star(\gamma_L^z) &= -J \delta_-^N x_N,
\\
\mathcal{L}^\star(\gamma_R^z) &= J \delta_+^N p_1,
\\
\mathcal{L}^\star([\gamma_L^s,A]) - [\gamma_L^s,\mathcal{L}^\star(A)] &= -J \delta_+^N[x_N,A], \,\,\forall A,
\\
\mathcal{L}^\star([\gamma_R^s,A]) - [\gamma_R^s,\mathcal{L}^\star(A)] &= J \delta_-^N[p_1,A], \,\,\forall A.
\end{align*}
Whenever $\delta_+$ or $\delta_-$ has modulus less than one, the corresponding $c/s$ mode is an approximate ZM / SG. Independence of $\delta_+$ and $\delta_-$ leads to the phase diagram of Fig.\,\ref{dbkcpds}(c). It is interesting to note that, unlike the exact normal-modes of the system (which are pinned to the edges via the NHSE), the localization lengths of the MBs, $\xi_\pm\equiv [\ln |\delta_\pm|]^{-1}$, diverge at the topological phase transition. In fact, divergence of the localization length explicitly define the phase boundaries, reminiscent of Majorana fermions in topological superconductors. 

\subsubsection{The parameter regime $\Gamma \ne 0$}

When the NNN damping is turned on and exceeds a certain threshold, the DBKC can exhibit non-topological metastability. The case 
where $\mu=0$ is illustrated in both Fig.\,\ref{dbkcpds}(a) and Fig.\,\ref{DBKCrapidities}(b). In this case, Eq.\,\eqref{eq:rapB} reveals that the two rapidity bands are degenerate; with respect to the elliptical bands of the BKC Hamiltonian, the onsite damping introduces an overall shift, while the NNN damping introduces more complex curvature. The bulk stability gap is found as 
\begin{align*}
\Delta_S^\text{BIBC} =\Delta_S^\text{SIBC} = \begin{cases}
-\kappa +\Delta - 2\Gamma, & \Gamma/\Delta <1/8,
\\
-\kappa +\Delta^2/16\Gamma + 2\Gamma, & \Gamma/\Delta > 1/8, 
\end{cases}
\end{align*}
with the region where $\Delta_S^\text{BIBC}>0$ being divided into a sector where both bands wind around the origin and one where neither do. The latter corresponds to a non-topological dynamically metastable phase, with no MBs. Nonetheless, transient amplification is observed \cite{Bosoranas}. As explained in Sec.\,\ref{DynSec}, this follows due to the fact that non-trivial pseudospectra (equivalently, the infinite-size limit spectra) emerges in the right-half complex plane. In Sec.\,\ref{MultiSec} we will see that this non-topological metastability can be distinguished from its topological counterpart using two-time correlation functions.

\subsection{A dissipative bosonic Kitaev chain with a pure steady state}\label{PureSSSec}

In all models considered thus far, we have confirmed by numerically determining the purity that the SS of the OBC configuration is always a mixed state, see Fig.\,\ref{dbkcpds}(b). This prevents us from exploring the potential interplay between topological metastability and pure SSs (so-called ``dark states" in quantum optics). Following the procedure detailed in Sec.\,\ref{PureSSApp}, however, it is always possible, given a dynamically stable Hamiltonian with QPV $\ket{\widetilde{0}}$, to engineer a dissipator that relaxes any initial condition to $\ket{\widetilde{0}}$. \rev{In the spirit of dissipative pure-state stabilization \cite{TicozziMarkov,Kraus2008},} such a procedure is especially interesting when \rev{such a} vacuum state possesses certain nontrivial properties such as non-zero squeezing, as is the case with the BKC. Can topological metastability arise in such a system?  The answer is Yes, as we will now show by applying the procedure to the BKC Hamiltonian under OBCs. 

Consider the Lindbladian defined via the BKC Hamiltonian (with $\mu=0$ for simplicity) and the site-local dissipator
\begin{equation}
\label{SqD}
\mathcal{D} = 2\kappa \sum_{j=1}^N\mathcal{D}[\alpha_j(r)] , \quad \kappa >0, 
\end{equation}
where we have introduced squeezed bosonic degrees of freedom according to
\begin{equation*}
\alpha_j(r) \equiv  \cosh(jr)a_j - \sinh(j r)a_j^\dag ,
\end{equation*}
and the squeezing parameter $r$ is fixed according to $\tanh(r) = \Delta/J$. This represents a local dissipative process of the squeezed degrees of freedom with constant loss rate $2\kappa$. A number of remarks are in order.

First, the dissipator is not bulk-translationally invariant. Explicitly, the squeezing of the mode $\alpha_j(r)$ increases with $j$. Remarkably, however, the dynamical matrix $\mathbf{G} = \mathbf{G}\big|_{\kappa=0} - i\kappa \mathds{1}_{2N}$ {\em is} translationally invariant, up to BCs. In fact, it is precisely equivalent to that of the DBKC of Sec.\,\ref{DBKCSec}. Mathematically, the translation invariance-breaking terms are entirely contained in the bosonic projection of the GKLS matrix $\mathbf{M}$, while $\mathscr{F}(\mathbf{M}) = \kappa \tau_3$. Thus, the Heisenberg equations of motion for the Nambu array, and thus all linear observables, are translationally invariant, up to BCs.  In this sense, we say this model has a \textit{restricted translational symmetry}. This allows for a straightforward computation of rapidities and pseudospectra. Since the dynamical matrix is equivalent to that of the DBKC, the chain is topologically metastable for $|\kappa/\Delta|<1$ and anomalously relaxing for $|\kappa/\Delta|> 1$. In particular, there are two MB pairs in the topologically metastable regime. 

Second, the dissipator has a particularly simple representation in terms of a set of normal modes of the BKC Hamiltonian. Recall that the set of normal modes of the BKC is non-unique (as a consequence of anti-commutation with time-reversal and a resulting $\pm$ symmetry in the many-body energies), with each choice of normal modes corresponding to a particular QPV \cite{Decon}. We may choose one particular set and, under OBCs, diagonalize the BKC Hamiltonian as $H_{\text{BKC}}=\sum_\mu \omega_\mu (\psi_\mu^\dag \psi_\mu+1/2)$, for bosonic quasi-particles $\{\psi_\mu\}$ that satisfy the CCRs \cite{ClerkBKC,Decon} and are `standing-waves' in the squeezed basis $\alpha_j(r)$,
\begin{equation*}
\psi_\mu = \sqrt{\frac{2}{N+1}}\,\sum_{j=1}^N i^j \sin\left(\frac{\mu \pi j}{N+1}\right) \alpha_j(r).
\end{equation*}
From here, the dissipator is given by
\begin{equation}
\label{Dpss}
\mathcal{D} = 2\kappa \sum_{\mu=1}^N \mathcal{D}[\psi_\mu].
\end{equation} 
That is, in the normal mode basis, the chain is a set of $N$ decoupled dissipative harmonic oscillators with mode-varying frequencies $\omega_\mu$ and uniform damping $\kappa$ \footnote{The fact that the dissipator has more than one diagonal representation, Eqs.\,\eqref{SqD} and \eqref{Dpss}, results from a large degeneracy in the GKLS matrix spectrum. In particular, $\mathbf{M}$ has an $N$-fold degenerate $0$ eigenvalue. Different diagonal representation of $\mathcal{D}$ are derived from different choices of bases for the $0$ eigenvalue eigenspace. }.  Equivalently, $L_\mu = \sqrt{2\kappa}\,\psi_\mu$. We further note that dissipators diagonal in the normal mode basis of the system Hamiltonian often arise in self-consistent derivations of quadratic master equations \cite{RossiniMicroQuad}.

\subsubsection{Relaxation dynamics}

Eq.\,\eqref{Dpss} makes the identification of the SS trivial. If $\ket{\widetilde{0}}$ is the QPV of the BKC, that is, $\psi_\mu \ket{\widetilde{0}} = 0$ for all $\mu$, then $\mathcal{L}(\ket{\widetilde{0}}\bra{\widetilde{0}})=0$. This state is a squeezed Gaussian state that we have constructed explicitly in Ref.\,\onlinecite{Decon}. Since it is defined as a QPV of $H_\text{BKC}$, it is necessarily a function of the Hamiltonian parameters $J$ and $\Delta$ only. In particular, it is insensitive to $\kappa$, and hence to the topology of the rapidity bands. This demonstrates that topological metastability need not be reflected in the structure of the SS. 

Given a system with both a pure SS and topological metastability, the manifold of quasi-SSs admits a simple characterization. Let $\gamma^s$ be one of two edge-symmetries $\gamma_L^s$ or $\gamma_R^s$. Following Sec.\,\ref{MBGen}, we construct the Weyl-displaced quasi-SSs, $\rho_\theta  \equiv e^{i\theta \gamma^s}\ket{\widetilde{0}}\bra{\widetilde{0}}e^{-i\theta\gamma^s}$, which are also pure. In fact, the states $\ket{\vec{\alpha}(\theta),\{\psi_\mu\}}\equiv e^{i\theta\gamma^s} \ket{\widetilde{0}}$, which are generically squeezed coherent states with respect to the physical degrees of freedom $\{a_j\}$, can be interpreted as coherent states with respect to the squeezed bosonic normal modes $\{\psi_\mu\}$. Explicitly,
\begin{equation*}
\psi_\mu \ket{\vec{\alpha}(\theta),\{\psi_\mu\}} = \alpha_\mu(\theta) \ket{\vec{\alpha}(\theta),\{\psi_\mu\}}, \quad \alpha_\mu(\theta)\in {\mathbb C}.
\end{equation*}  
The amplitudes $\alpha_\mu(\theta)$ encode the extent to which the normal modes commute with the SG, i.e.,  $[\psi_\mu,\gamma^s] = i\theta\vec{\psi}_\mu^\dag \bm{\tau}_3 \vec{\gamma}^s\,1_{\mathcal{F}} $ $= \alpha_\mu(\theta) 1_{\mathcal{F}}$, where $\vec{\psi}_\mu$ and $\vec{\gamma}^s$ are the single-particle vectors associated to $\psi_\mu$ and $\gamma^s$, respectively. As in the general case, the mean vector $\vec{m}(t) = \braket{\vec{\alpha}(\theta),\{\psi_\mu\}|\Phi(t)|\vec{\alpha}(\theta),\{\psi_\mu\}} = i\theta\vec{\gamma}^s(t)$ is long-lived (recall Eq.\,\eqref{mbound}). Due to the correspondence between dynamical matrices, this is the same mean vector as the mixed quasi-SSs in the DBKC. 

Identifying the quasi-SSs as coherent states evolving under decoupled dissipative harmonic dynamics (in the normal-mode basis) affords us the ability to compute their dynamics exactly \cite{FujiiDissQHO,KorschDissQHO} and analyze their relaxation dynamics more precisely. As they evolve, the states remain coherent, i.e., $\rho_\theta(t) = \ket{\vec{\alpha}(\theta,t),\{\psi_\mu\}}\bra{\vec{\alpha}(\theta,t),\{\psi_\mu\}}$, with amplitudes
\begin{equation*}
\alpha_\mu(\theta,t) =\alpha_\mu(\theta) e^{-(\kappa+i\omega_\mu)t} = \braket{\psi_\mu}(t),  
\end{equation*}
which relax to equilibrium {\em exponentially fast} in time. This apparent paradox is resolved by noting that these amplitudes correspond to the expectation values of $\psi_\mu$ which, due to squeezing, have exponentially large coefficients when expressed in the basis of the physically relevant degrees of freedom $\{a_j\}$. This explains why the normal-mode amplitudes decay exponentially, despite the mean vector being long-lived, with a lifetime increasing in system size. While this property holds independently of which SG is chosen (left- or right-localized), the exact nature of the relaxation dynamics do. 

\begin{figure}
\includegraphics[width=.75\columnwidth]{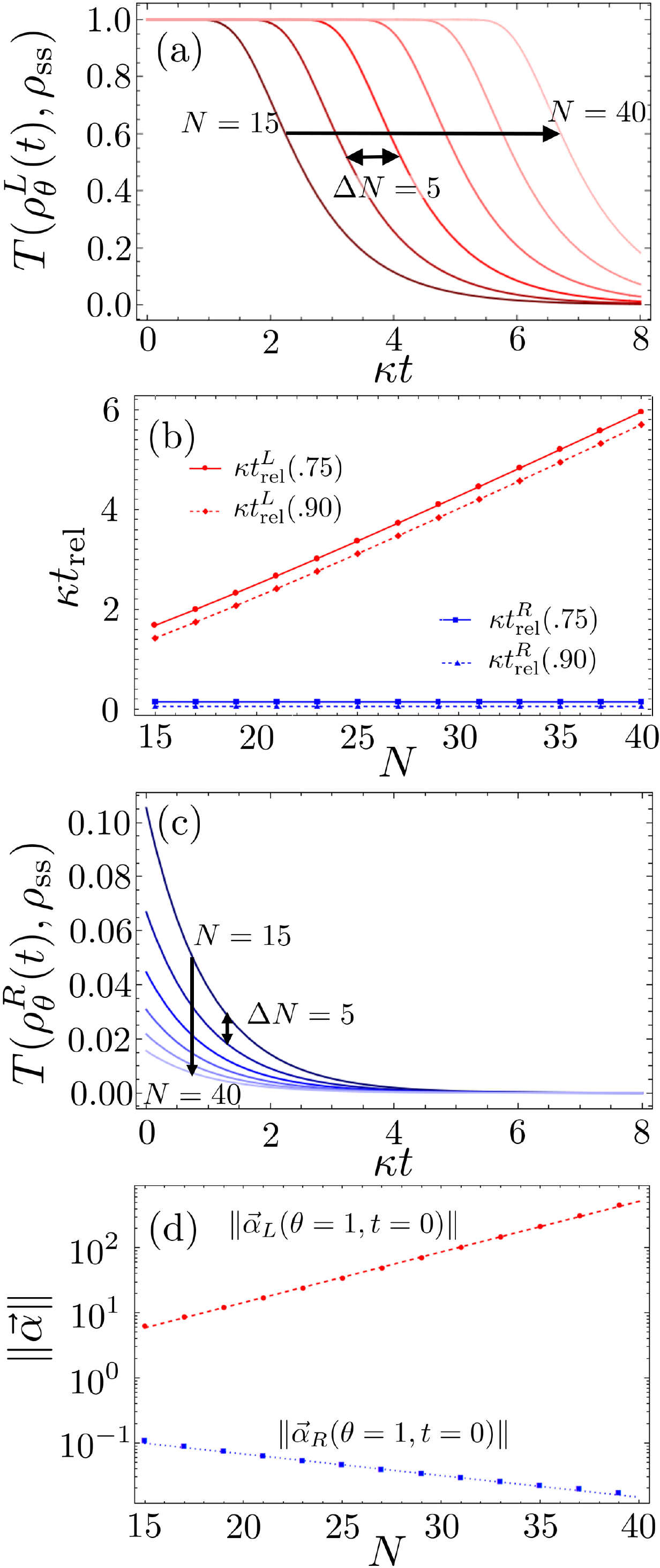}
\vspace*{-1mm}
\caption{{\bf (a)} Relaxation dynamics of the left-localized quasi-SS as $N$ increases. {\bf (b)} Same as in {\bf (a)}, but for the right-localized quasi-SS. {\bf (c)} Relaxation times $t_\text{rel}(\delta)$ for the left- (red) and right- (blue) localized quasi-SSs for accuracies $\delta=0.75$ (solid) and $\delta=0.90$ (dashed). {\bf (d)} Scaling of the initial amplitude vector norms for the left- (red) and right- (blue) localized quasi-SSs with system size. In all cases, we take $\theta=1$, symmetrically normalize the MBs, and set $J=2$, $\Delta=0.5$, and $\kappa=0.3$. 
}
\label{RelDyns}
\end{figure}

Consider the trace distance between two quantum states, $T(\rho,\sigma) = (1/2)\norm{\rho-\sigma}_\text{tr}$, which provides a measure of their distinguishability. Thanks to our knowledge of the exact dynamics, we can study the relaxation time of the quasi-SSs via the distance from equilibrium,
\begin{eqnarray*}
T(\rho_\theta(t) , \rho_\text{ss}) &=& \sqrt{1-|\braket{\vec{\alpha}(\theta,t),\{\psi_\mu\}|\widetilde{0}}|^2}
\\
&=& \sqrt{1-\exp\left(-\norm{\vec{\alpha}(\theta,0)}^2e^{-2\kappa t}\right)},
\end{eqnarray*} 
where $\vec{\alpha}(\theta,t)$ is either $\vec{\alpha}_L(\theta,0)$ or $\vec{\alpha}_R(\theta,0)$ depending which SG is used to generate the quasi-SSs. We then define the relaxation time to be the time $t_\text{rel}(\delta)$ such that the relative distance from equilibrium $T(\rho_\theta(t) , \rho_\text{ss})/T(\rho(0) , \rho_\text{ss})$ falls (and remains) below a prescribed accuracy $\delta>0$. It follows that 
\begin{equation*}
2 \kappa \,t_\text{rel}(\delta) = 
\ln\left[
\frac{\norm{\vec{\alpha}(\theta,0)}^2}{\ln\left[\left(1-\delta(1-e^{-\norm{\vec{\alpha(\theta,0)}}^2})
\right)^{-1}\right]}
\right] .
\end{equation*}
The system-size scaling of $t_\text{rel}(\delta)$ is thus explicitly tied to the system-size scaling of 
the norm of the initial amplitude vector, $\norm{\vec{\alpha}(\theta,0)}$. Remarkably, the two manifolds of quasi-SSs display dramatically different relaxation dynamics, as inferred from their relaxation times. We find, numerically, that $\norm{\vec{\alpha}_L(\theta, 0)}$ increases exponentially with $N$, whereas $\norm{\vec{\alpha}_R(\theta,0)}$ decreases exponentially. The consequences of this observation are displayed in Fig.\,\ref{RelDyns}: the left-localized quasi-SS exhibits a increasingly long relaxation time, while the right-localized shows just the opposite. The asymmetry in the dynamics may be understood by noting that, while the dynamical matrix is translationally invariant (up to boundaries), the full generator is not. Specifically, the Lindblad operators $\alpha_j(r)$ are right-localized. 

The boundedness of the relaxation times for the right-localized state does not contradict our claim of long-livedness. Per the general theory, the lifetime of the {\em physically accessible} degrees of freedom, as captured by the macroscopic mean vector $\vec{m}_\theta^R(t)$, remains indistinguishable from its initial value $\vec{m}_\theta^R(0)$, as exemplified by Eq.\,\eqref{mbound}. On the other hand, the quasi-SS and the true SS have overlap exponentially close to one as $N$ increases. The resolution comes by noting that $\vec{m}^R_\theta(t)$ is obtained from the exponentially small normal mode amplitudes $\vec{\alpha}_R(\theta,t)$ via a squeezing transformation. This transformation dilates the exponentially small amplitudes into the physical amplitudes. Similarly, the same transformation squeezes the exponentially large normal mode amplitudes $\vec{\alpha}_L(\theta,t)$ into the physical amplitudes $\vec{m}_\theta^L(\theta,t)$.

\subsubsection{Transient odd-parity behavior}

To conclude our analysis of this model, we pose the question: Do the pure quasi-SSs span a subspace of long-lived states? That is, do linear combinations of these pure states provide long-lived states themselves? This question is interesting from two perspectives. Firstly, from the perspective of continuous-variable quantum information, linear combinations of coherent states provide the simplest realization of bosonic cat-codes \cite{LiangCatCodes,CatCodesReview}. Thus, the ability to potentially dissipatively prepare and sustain such states could have unforeseen implications, for instance in the context of robust encodings. Secondly, from the perspectives of condensed-matter physics and quantum optics, ground states and SSs are intrinsically {\em averse to odd bosonic parity}. Specifically, if a QBH (QBL) has a unique ground (steady) state, then it must have {\em even} bosonic parity. In the Hamiltonian case, this eliminates the possibility for the parity switching behavior characteristic of topological superconductors. Moreover, negative expectation values of parity signify quantum non-Gaussianity.

\begin{figure}
\includegraphics[width=.85\columnwidth]{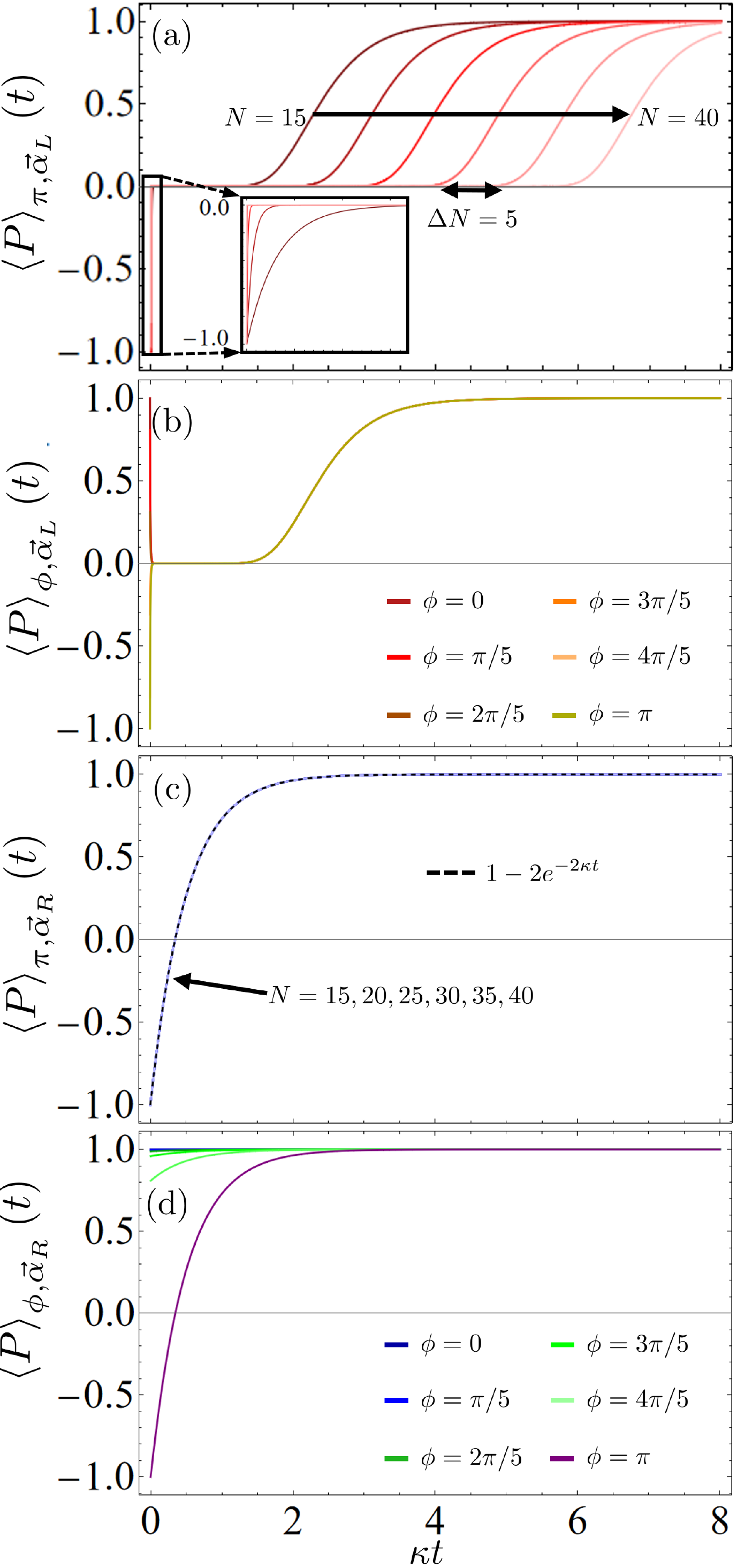}
\vspace*{-2mm}
\caption{{\bf  (a)} Parity dynamics of the initial odd-parity cat state formed from the left-localized quasi-SS for increasing $N$. Inset: Short time behavior. {\bf (b)} Parity dynamics of the initial cat state formed from the left-localized quasi-SS for $N=25$ and varying $\phi$. {\bf (c)} Parity dynamics of the initial odd-parity cat state formed from the right-localized quasi-SS for increasing $N$. The dashed line indicates the $N\to\infty$ limit. {\bf (d)} Parity dynamics of the initial cat state formed from the right-localized quasi-SS for $N=25$ and varying $\phi$. In all cases, we take $\theta=1$, symmetrically normalize the MBs, and set $J=2$, $\Delta=0.5$, and $\kappa=0.3$.
}\label{ParityPlot}
\end{figure}

In general, the question is difficult to answer. However, we can make interesting statements about particularly relevant linear combinations. Concretely, let us consider bosonic ``cat states'' of the form 
\begin{equation*}
\ket{\mathcal{C}_\phi(\vec{\alpha})} = \mathcal{N}_{\phi}(\vec{\alpha}) \Big( \ket{\vec{\alpha}(\theta),\{\psi_\mu\}} + e^{i\phi} \ket{-\vec{\alpha}(\theta),\{\psi_\mu\}} \Big),
\end{equation*}
where $\mathcal{N}_\phi(\vec{\alpha})$ is a normalization constant, and $\phi\in [0,2\pi]$. Being generally non-Gaussian, these states are not simply characterized by their first and second cumulants. Moreover, computation of overlaps becomes substantially more difficult than in the previous case. Instead of a direct analysis of relaxation times, we focus explicitly on parity dynamics,
\begin{align*}
\braket{P}_{\phi,\vec{\alpha}}(t) = \braket{\mathcal{C}_\phi(\vec{\alpha})|P(t)|\mathcal{C}_\phi(\vec{\alpha})},
\end{align*}
with $P$ the bosonic parity operator introduce in Sec.\,\ref{QBLs}. It is then possible to derive the following exact analytical result (see App.\,\ref{CatParApp} for detail):
\begin{equation}
\label{Monster}
\braket{P}_{\phi,\vec{\alpha}}(t) = \frac{e^{-2\norm{\vec{\alpha}}^2}+\cos(\phi) e^{-2 \norm{\vec{\alpha}}^2(1-e^{-2\kappa t})}}{1+\cos(\phi)e^{-2\norm{\vec{\alpha}}^2}}.
\end{equation}
Again, which quasi-SS manifold is chosen strongly affects the ensuing behavior. If $\vec{\alpha} = \vec{\alpha}_L(\theta,0)$, the parity shows a dramatic dependence on both $N$ and $\phi$, as seen in Fig.\,\ref{ParityPlot}. Unless $\phi=\pm \pi/2$, the parity drops incredibly fast to zero, remains zero for a transient time that scales linearly with $N$, and then eventually rises up to the asymptotic value of $1$. The extremely fast initial drop to zero corresponds to a singularity in the derivative
\begin{equation*}
\braket{\dot{P}}_{\phi,\vec{\alpha}}(0) = -4\norm{\vec{\alpha}_L(\theta,0)}^2 
\bigg( \frac{e^{-2\norm{\vec{\alpha}}^2}-\cos(\phi)}{1+e^{-2\norm{\vec{\alpha}}^2}\cos(\phi)}\bigg),
\end{equation*}
as $N$ (and hence $\norm{\vec{\alpha}}$) goes to infinity. The transient state of zero parity that follows is interesting, as it represents a long-lived period where the measurement statistics of parity are split evenly between the $+1$ and $-1$ outcomes. While we do not sustain a state of odd parity explicitly, we do, in fact, sustain an even mixture between even and odd parity sectors, for a transient that grows linearly with system size.

In sharp contrast, if $\vec{\alpha}=\vec{\alpha}_R(\theta,0)$, the parity dynamics are much more well-behaved. The exponentially small norms of $\vec{\alpha}_R(\theta,0)$ ensure that, for sufficiently large, $N$, 
\begin{equation*}
\lim_{N\to\infty}\braket{P}_{\phi,\vec{\alpha}_R}(t) = \begin{cases}
1 & \phi\neq \pi,
\\
1-2e^{-2\kappa t} & \phi = \pi .
\end{cases}
\end{equation*}
Accordingly, unless $\phi=\pi$, the parity approaches $1$ for all $t$ as $N$ increases. Moreover, when $\phi=\pi$, the parity is indistinguishable from $1-e^{-2\kappa t}$ for sufficiently large $N$. Regardless, unlike their left-localized partners, these right-localized cat states fail to support any semblance of an odd-parity state for any meaningful amount of time. 

\section{Topological dynamical metastability in number-symmetric QBLs}
\label{TopoMS2}

\subsection{Dirac edge bosons: General properties}
\label{numsymtopmet}

The topological modes in all the model systems of Sec.\,\ref{TopoMS1} bear the closest resemblance to the Majorana fermions of topological superconductivity. Thus, one could suspect that our analysis may fail to produce topological modes analogous to the ones hosted by topological insulators. To see why this is false, we consider QBLs with number symmetry to provide the natural dissipative-bosonic analogues of insulators.

Number-symmetric QBLs possess a weak U(1) symmetry \rev{of the form} $a_j\mapsto e^{i\phi}a_j$, generated by the total number operator $\sum_{j} a_j^\dag a_j$. The corresponding Bogoliubov transformation is $\Phi\mapsto e^{i\bm{\tau}_3\theta}\Phi$. Thus, number symmetry implies $[\mathbf{G},\bm{\tau}_3]=0$, and the dynamical matrix may be written as 
\begin{equation}
\label{GK}
\mathbf{G} = \mathbf{K} \otimes \begin{bmatrix}
1 & 0 \\ 0 & 0
\end{bmatrix} +  (-\mathbf{K}^*)\otimes \begin{bmatrix}
0 & 0 \\ 0 & 1
\end{bmatrix},
\end{equation}
with $\mathbf{K}$ an arbitrary $N\times N$ complex matrix. Just as the spectra of $\mathbf{G}$ is determined by that of $\mathbf{K}$, so is the pseudospectra. In particular, if the system is topologically metastable, both $\mathbf{G}$ and $\mathbf{K}$ possess approximate kernel vectors. 

To be more concrete, let us suppose the system is topologically metastable. This affords us at least one MB pair $(\gamma^z_1,\gamma^s_1)$. Expanding in the bosonic basis yields
\begin{align*}
\gamma^z_1 = \sum_{j=1}^N (u_j^* a_j + u_j a^\dag_j),\quad \gamma^s_1 = \sum_{j=1}^N (v_j^* a_j + v_j a^\dag_j ),
\end{align*}
where $u_j$ and $v_j$ are related to the Nambu vectors via $\vec{\gamma}^z_1 = [u_1,-u_1^*,\ldots,u_N,-u_N^*]^T,$ 
$\vec{\gamma}^{s}_1 = [v_1,-v_1^*,\ldots,v_N,-v_N^*]^T.$ The HWRs imply that $\text{Im} \sum_{j=1}^N u_j^* v_j=1/2$, since
\begin{align*}
i = [\gamma_1^z,\gamma_1^s] = \sum_{j=1}^N (u_j^* v_j - v_j^* u_j ) = 2i\,\text{Im} \sum_{j=1}^N u_j^* v_j.
\end{align*}
Number symmetry additionally implies that an approximate (or exact) ZM or SG remains so after a phase rotation $a_j\mapsto e^{i\phi}a_j$. In particular, we may rotate each of our MBs to construct a {\em second, linearly independent} MB pair. Specifically, if we fix $\phi=\pi/2$, then we immediately find another linearly independent MB pair given by 
 \begin{align*}
\gamma^z_2 = i\sum_{j=1}^N (u_j^* a_j-u_j a^\dag_j),\quad
\gamma^s_2 = i\sum_{j=1}^N (v_j^*a_j - v_j a^\dag_j).
\end{align*}

We have not yet encountered anything resembling the edge modes of topological insulators. For that, we need to utilize {\em both} MB pairs. Explicitly, consider the operators
\begin{eqnarray*}
\label{ab2mb}
\alpha &\equiv & \frac{1}{2\sqrt{C^z}}(\gamma^z_1 - i \gamma^z_{2}) = \frac{1}{\sqrt{C^z}}\sum_{j=1}^N u_j^* a_j
,\\
\beta &\equiv & \frac{1}{2\sqrt{C^s}}(\gamma^s_1 - i \gamma^s_{2}) = \frac{1}{\sqrt{C^s}}\sum_{j=1}^N v_j^* a_j,
\end{eqnarray*}
where $C^z = \sum_{j} |u_j|^2$ and $C^s = \sum_j |v_j|^2$ are positive real numbers chosen to ensure that the {\em Dirac boson modes} $\alpha$ and $\beta$ satisfy the following properties:
\begin{itemize}
\item They are bosonic: $[\alpha,\alpha^\dag] = [\beta,\beta^\dag] = 1_{\mathcal{F}}$.\vspace*{-2mm}
\item They are algebraically related via $[\alpha,\beta]=0$ and $[\alpha,\beta^\dag] \neq  0$.\vspace*{-2mm}
\item They are edge-localized according to the localization of the constituent MBs.\vspace*{-2mm}
\item The bosonic mode $\alpha$ is approximately conserved, while the real and imaginary quadratures of $\beta$ generate two (non-commuting) approximate symmetries. 
\end{itemize}
Algebraically speaking, we have constructed edge modes on which the number symmetry acts trivially. The necessary and sufficient ingredients for this construction are topological metastability and number symmetry, and so it applies broadly.

The normalization of these operators is far more straightforward than in the number non-symmetric case. To see this, consider a left-localized, approximately conserved, bosonic operator $\alpha$, i.e.,
\begin{equation}\label{alphagen}
\alpha = \mathcal{M}(N)\sum_{j=1}^N \delta^{j-1} a_j , \quad |\delta|<1.
\end{equation}
Taking $\delta >0$ without loss of generality, canonical commutation relations require
\begin{align}
\mathcal{M}(N) = \sqrt{\frac{1-\delta^{2}}{1-\delta^{2N}}}
\label{DBNorm}
\end{align}
which, unlike in the case of the MBs, converges to a finite value as $N\to\infty$. Namely, $\lim_{N\to\infty}\mathcal{M}(N) = \sqrt{1-\delta^2}$.
With this, the exact ZM in the $N\to\infty$ limit is
\begin{equation*}
\alpha = \sqrt{1-\delta^2} \,\sum_{j=1}^\infty \delta^{j-1}a_j .
\end{equation*}
While, in general, the exact expression for $\alpha$ (and, similarly, $\beta$) could be more complicated than Eq.\,\eqref{alphagen}, this argument demonstrates the unambiguous normalizability of the modes in the infinite size limit. 

It is interesting to ask whether a number-symmetric analogue to the PDMC may exist, namely, a purely dissipative, number-symmetric chain that exhibits dynamical metastability.  In the purely dissipative case, we have $\mathbf{G} = -i\bm{\tau}_3\mathcal{F}(\mathbf{M})$, and the number symmetry property $[\mathbf{G},\bm{\tau}_3]=0$ manifests as $[\mathcal{F}(\mathbf{M}),\bm{\tau}_3]= 0$. Thus, $\mathbf{G}^\dag = (-i\bm{\tau}_3\mathcal{F}(\mathbf{M}))^\dag = i\bm{\tau}_3\mathcal{F}(\mathbf{M}) = -\mathbf{G}.$ That is, a purely dissipative number-symmetric chain must have an anti-Hermitian (therefore, normal) dynamical matrix $\mathbf{G}$. We conclude that no such model can exhibit dynamical metastability.

\subsection{A number-symmetric dissipative chain}
\label{Dirac}

Let us explore the interplay between number symmetry and topological metastability in a concrete example. The dynamical matrix of a number-symmetric QBL has the general form Eq.\,\eqref{GK}, where $\mathbf{K}$ an arbitrary $N\times N$ complex matrix. Topological metastability can then be engineered through an appropriate choice of $\mathbf{K}$. Specifically, let us consider 
\begin{equation}
\label{HNK}
\mathbf{K} = -i\kappa \mathds{1}_N + J_L \mathbf{S} + J_R \mathbf{S}^\dag ,
\end{equation}
with $\kappa\geq 0$, $J_L,J_R\in\mathbb{R}$, and $\mathbf{S}$ the usual BC-dependent shift operator. We identify this matrix as that of the Hatano-Nelson (HN) asymmetric hopping model \cite{HatanoNelson}, with an identity shift that will ultimately serve to stabilize the QBL. For convenience, we define $J_\pm \equiv  (J_L\pm J_R)/2$. 
The Hamiltonian can be unambiguously determined from $\mathbf{G}$ and is given by
\begin{equation*}
H_\text{NS} = \frac{J_+}{2}\sum_{j=1}^N \left( a_j^\dag a_{j+1} + a_{j+1}^\dag a_j\right) .
\end{equation*}
Per usual, the QBL is not fully determined until we specify $\mathcal{B}(\mathbf{M})$. Moreover, the second necessary condition for the U(1) symmetry is that $\bm{\tau}_3 \mathcal{B}(\mathbf{M})$ commutes with $\bm{\tau}_3$. Together, $[\mathbf{G},\bm{\tau}_3]=0$ and $[\bm{\tau}_3\mathcal{B}(\mathbf{M}),\bm{\tau}_3]=0$ are necessary and sufficient for the \rev{requisite} U(1) symmetry. We specify $\mathcal{B}(\mathbf{M})$ implicitly, by defining the dissipator $\mathcal{D}_\text{NS} \equiv \mathcal{D}_{-,0} +\rev{\mathcal{D}_{+,0}+ } \mathcal\mathcal{D}_{-,1}$, with
\begin{eqnarray*}
\mathcal{D}_{-,0} &=& 2\kappa_- \sum_{j=1}^N\mathcal{D}[a_j],
\\
\mathcal{D}_{+,0} &=& 2\kappa_+ \sum_{j=1}^N\mathcal{D}[a_j^\dag] ,
\\
\mathcal{D}_{-,1} &=& 2i J_-\sum_{j=1}^N\mathcal{D}[a_j,a_{j+1}^\dag]  - \mathcal{D}[a_{j+1},a_{j}^\dag].
\end{eqnarray*}
Here, $2\kappa_-\geq 0$ and $2\kappa_+\geq 0$ are the onsite loss and gain rates, respectively, while $2J_-$ takes the role of the NN loss rate. The GKLS matrix is positive-semidefinite for OBCs and PBCs, and for all $N$, if $\kappa_- \geq 2|J_-|$. This QBL has a dynamical matrix specified by Eq.\,\eqref{HNK} if we further identify $\kappa \equiv \kappa_--\kappa_+$. We refer to this model as the dissipative number-symmetric (DNS) chain. The SS behavior of a related model has been considered in Ref.\,\onlinecite{ClerkNESS}.

\begin{figure}
\includegraphics[width=.65\columnwidth]{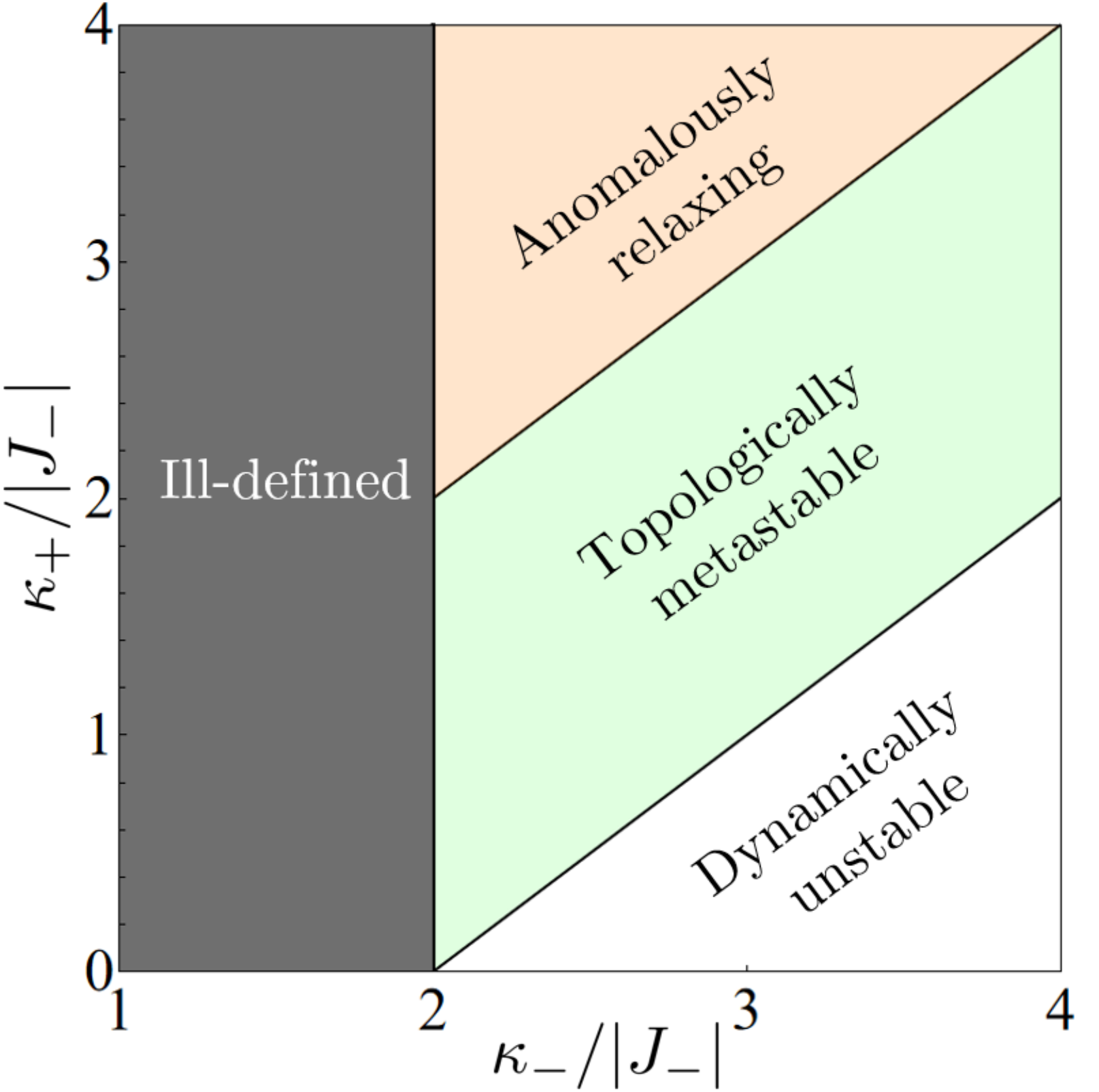}
\vspace*{-3mm}
\caption{The stability phase diagram for the DNS under OBCs with $|J_+|> |J_-|$ under OBCs. The ``Ill-defined" region corresponds to the parameter regime where $\mathbf{M}$ is no-longer positive semidefinite.}\label{u1pd}
\end{figure}

The rapidities can be easily obtained from the well-known HN spectrum, and closely resemble that of the DBKC. For BIBCs, the bands are given by $\{\lambda(k), \lambda(k)^*\}$, with
\begin{equation*}
\lambda(k) = -\kappa + 2J_-\sin(k) +i 2J_+\cos(k).
\end{equation*}
These bands trace out an ellipse centered at $-\kappa$ in the complex plane. Winding about the origin requires $2|J_-|>\kappa$. 
For (finite-size) OBCs, the eigenvalues are given by $\lambda_m$,  $m=0,\ldots,2N-1$, where
\begin{equation*}
\lambda_m = -\kappa + 2i\sqrt{J_+^2-J_-^2} \cos\left(\frac{m\pi}{N+1}\right).
\end{equation*}
To simplify the discussion (and fix the OBC Lindblad gap $\Delta_\mathcal{L} = \kappa$, for all $N$), we focus on $|J_+|\geq |J_-|$. Combining the GKLS-matrix positivity condition and the rapidity band-winding condition, we identify a topologically metastable regime whenever $\kappa_-/|J_-|\geq 2$ and $0 \leq \kappa/|J_-|\leq 2$. The OBC stability phase diagram is shown in Fig.\,\ref{u1pd}.

In the special case $J_R=0$, the pseudoeigenvectors of $\mathbf{K}$ (and $\mathbf{K}^\dag$) with zero pseudoeigenvalue may be computed analytically. Such pseudoeigenvectors can be used to build approximate kernel vectors of $\mathbf{G}$ and $\widetilde{\mathbf{G}} = \mathbf{G}^\dag$, which correspond to approximate SGs and ZMs, respectively. Specifically, consider the bosonic modes
\begin{align*}
\alpha \equiv \mathcal{M}(N)\sum_{j=1}^N (i\delta)^{N-j} a_j,
\quad
\beta \equiv \mathcal{M}(N)\sum_{j=1}^N (-i\delta)^{j-1} a_j ,
\end{align*}
where $\delta \equiv -\kappa/2J_-$ and $\mathcal{M}(N)$ is given in Eq.\,\eqref{DBNorm}. First, we have $[\alpha,\alpha^\dag]=[\beta,\beta^\dag]=1_\mathcal{F}$ and $[\alpha,\beta]=0$. Second, these operators satisfy
\begin{align}
\mathcal{L}^\star(\alpha) &= -\kappa \mathcal{M}(N)\, (-i\delta)^{N-1}a_1,
\label{alphaZM}
\\
\mathcal{L}^\star([\beta,A]) - [\beta,\mathcal{L}^\star(A)] &= -\kappa \mathcal{M}(N) \,(i\delta)^{N-1}[a_N,A], \; \forall A.
\label{betaSG}
\end{align}
Utilizing $\mathcal{L}^\star(A^\dag) = [\mathcal{L}^\star(A)]^\dag$ yields similar expressions for $\alpha^\dag$ and $\beta^\dag$. 
For $|\delta|<1$, the right hand-side of each equation goes to zero as $N\to\infty$. Physically, this means $\alpha$ is a bosonic approximate ZM while the real and imaginary quadratures of $\beta$ generate approximate symmetries. That is, $\alpha$ and $\beta$ are DBs. 

We may trace these two bosonic modes back to two pairs of MBs using Eqs.\,\eqref{ab2mb}. Specifically, we have
\begin{eqnarray*}
\gamma_{1}^z &=& \mathcal{M}_z(N)\sum_{j=1}^N\left( (i\delta)^{N-j} a_j + (-i\delta)^{N-j} a_j^\dag\right), 
\\
\gamma_{1}^s &=& \mathcal{M}_s(N)\sum_{j=1}^N\left( (-i\delta)^{j-1} a_j + (i\delta)^{j-1} a_j^\dag\right),
\\
\gamma_{2}^z &=& i\mathcal{M}_z(N)\sum_{j=1}^N\left( (i\delta)^{N-j} a_j - (-i\delta)^{N-j} a_j^\dag\right), 
\\
\gamma_{2}^s &=& i\mathcal{M}_s(N)\sum_{j=1}^N\left( (-i\delta)^{j-1} a_j - (i\delta)^{j-1} a_j^\dag\right),
\end{eqnarray*}
for normalization constants $\mathcal{M}_z(N)$ and $\mathcal{M}_s(N)$ chosen to ensure that $[\gamma_j^z,\gamma_j^s]=i1_{\mathcal{F}}$ for $j=1,2$. The relevant equations of motion follow from Eqs.\,\eqref{alphaZM}-\eqref{betaSG}, that is, 
\begin{align*}
\mathcal{L}^\star(\gamma_1^z) &= -\kappa \mathcal{M}_z(N) \delta^{N-1}Q_L ,
\\
\mathcal{L}^\star(\gamma_2^z) &= \kappa \mathcal{M}_z(N) \delta^{N-1}P_L ,
\\
\mathcal{L}^\star([\gamma_{1}^s,A]) - [\gamma_{1}^s,\mathcal{L}^\star(A)] &= -\kappa \mathcal{M}_s(N) \delta^{N-1}[Q_R,A] ,
\\
\mathcal{L}^\star([\gamma_{2}^s,A]) - [\gamma_{2}^s,\mathcal{L}^\star(A)] &= \kappa \mathcal{M}_s(N) \delta^{N-1}[P_R,A] ,
\end{align*}
with $Q_L \equiv \chi_N^* a_1 + \chi_N a_j^\dag$, $P_L\equiv -i(\chi_N^*a_1 - \chi_N a_1^\dag)$, $Q_R\equiv \chi_N a_N+\chi_N^*a_N^\dag$, $P_R\equiv -i(\chi_N a_N-\chi_N^* a_N^\dag)$, and $\chi_N\equiv i^{N-1}$. Following the discussion of Sec.\,\ref{MBGen}, there exists normalization schemes \rev{ensuring that} the right hand-sides of these four equations vanish as $N\to\infty$, while keeping each pair canonically conjugate (for instance, the symmetric normalization scheme we have previously employed). 

We remark that these pairs of MBs are {\em not} simply the real and imaginary quadratures of $\alpha$ and $\beta$, i.e., the operators $x_{\alpha}\equiv (\alpha+\alpha^\dag)/\sqrt{2}$, $p_\alpha\equiv i(\alpha-\alpha^\dag)/\sqrt{2}$ (and similarly for $\beta$). Instead, the MBs are {\em proportional} to these quadratures, according to the expressions above. The system-size-dependent proportionality constants are key to ensuring that the macroscopically separated pairs $(\gamma_1^z,\gamma_1^s)$ and $(\gamma_2^z,\gamma_2^s)$ are canonically conjugate. We further remark that each edge supports two Noether modes of the {\em same type}: the left edge supports two SGs, whereas the right edge supports two ZMs. This is to be contrasted with the previous models, which all featured at most one of each type on each edge. A similar phenomena is uncovered in a model without the additional constraint of number symmetry in App.\,\ref{DDW}.

\section{Observable signatures of topological metastability}
\label{MultiSec}

Just like spatial, equal-time correlation functions of physical observables play a central role in characterizing equilibrium phases of matter and phase transitions, multi-time correlation functions provide an essential tool for probing dynamical response properties of many-body systems away from equilibrium \cite{Politi,SciollaCorrelations}. In particular, standard material characterization techniques ranging from neutron scattering to reflectivity and photoemission spectroscopy rely on access to two-time correlation functions of the form $\langle A(t)B(t')\rangle$  \cite{BrownCorrelation,Freericks}; likewise, in the context of quantum optics, statistical properties of the photon field are directly related to first- and second-order {\em coherence functions}, $g_{ij}^{(1)}(t,t')\propto \braket{a_i^\dag(t) a_j(t')}$, and $g_{ij}^{(2)}(t,t') \propto \braket{a_i^\dag(t) a^\dag_j(t') a_i(t') a_j(t)}$,  
which can be inferred \rev{through} photon counting and \rev{quantum} interference experiments \cite{GardinerNoise,WallsOptics}. 

Two-time correlation functions have been suggested as a means for operationally characterizing multi-step relaxation dynamics in Markovian quantum systems \cite{GarrahanMeta,MacieszczakMeta}. Since, for the QBL systems we are interested in, dynamical metastability is inherently a {\em transient} phenomena, it is natural to expect that they may also serve as an experimentally accessible diagnostic tool, in principle. As we will show next, it is indeed possible to identify response functions whose behavior can not only distinguish topologically trivial from non-trivial dynamical metastability but, in the non-trivial case, further provide distinctive signatures of the different types of topological bosonic edge modes the system may host.

\subsection{Two-time correlation functions and power spectra}

Since systems in equilibrium are time-translation invariant, any two-time response function can only depend upon the relative time, say, $\tau\equiv t' - t$. In non-equilibrium situations, in contrast, the full dependence on both $t$ and $\tau$ becomes important in general. For Markovian dynamics, two-time averages may be computed under the \rev{assumption that the quantum regression theorem (QRT) holds} \cite{GardinerNoise}. Specifically, given two operators $A$ and $B$ and a state $\rho$, \rev{their two-point correlation function may then} be expressed as 
\begin{eqnarray*}
C^+_{A,B}(t,\tau) & \equiv &\braket{A(t+\tau)B(t)} \rev{ \overset{\text{QRT}}{=} \text{tr} {\{A(0)e^{\tau{\mathcal L}} [B(0)\rho(t)]\}}} \\
& =& \text{tr}[A(\tau)B(0)\rho(t)],\quad  t,\tau\geq 0 ,
\end{eqnarray*}
where $A(\tau) = e^{\tau \mathcal{L}^\star }(A)$ and $\rho(t) = e^{t\mathcal{L}}(\rho)$ and \rev{the superscript $+$ signifies that $B$ is measured first. In this way, under the same assumptions,} we also have
\begin{eqnarray*}
C^{-}_{A,B}(t,\tau ) &\equiv& \braket{A(t)B(t+\tau)} \overset{\text{QRT}}{=} \text{tr}[A(0)B(\tau)\rho(t)] \\
&=& [C^+_{B,A}(t,\tau)]^* , \quad t, \tau\geq 0.
\end{eqnarray*}

In the case of a unique SS $\rho_\text{ss}$, we consider the SS two-time correlation functions $\lim_{t\to\infty} C^\pm_{A,B}(t,\tau)$, which we can express compactly as
\begin{equation}
C^\text{ss}_{A,B}(\tau) = \begin{cases}
\text{tr}[A(\tau)B(0)\rho_\text{ss}], & \tau\geq 0, \\
\text{tr}[A(0)B(|\tau|)\rho_\text{ss}], & \tau<0.
\end{cases}
\label{SScorr}
\end{equation}
As it may be more practically significant in certain situations, we additionally define the (two-sided) SS power spectrum
\begin{equation*}
S^{\text{ss}}_{A,B}(\omega) = \int_{-\infty}^\infty e^{i\omega \tau}C^{\text{ss}}_{A,B}(\tau)\,d\tau .
\end{equation*}
Long-lived correlations are then revealed through large power-spectral peaks at zero frequency. To more appropriately captures the relative decay of correlations, we define the normalized correlation functions and power spectra \cite{NoriEPCohfn},
\begin{equation}
\label{NormCorr}
\widetilde{C}^\text{ss}_{A,B}(\tau) \equiv  \frac{C^\text{ss}_{A,B}(\tau)}{C^\text{ss}_{A,B}(0)},\quad
\widetilde{S}^\text{ss}_{A,B}(\omega) \equiv \frac{S^\text{ss}_{A,B}(\omega)}{C^\text{ss}_{A,B}(0)}.
\end{equation}

For QBLs, we will primarily focus on the case where $A$ and $B$ are linear forms, i.e., $A = \widehat{\vec{\alpha}}$ and $B=\widehat{\vec{\beta}}\mbox{}^\dag$, with $\vec{\alpha},\vec{\beta}\in\mathbb{C}^{2N}$. In this case, the unnormalized SS correlation functions and power spectra take on a simple closed form:
\begin{eqnarray}
C^\text{ss}_{\alpha,\beta^\dag}(\tau) = \begin{cases}
\vec{\alpha}{}^\dag \bm{\tau}_3 e^{-i\mathbf{G} \tau}\mathbf{Q}_\text{ss}\bm{\tau}_3\vec{\beta} & \tau\geq 0, \\
\vec{\alpha}{}^\dag \bm{\tau}_3 \mathbf{Q}_\text{ss}e^{i\mathbf{G}^\dag \tau}\bm{\tau}_3\vec{\beta} & \tau<0, 
\end{cases} \notag \\
S^{\text{ss}}_{\alpha,\beta^\dag}(\omega) = \vec{\alpha}\bm{\tau}_3[\bm{\chi}(\omega) \mathbf{Q}_\text{ss} + \mathbf{Q}_\text{ss} \bm{\chi}^\dag(\omega)]\bm{\tau}_3\vec{\beta}, 
\notag
\end{eqnarray}
in terms of a {\em susceptibility matrix} \cite{NunnenkampTopoAmp,NunnenkampRestore}
$$\bm{\chi}(\omega) \equiv i(\omega \mathds{1}_{2N} - \mathbf{G})^{-1}.$$ 
Mathematically, $\bm{\chi}(\omega)$ is resolvent of $-i\mathbf{G}$ evaluated at $-i\omega$. 

The restriction to linear forms further yields a {\em state-independent} notion of correlation functions. Note that
\begin{eqnarray*}
\alpha(\tau)\beta^\dag(0) &=& \frac{1}{2}\{\alpha(\tau),\beta^\dag(0)\} + \frac{1}{2}[\alpha(\tau),\beta^\dag(0)]\\
&=& \frac{1}{2}\{\alpha(\tau),\beta^\dag(0)\} + \frac{1}{2}\vec{\alpha}^\dag e^{-i\widetilde{\mathbf{G}}^\dag\tau}\bm{\tau}_3\vec{\beta}\,1_{\mathcal{F}}.
\end{eqnarray*}
Since quantum states have unit trace, we have
\begin{equation*}
C^+_{\alpha,\beta^\dag}(t,\tau) = C^{+,\text{cl}}_{\alpha,\beta^\dag}(t,\tau) + \frac{1}{2}\vec{\alpha}^\dag e^{-i\widetilde{\mathbf{G}}^\dag\tau}\bm{\tau}_3\vec{\beta},
\end{equation*}
where $C^{+,\text{cl}}_{\alpha,\beta^\dag}(t,\tau) = \frac{1}{2}\tr[\{\alpha(\tau),\beta^\dag(0)\}\rho(t)]$ is the classical (symmetrized) correlation function. We identify the quantum (antisymmetrized) correlation functions as
\begin{equation*}
C^{+,\text{qu}}_{\alpha,\beta^\dag}(\tau)= \frac{1}{2}\vec{\alpha}^\dag e^{-i\widetilde{\mathbf{G}}^\dag\tau}\bm{\tau}_3\vec{\beta},
\end{equation*}
whose state- and $t$-independence follows from the state independent nature of $\braket{[\alpha(\tau),\beta^\dag(0)]}$ when $\alpha$ and $\beta$ are linear forms. We define $C^{-,\text{cl}}_{\alpha,\beta^\dag}(t,\tau)$ and $C^{-,\text{qu}}_{\alpha,\beta^\dag}(\tau)$ analogously. 
In the case where $\alpha$ and $\beta$ are observables (self-adjoint), we have 
\begin{equation*}
i \text{Im} \,C_{\alpha,\beta^\dag}^{\pm}(t,\tau) = C^{\pm,\text{qu}}_{\alpha,\beta^\dag}(\tau), \quad \forall t, 
\end{equation*}
which directly relates the quantum correlation function to the imaginary part of the full one, {\em independently} of the state. Similar to Eq.\,\eqref{SScorr}, we can drop the $\pm$ by defining
\begin{equation*}
C^{\text{qu}}_{\alpha,\beta^\dag}(\tau) \equiv  \begin{cases}
C^{+,\text{qu}}_{\alpha,\beta^\dag}(\tau) & \tau\geq 0,
\\
C^{-,\text{qu}}_{\alpha,\beta^\dag}(|\tau|)& \tau< 0 ,
\end{cases}
\end{equation*}
The corresponding frequency-space quantum power spectrum may be defined and expressed as
\begin{eqnarray*}
S_{\alpha,\beta^\dag}^\text{qu}(\omega) &=& \int_{-\infty}^\infty e^{i\omega \tau} C^{\text{qu}}_{\alpha,\beta^\dag}(\tau)\,d\tau
\\
&=& \vec{\alpha}^\dag\left(\bm{\tau}_3\bm{\chi}(\omega) + \bm{\chi}^\dag(\omega)\bm{\tau}_3\right)\vec{\beta},
\end{eqnarray*}
which is, again, a state-independent quantity. 

\subsection{Distinguishing topologically non-trivial from trivial dynamical metastability}

Dynamically metastable systems can be either topologically trivial or non-trivial. Starting from 
the susceptibility matrix, one can predict certain properties of the SS power spectrum
that can distinguish the two regimes. We will focus on the normalized power spectra [Eq.\,\eqref{NormCorr}] in order to 
capture the relative behavior of correlations. This eliminates the influence of exponentially large 
\rev{steady state} second moments $\mathbf{Q}_\text{ss}$ (e.g., occupation numbers) that may arise
in systems displaying transient amplification.

\begin{figure*}
\includegraphics[width=\textwidth]{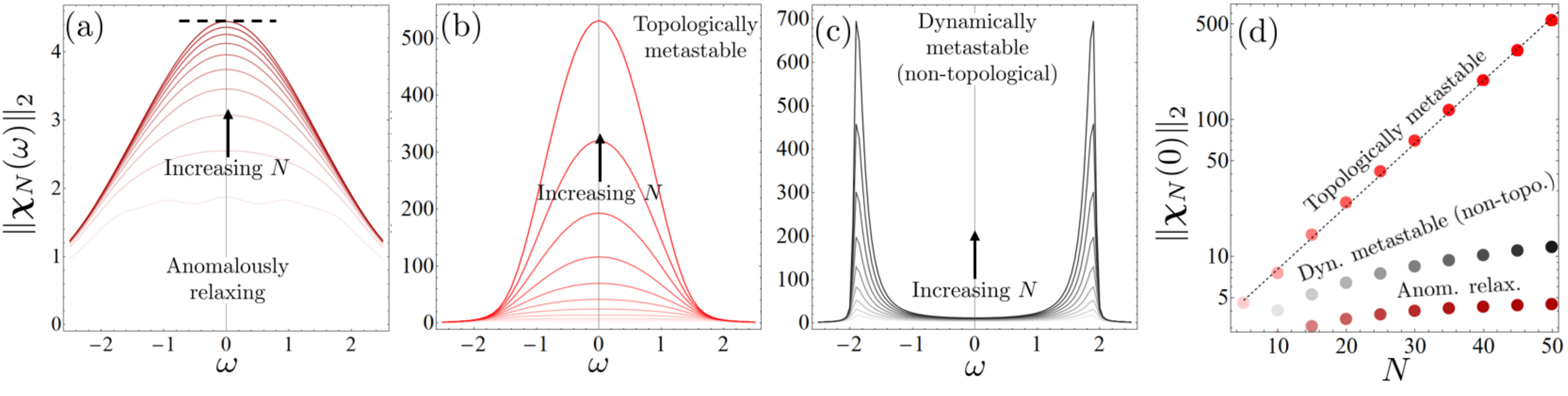}
\vspace*{-5mm}
\caption{{\bf (a-d)} The 2-norm of the DBKC's OBC susceptibility matrix in the anomalously relaxing, topologically metastable, and non-topological dynamically metastable regimes, respectively. {\bf (a)} and {\bf (b)} corresponds to the regimes whose rapidities are the open and filled markers in Fig.\,\ref{DBKCrapidities}(a), respectively. {\bf (c)} corresponds to the regime whose rapidities are shown in Fig.\,\ref{DBKCrapidities}(b). {\bf (d)} shows the zero-frequency susceptibility norm in the three aforementioned regimes. The dashed line is a linear fit.  
}\label{chiomega}
\end{figure*}

Let $\bm{\chi}_N(\omega)$ denote the susceptibility matrix/resolvent of the dynamical matrix 
$-i\mathbf{G}_N$ for an open chain of length $N$. On the one hand, if the chain is anomalously relaxing, 
then necessarily $\bm{\chi}_N(\omega)$ is bounded (in norm) for all $\omega$. The reason is that
the rapidity bands of anomalously relaxing systems are strictly bound to the left-half of the complex plane
and so $-i\omega$ is not in the SIBC spectrum. Thus, we have a system-size independent upper 
bound on $\norm{{\color{revred}\bm{\chi}}_N(\omega)}$. On the other hand, if the chain is dynamically metastable, then 
there is necessarily a subset of the imaginary axis contained within the SIBC rapidity spectrum. 
Equivalently, there are intervals on the imaginary axis about which the rapidity bands wind. Ultimately, 
the restriction of $\bm{\chi}_N(\omega)$ to these intervals will necessarily grow without bound as $N\to\infty$. 
Since topological metastability is characterized by the presence of zero in these non-trivial intervals, 
we conjecture that it generically elicits a {\em peak of the power spectrum at zero frequency}, which grows without 
bound with system size. There should be no such peak in a dynamically metastable system that is
topologically trivial.  

The distinctive behavior of $\bm{\chi}_N(\omega)$ in these regimes is exemplified in Fig.\,\ref{chiomega}. In (a), the $2$-norm of the susceptibility matrix converges for all values of $\omega$ considered. In (b) and (c), we see a divergence of the norm at frequencies $\omega$ such that $i\omega$ is contained in the non-trivial interior of the rapidity bands. The zero frequency behavior of $\norm{\bm{\chi}_N(\omega)}_2$ is shown in (d). In particular, the topological regime is distinguished from both the non-topological dynamically metastable regime and the anomalously relaxing regime by an exponential divergence in system size. We will further see that this manifests directly in certain quantum power spectra.

\subsubsection{Signatures of Majorana edge bosons}

Within the class of topological metastable QBLs with no U(1) symmetry, models may be distinguished based on whether their MBs are split or non-split. Let $(\gamma^z,\gamma^s)$ denote a split MB pair whose constituent modes satisfy $[\gamma^z,\gamma^s] = i$ and are localized on opposite sides of a chain. During the transient timescale ($t<t_N$, for some $t_N$ increasing with system size $N$), we have $\gamma^z(t) \simeq \gamma^z(0)$. However, because the MBs are split, $\gamma^s(t)$ deviates \rev{significantly} from $\gamma^s(0)$ over the same timescale. Consider the associated correlation function $C^\text{ss}_{\gamma^z,\gamma^s}(\tau)$, obtained from Eq.\,\eqref{SScorr}. Firstly, we note that canonical commutation implies the existence of non-zero \rev{equal-time quantum correlations,} at $\tau=0$. Explicitly, $C^\text{qu}_{\gamma^z,\gamma^s}(0) =i/2$. Remarkably, this persists in spite of the macroscopic spatial separation of the two modes. 

However, non-split MBs satisfy the same identity. To distinguish between split and non-split pairs, it is necessary to go beyond $\tau=0$. For  $0<\tau<t_N$, we have
\begin{eqnarray*}
C^\text{ss}_{\gamma^z,\gamma^s}(\tau) &=& \tr[\gamma^z(\tau)\gamma^s(0)\rho_\text{ss}] 
\\
&\simeq& \tr[\gamma^z(0)\gamma^s(0)\rho_\text{ss}]  = C^\text{ss}_{\gamma^z,\gamma^s}(0).
\end{eqnarray*}
On the other hand, for $\tau<0$ we have 
\begin{equation*}
C^\text{ss}_{\gamma^z,\gamma^s}(\tau) = \tr[\gamma^z(0)\gamma^s(|\tau|)\rho_\text{ss}]  \not\simeq C^\text{ss}_{\gamma^z,\gamma^s}(0).
\end{equation*}
If instead the MBs were non-split, we would additionally have $\gamma^s(t) \simeq \gamma^s(0)$ for $t<t_N$. So, $C^\text{ss}_{\gamma^z,\gamma^s}(\tau) \simeq C^\text{ss}_{\gamma^z,\gamma^s}(0)$. Therefore, split and non-split MBs may be distinguished by {\em asymmetries} in the associated \rev{steady state} correlation function around $\tau=0$, reflecting the fact that split MBs are approximately stationary for $0<\tau<t_N$, whereas non-split MBs are approximately stationary for $-t_N<\tau<t_N$. Two remarks are in order: (i) While we have treated the full \rev{steady state} correlation function explicitly, the same conclusions hold for both the classical and the (state-independent) quantum contributions of the \rev{two-time} correlation function. (ii) Per the properties of MBs, the stationarity of the correlation functions in each case will become more pronounced as system size is increased.

To exemplify the distinctions, we focus on the DBKC and the PDMC as representative examples of these two classes. Moreover, by leveraging the mapping described in App.\,\ref{PDCdetails} between these two models [Eq.\,\eqref{DBKCPDC}] allows us to directly compare topological phases. First, to distinguish the two models, we relabel the quantities $J$, $\Delta$, and $\mu$ in the PDMC by $J_F$,  $\Delta_F$, and $\mu_F$, respectively; then, we identify $J_F = \Delta/2,$ $\Delta_F = J/2$,  $\mu_F = -\kappa$, and take $\mu=\Gamma=0$. Let $(\gamma_L^{c},\gamma_R^s)$ and $(\gamma_L^s,\gamma_R^z)$ denote the (split) MB pairs of the DBKC, and $(\gamma_L,\gamma_R)$  the (non-split) MB pair of the PDMC. The above parameter identification yields
\begin{equation*}
\gamma_L = \gamma_L^z,\quad\quad \gamma_R = \gamma_R^s ,
\end{equation*}
which may be directly verified in the case $J=\Delta$. In particular, the second MB pair of the DBKC $(\gamma_L^s,\gamma_R^z)$ are not approximate ZMs, nor Weyl SGs, in the PDMC. This has several implications for certain two-time correlation functions. Since the \rev{steady states} of these two models may differ in meaningful ways, we can directly compare the state-independent quantum correlations of the MBs. Our general analysis above predicts that the DBKC correlation function $C^\text{qu}_{\gamma_L^z,\gamma_R^s}(\tau)$ will be {\em asymmetric about zero} and increasingly stationary in the positive $\tau$ direction as $N$ increases. On the contrary, the correlation function $C^\text{qu}_{\gamma_c,\gamma_R}(\tau)$ for the PDMC should be symmetric and increasingly stationary {\rev{in both} the positive and negative $\tau$ direction as $N$ increases. \rev{All of these} predictions are numerically verified in Fig.\,\ref{SplitVNonSplit}. 

\begin{figure}
\includegraphics[width=\columnwidth]{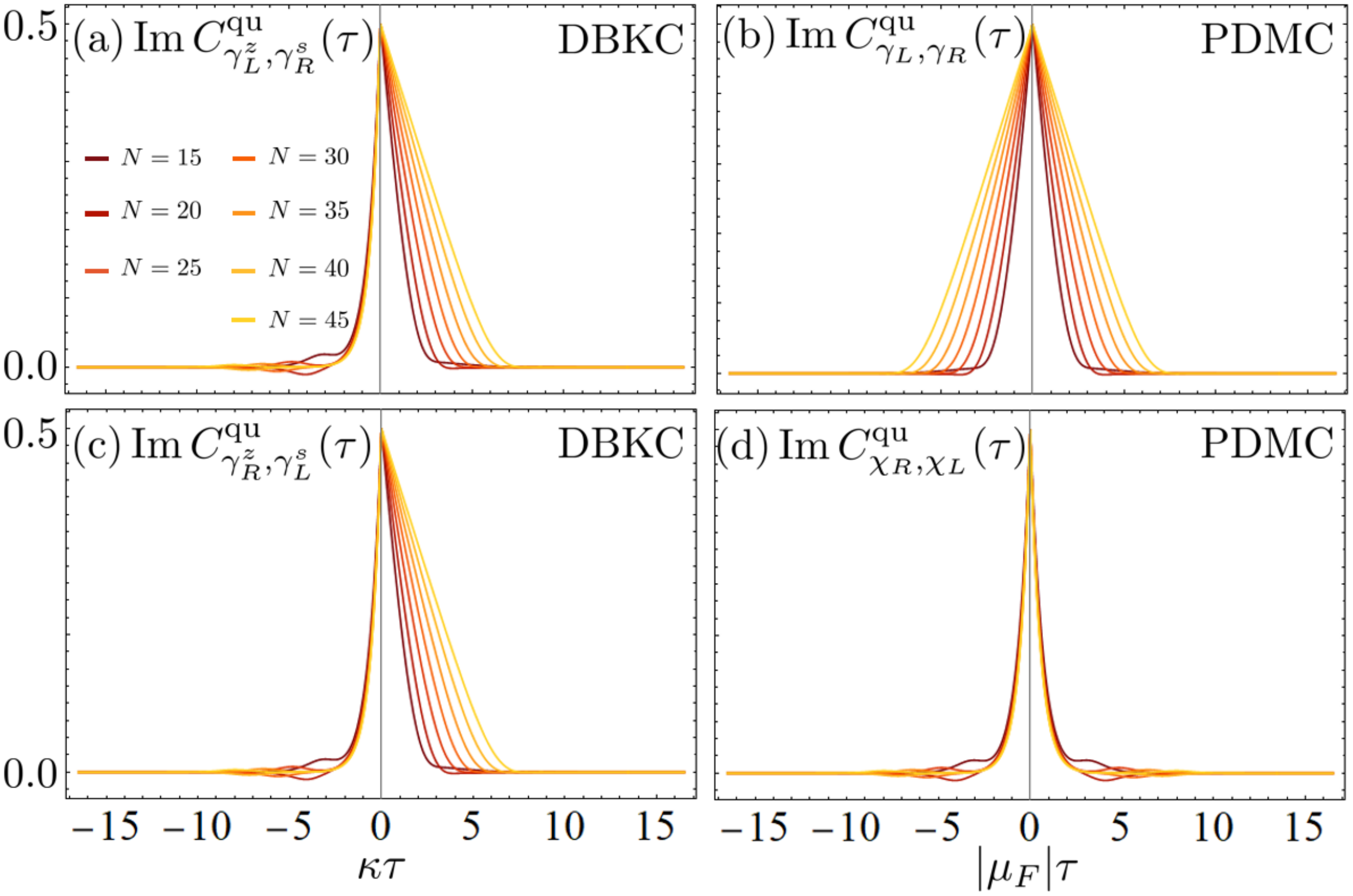}
\vspace*{-3mm}
\caption{{\bf (a)} A MB correlation function for the DBKC. {\bf (b)} A MB correlation function for the PDMC. {\bf (c)} A different MB correlation function for the DBKC. {\bf (d)} A correlation function for the same operators in {\bf (c)} but for the PDMC. In all cases, the modes are normalized so that canonical commutation relations hold at $\tau=0$. Note that $\gamma_L^z = \gamma_L$,  $\gamma_R^s = \gamma_R$, $\gamma_R^z=\chi_R$, and $\gamma_L^s=\chi_L$ at $\tau=0$, when the parameter mapping discussed in the main text is applied.}
\label{SplitVNonSplit}
\end{figure}

A further distinction between these two models is identified by noting that the second MB pair in the BKC has no analogue in the PDMC. Explicitly, the operators $\gamma_L^s$ and $\gamma_R^z$, when mapped to the PDMC, are neither approximate ZMs nor Weyl SGs. To distinguish which model we are working in, let $\gamma_L^s\mapsto \chi_L$ and $\gamma_R^z\mapsto\chi_R$ denote the image of the BKC's second MB pair in the PDMC. As argued above, the correlation function $C_{\gamma_R^z,\gamma_L^s}(\tau)$ will become more and more stationary for $\tau\geq 0$ as $N\to\infty$. On the contrary, no such argument applies to $C_{\chi_R,\chi_L}(\tau)$ and so we generally expect exponentially decaying correlations.

\subsubsection{Signatures of Dirac edge bosons}

We have thus far focused on observable signatures of MBs. How, if at all, do the correlation signatures we identified above change in the presence of number symmetry? That is, can we detect the \rev{DBs} of Sec.\,\ref{Dirac}? Consider the elementary \rev{steady state} correlator $C_{i,j}^\text{ss}(\tau) \equiv C_{\Phi_i,\Phi_j^\dag}^\text{ss}(\tau).$ This correlator is elementary in the sense that any correlation function of linear observables can be written as a linear combination of the above:
\begin{equation*}
C_{\alpha,\beta^\dag}^\text{ss} (\tau) = \sum_{i,j}c^{i,j}_{\alpha,\beta^\dag}C_{i,j}^\text{ss}(\tau),
\end{equation*}
with $c^{i,j}_{\alpha,\beta^\dag}$ determined by the coefficients of $\Phi_i$ and $\Phi_j^\dag$ in the definitions of $\alpha$ and $\beta^\dag$. Explicitly, $c^{i,j}_{\alpha,\beta^\dag} = \alpha_i\beta_j^*$, with $\alpha_i$ and $\beta_j$ elements of $\vec{\alpha}$ and $\vec{\beta}$. 

Number symmetry ensures that $\mathcal{L}$ commutes with the superoperator defined by the action $\rho \mapsto e^{i\theta\widehat{N}}\rho e^{-i\theta \widehat{N}}$, with $\widehat{N}$ the total boson number operator and $\theta\in\mathbb{R}$. Combining this with uniqueness of the \rev{steady state} immediately yields $[\rho_\text{ss},\widehat{N}] = 0$. This guarantees that the ``off-diagonal" elementary correlators, i.e., correlators of the form $\braket{a_i^\dag(\tau)a_j^\dag(0)}_\text{ss}$ and their Hermitian conjugate counterparts, vanish. Mathematically, $C_{i,j}^\text{ss}(\tau)=0$ for $i\leq N < j$ and $j\leq N < i$. In fact, the off-diagonal, state-independent quantum correlations always vanish. This follows because number symmetry guarantees 
\begin{align*}
a_i(\tau) = \sum_{j=1}^N d_{ij}(\tau)a_j(0),
\end{align*}
for some time-dependent coefficients $d_{ij}(\tau) = 2C_{i,j}^{\text{qu}}(\tau)$. This observation, combined with canonical commutation relations, ensures that $[a_i(\tau),a_j(0)]=0$, for all $\tau$. The equivalent statement in the quadrature basis is $C^\text{qu}_{x_j,p_i}(\tau) = C_{x_i,p_j}^\text{qu}(-\tau)$. Since this restatement is in terms of correlations between observables, it is directly accessible in principle. 
 
With this, we can characterize topologically metastable, number-symmetric chains through the vanishing of their off-diagonal \rev{two-time} correlation functions, in conjunction with long-lived correlations / divergent zero frequency power-spectral peaks. This behavior is reflected in Fig.\,\ref{DNSZFPs}. The quantum power spectra $S_{a_1^\dag,a_N}(\omega)$ display exponential divergence at zero frequency in both the DNS chain and the DBKC. However, the off-diagonal spectra $S_{a_1^\dag,a_N^\dag}(0)$ are exactly zero for the DNS chain and diverging exponentially for the DBKC. 
 
\begin{figure}
\includegraphics[width=.7\columnwidth]{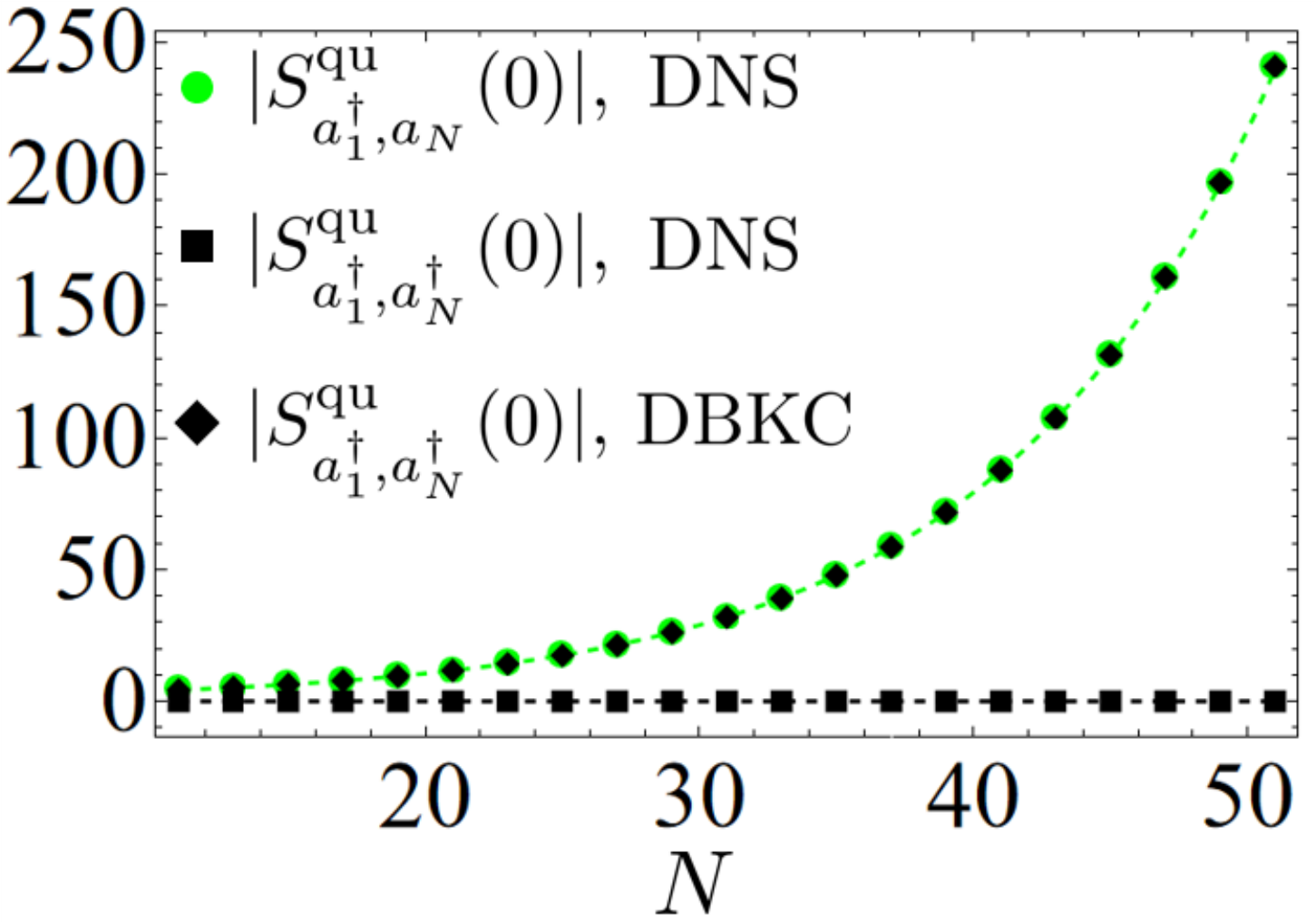}
\vspace*{-2mm}
\caption{(Color online) Various quantum power spectra for the DNS chain and the DBKC in their respective topologically-metastable regimes. The DNS shows exponential divergence of the zero frequency peak $S^\text{qu}_{a_1^\dag,a_N}(0)$ (green circles) and the vanishing of the off-diagonal spectra $S^\text{qu}_{a_1^\dag,a_N^\dag}$ (black squares). To contrast, the DBKC exhibits exponential growth of the off-diagonal spectra (black diamonds). The parameters for the DNS are $J_+=1$, $J_-=0.25$, and $\kappa=0.3$, while $J=2$, $\Delta=0.5$, $\mu=0$, $\kappa=0.3$, and $\Gamma=0$ for the DBKC. These choices ensure an isospectral relationship between the two models. An exponential fit is shown as a green dashed line while $0$ is emphasized with a black dashed line.}
\label{DNSZFPs}
\end{figure}

\section{Conclusions and outlook}
\label{ConcSec}

In this work, we have uncovered several convincing signatures of SPT-like physics in a class of  quadratic bosonic systems undergoing Markovian dissipation. In this sense, we have progressed the search for SPT phases in free bosonic systems by a great deal. The phenomenon of topologically non-trivial dynamical metastability provides clear evidence that SPT phases may be a \textit{transient, dynamical} phenomena for non-interacting bosons. Specifically, topologically metastable systems (i) host pairs of edge modes that bear an extremely close resemblance to the fermionic edge modes of topological superconductors and insulators, in a sense we have made mathematically precise; and (ii) exhibit a sort-of ``ground state degeneracy" in the form of a manifold of Weyl-displaced quasi-steady states. \rev{Remarkably, a key ingredient for topological metastability -- namely, bulk-instability -- is \textit{explicitly forbidden} in quadratic fermionic systems.} Utilizing various new techniques, we have produced and characterized several non-trivial models that sample the expansive realm of possible instances of topological metastability. 

With a robust theory of topological metastability at our fingertips, there are a number of natural directions for future research. The most compelling next step is to address the experimental accessibility of the phenomena we predict. Potentially, there exists a number of promising proposals and experimental implementations of related systems in cavity- and circuit-QED platforms  \cite{PeanoTopoSqueeze,ClerkBKC,GirvincQED,Devoret2011,Devoret2013, Devoret2015,ClerkNonRec,SimonCavity,MartinisChiral, MattiasDPT,SchusterHofst,GirvinBilinear,PainterTopo,SoonOhCavity}, microlasers and ring resonators \cite{FengMicrolaser,KhajEdge}, optomechanical systems \cite{NunnenkampOpto,KippenbergOpto}, and vibronic lattices \cite{PeanoSound}. While most of the contributions entering the driven-dissipative models we considered have been individually realized, the dissipative hopping and the (coherent and dissipative) bosonic pairing mechanisms are, arguably, more exotic and challenging from a practical standpoint. Notably, Ref.\,\onlinecite{PorrasTopoAmp} proposes a method for realizing dissipative hopping, while realizations of both coherent and dissipative pairing have been proposed via three-wave mixing with suitably tuned couplings with auxiliary modes \cite{Devoret2013TWM,Devoret2017,Devoret2019,ClerkDiss}. Once a topologically metastable chain is successfully implemented, the two-time correlation signatures we identified can, in principle, be accessed via their connection to coherence functions. While significant  challenges are likely to arise, it is our hope that the possibility of validating our predictions of tight analogues of fermionic zero modes in free bosons will spur a focused effort across the community.

Beyond experimental detection of topological metastability, several outstanding theoretical questions are also well-worth of further investigation. One avenue involves extending the dynamical theory of metastability into the realm of fermionic systems.  While topological features of said systems are fairly well-understood, we believe that our pseudospectral-based metastability framework can provide major utility in explaining anomalous relaxation \cite{UedaSkin} and cutoff phenomena \cite{VernierCutoff}, among other things. Specifically, while dynamically metastable phases  are explicitly forbidden in fermions (since dynamical instabilities are \cite{Prosen3QFermion,BarthelQuadLindblad}, anomalously relaxing phases are not. Further exploration into the splitting of zero modes and symmetry generators in fermionic systems are also of interest. Still concerning bosons, addressing higher dimensions is a major necessary extension of our framework. For example, what is the dissipative bosonic analogue of the surface bands so prevalent in two-dimensional topological insulators and superconductors, such as quantum Hall systems or the $p+ip$ superconductor? More pressingly, to what extent are the signatures we have uncovered actual consequences of a genuine (one-dimensional) SPT phase? That is, how do many-body symmetries factor into the theory of topological metastability and to what extent do transitions into a topological metastable regimes constitute genuine (dissipative) phase transition? This final question is particularly interesting in light of recent work \cite{BarthelDPT}, which appears to exclude criticality for {\em stable} 1D QBLs. \rev{Moreover, answering it would further elucidate precisely how closely the parallel we are drawing between topologically metastable QBLs and topologically non-trivial QFHs holds. }

Finally, drawing analogy with Majorana-based quantum computing proposals, it is natural to ask whether Majorana bosons could find utility in the context of continuous-variable quantum information processing. In particular, central to quantum computation schemes employing Gottesman-Kitaev-Preskill \cite{GKP,HomeGKP,DevoretGKP} is a pair of canonically conjugate quadratures. From here, the Gottesman-Kitaev-Preskill code is built from fixed displacement operators within the associated phase space and the logical states are built from the (ideally) infinitely squeezed eigenstates of these operators. In our topological metastability paradigm, we are provided with two canonically conjugate quadratures that have an additional non-trivial property: they can be \textit{macroscopically separated} in space. This fact, which arises due to their topological origin, affords an extra degree of robustness that may prove beneficial for such applications. 
Notably, in the non-split case, the MBs generate (orthogonal) phase-space displacements that leaves the overall dynamics invariant to an arbitrarily high degree of precision (as set by the system size). We look forward to address some of these questions in future studies.

\section*{Acknowledgements}

It is a pleasure to thank Mariam Ughrelidze, Michiel Burgelman, Mattias Fitzpatrick, and Joshuah Heath for valuable discussions. Work at Dartmouth was partially supported by the US National Science Foundation through Grants No. PHY-2013974 and No. OIA-1921199, the US Department of Energy, Office of Science, Office of Advanced Scientific Computing Research, under the Accelerated Research in Quantum Computing (ARQC) program, and the Constance and Walter Burke Special Projects Fund in Quantum Information Science. 


\appendix 

\section{Symmetry-protected topological phases \\ of free fermions}
\label{QFHBG}

Given their core relevance in motivating our work, in this appendix we review the basics of SPT phases of fermionic matter, with emphasis on the distinctive properties that topological zero-energy edge states enjoy in topological superconductor and insulator models. 

Focusing on the most basic topological classification, the tenfold way \cite{RyuClassification}, there are four protecting many-body symmetries (electron number, spin rotations, the physical time-reversal operation for electrons, and a particle-hole anti-unitary transformation that exchanges creation and annihilation operators) and ten (out of sixteen) symmetry classes that can be topologically non-trivial depending on the space dimension. In any space dimension, five out of the ten classes are topologically non-trivial. What changes with the dimension is which ones are, and the change happens in a predictable, periodic pattern. 

Let us focus on the non-trivial classes in 1D. From the point of view of the bulk, two classes break up into two disconnected components characterized by a topological Pfaffian invariant in one case, and an invariant that mixes the Pfaffian and the winding number in the other case; three other classes break up into a countable infinite of components labeled by a winding number. From the point of view of the edge, the Pfaffian invariant is associated to a boundary invariant that counts the number of Majorana ZMs per edge modulo two. This is the class of the Majorana chain of Kitaev. The ``mixture'' invariant is associated to a boundary invariant that counts the number of Dirac ZMs per edge modulo two or, equivalently, the number of pairs of Majorana modes per edge, modulo two. And finally, the winding number invariant is associated to a boundary invariant that counts the number of Dirac ZMs per edge. Loosely speaking, a Dirac ZM is a fermionic degree of freedom that is not Hermitian, that is, described by a creation and conjugate annihilation pair of operators. In contrast, a Majorana mode is a Hermitian fermionic degree of freedom. 

The general picture is well-captured by two models: The Majorana chain of Kitaev \cite{KitaevMajorana} (FKC), and the mean-field model of the electronic structure of trans-polyacetylene introduced by Su, Schrieffer, and Heeger \cite{SuSSH} (SSH). Let us consider the Majorana chain first from the point of view of the boundary. The model is characterized by the Hamiltonian
\begin{equation*}
\label{FKCham}
H_\text{FKC} = -\sum_{j=1}^N \mu c_j^\dag c_j - \sum_{j=1}^{N-1}\left(J c_j^\dag c_{j+1} - \Delta c_j^\dag c_{j+1}^\dag + \text{H.c.}\right),
\end{equation*} 
with $\mu,J,\Delta\in\mathbb{R}$. In the language of the previous paragraph, there is one Dirac fermion per site. However, since the symmetry of fermion number is broken anyways, it is more natural to analyze the model in terms of Majorana fermions \(\gamma_l\), \(l=1,\dots, 2N\), such that \( 2c_j\equiv \gamma_{2j-1}+i\gamma_{2j}. \) In this language, the Hamiltonian becomes (up to an additive constant)
\begin{align*}
&H_\text{FKC}=-\frac{\mu}{2}\sum_{j=1}^Ni\gamma_{2j-1}\gamma_{2j}\\
&-\frac{J+\Delta}{2}\sum_{j=1}^{N-1}i\gamma_{2j-1}\gamma_{2j+2}
-i\frac{\Delta - J}{2}\sum_{j=1}^{N-1}\gamma_{2j}\gamma_{2j+1}.
\end{align*}
Looking individually at the three terms, the first term is topologically trivial and the other two are non-trivial in the sense that they support topologically mandated edge modes. Whether their sum is trivial or not depends on the competition among these three terms. Let us focus for concreteness on the case \(J=\Delta\). Then, the model is non-trivial for \( |\mu|<|2J|\) and trivial otherwise. The lines \(|\mu|=|2J|\) are critical. 

Let \(\delta=\mu/2J\). In the non-trivial regime, \(|\delta| <1\) and the operators
\begin{align*}
\gamma_L\equiv{\cal M}(N)\sum_{j=1}^N\delta^{j-1}\gamma_{2j-1},\quad
\gamma_R\equiv {\cal M}(N)\sum_{j=1}^N \delta^{N-j}\gamma_{2j},
\end{align*}
are candidate edge ZMs because 
\begin{align}
\label{gammaL}
i[H_\text{FKC},\gamma_L]&=\delta^N {\cal M}(N) 2J \,\gamma_{2N},\\
\label{gammaR}
i[H_\text{FKC},\gamma_R]&=\delta^N {\cal M}(N) 2J \,\gamma_{1}.
\end{align}
may potentially vanish exponentially fast with $N$ However, the normalization factor \({\cal M}(N)\) depends on the size of the chain and so its behavior must be ascertained. To determine it, notice that 
\begin{align*}
\{\gamma_L,\gamma_R\}=0, \quad
\gamma_L^2=\gamma_R^2
={\cal M}(N)^2\sum_{j=1}^N \delta^{2(j-1)}.
\end{align*}
Hence, this pair of operators will satisfy the algebra of Majorana fermions provided that
\begin{align}
\label{calmf}
{\cal M}(N)^{-1}=\sqrt{\frac{1-\delta^{2(N-1)}}{1-\delta^2}}.
\end{align}
In particular, \({\cal M}(N)\) is bounded as a function of system size if \(|\mu|<2J\). It follows that the commutators of Eqs.\,\eqref{gammaL}-\eqref{gammaR} do indeed vanish exponentially fast in the size of the system. 

The two Majorana fermions at zero energy can be combined to form a {\em single} Dirac fermion at the expense of locality. The resulting Dirac fermion is very peculiar: it is neither localized nor uniformly spread over the length of the chain. The degeneracy of the ground energy level is two, depending on whether this Dirac fermion is or not present in the ground state. As a consequence, the fermion number operator is distributed over the odd or even integers only in one or the other ground state: the two ground states differ by their fermionic parity. 

Let us consider next the SSH model. The Hamiltonian is
\begin{align*}
\nonumber
H_{\text{SSH}}=& \,t_2\sum_{\sigma}\sum_{j=1}^M (c_{2j-1,\sigma}^\dagger c_{2j,\sigma}+\text{H.c.})\\
\label{SSHham}
 +& \,t_1\sum_{\sigma}\sum_{j=1}^{M-1}(c_{2j,\sigma}^\dagger c_{2j+1,\sigma}+\text{H.c.}).
\end{align*}
Here, \(\sigma\) is the label associated to the spin degree of freedom of the electron. This Hamiltonian
belongs to the most symmetric symmetry class, BDI. The structure of the edge is identical for spin up and spin down electrons.
Hence, let us relabel \(c_{2j-1,\sigma}\) as \(a_j\) and \(c_{2j, \sigma}\) as \(b_j\), write \(N\) for \(M/2\), 
and focus on any one spin direction. The Hamiltonian of interest is then
\[H=\ t_2\sum_{j=1}^N (a_j^\dagger b_j+\text{H.c.})+
t_1\sum_{j=1}^{N-1}(b_j^\dagger a_{j+1}+\text{H.c.}). \]
This Hamiltonian is sometimes also called the Peierls model. The first term is topologically trivial and the second one is non-trivial, in the same sense as before. Their sum is non-trivial if \(|t_1|>|t_2|\). The line \(|t_1|=|t_2|\) is critical, as one expects from the fact that it marks the transition from one topological phase to another within a symmetry class.

Let \(\delta=-t_2/t_1\). In the nontrivial regime, \(\delta<1\) and the operator
\begin{align}
\nonumber
a_L={\cal M}(N)\sum_{j=1}^N \delta^{j-1}a_{j}
\end{align} 
and its Hermitian conjugate are candidate ZMs, because 
\[
[H, a_L]=  \delta^N {\cal M}(N) t_1 b_N \]
may vanish exponentially fast as a function of the size of the system. Let us confirm that the normalization factor \({\cal M}(N)\) is bounded as a function of $N$. Notice that 
\begin{align*}
a_L^2=a_L^{\dagger\, 2}=0,\quad
\{a_L,a_L^\dagger\}={\cal M}(N)^2\sum_{j=1}^N\delta^{2(j-1)}.
\end{align*}
Hence, the pair \(a_L, a_L^\dagger\) constitutes a Dirac ZM localized on the left edge provided that 
\({\cal M}(N)\) is given by Eq.\,\eqref{calmf}. A similar calculation shows that the operator 
\begin{align}
\nonumber
b_R={\cal M}(N)\sum_{j=1}^N \delta^{j-1}b_{N-j+1}
\end{align} 
and its Hermitian conjugate constitutes a Dirac ZM localized on the right edge. Since, in addition,
\begin{align*}
\{a_L,b_R\}=0, \quad \{a_L,b_R^\dagger\}=0,
\end{align*} 
the left and right ZMs taken together are two Dirac fermions. Hence, the degeneracy of the ground level is four. Notice that, unlike the zero-energy Dirac fermions of the FKC, those of the SSH chain are localized on the edges of the chain.

\section{Additional technical results}
\label{Proofs}

\subsection{Pseudospectra, singular values, and the \\
doubled-matrix approach}
\label{dblmatps}

In this appendix, we will recount the connection between the 2-norm pseudospectra and singular values, and establish the bridge to the so-called {\em doubled-matrix approach} utilized prominently in topological amplification contexts \cite{PorrasTopoAmp,PorrasIO}. 

Let $s_j(\mathbf{X})$, $j=1,\ldots,n$, denote the $n$ singular values of an $n\times n$ matrix $X$. If $\bf{X}$ is invertible, it is well-known that $\norm{\mathbf{X}^{-1}}_2 = 1/s_\text{min}(\mathbf{X})$ with $s_\text{min}(\mathbf{X})$ the minimal singular value of $\mathbf{X}$. Immediately we can see that the 2-norm $\epsilon$-pseudospectrum of a matrix $\mathbf{X}$ is precisely the set of complex numbers $\lambda$ such that $s_\text{min}(\lambda\mathds{1}-\mathbf{X}) < \epsilon$.  

The doubled matrix approach starts by defining the ``doubled-Hamiltonian"
\begin{equation*}
\overline{\mathbf{X}}(\omega) = \begin{bmatrix}
0 & \omega\mathds{1} - \mathbf{X} \\
\omega\mathds{1}-\mathbf{X}^\dag & 0
\end{bmatrix},
\end{equation*}
with $\omega$ a real frequency and $\mathbf{X}$ typically taken to be the dynamical matrix of interest. One may show that the built-in chiral structure of $\overline{X}(\omega)$ implies a $\pm$ symmetry of the spectrum, and moreover the eigenvalues are precisely $\pm s_j(\omega \mathds{1}-\mathbf{X})$. Often, ``midgap eigenvalues", i.e. exponentially-small-in-$N$ eigenvalues sandwiched between the positive and negative bulk-bands,  are of interest. Importantly, these correspond to the minimal singular values of $\omega\mathds{1}_n-\mathbf{X}$. In particular, if $\overline{\mathbf{X}}(\omega)$ has an exponentially small midgap eigenvalue, then $\omega\mathds{1}-\mathbf{X}$ has a vanishing minimal singular value indicating that $\omega$ exists within the pseudospectrum of $\mathbf{X}$ at sufficiently large system size. In short, if $\overline{\mathbf{X}}(\omega)$ has midgap eigenvalues, then $\omega$ is in the pseudospectrum of $\mathbf{X}$.

\subsection{A canonical correspondence \\ between approximate ZMs and SGs}
\label{ApproxThm}

Here, we prove that, under generic conditions, for each approximate SG there is a canonically conjugate ZM, thereby generalizing Theorem 1 to the approximate case. 

\medskip

{\bf Theorem 2.} {\em Given \rev{an arbitrary} QBL with dynamical matrix $\mathbf{G}$, let $\gamma^s=\widehat{\vec{\gamma}{}^s}$ be an approximate symmetry generator of accuracy $\epsilon$ (in the sense that $\vec{\gamma}^s$ satisfies $\normnoscl{\mathbf{G}\vec{\gamma}^s}<\epsilon$ for some $\epsilon>0$). Then, if the matrix $\mathbf{G}'\equiv \mathbf{G}-\vec{\alpha}\vec{\gamma}^s{}^\dag/\norm{\vec{\gamma}^z}$ hosts only Jordan chains of length one at $0$, there exists a canonically conjugate approximate zero mode $\gamma^z=\widehat{\vec{\gamma}^z}$ of accuracy $\epsilon'=\epsilon \normnoscl{\vec{\gamma}^z}^2$. The converse holds as well.}

\begin{proof}
We will assume that the approximate symmetry generator is not exact. If it is, apply Theorem 1. By the assumptions of the theorem, we have that $\mathbf{G}\vec{\gamma}^s = \vec{\alpha}$ for some vector $\vec{\alpha}$ with $\normnoscl{\vec{\alpha}} <\epsilon$. Hermiticity of $\gamma^s$ implies that the Nambu representation satisfies $\vec{\gamma}^s = -\bm{\tau}_1\vec{\gamma}^s{}^*$. Using this, and the fact that $\bm{\tau}_1\mathbf{G}^*\bm{\tau}_1 = -\mathbf{G}$, we have
\begin{align}
\label{alphaAH}
\bm{\tau}_1\vec{\alpha}^* = \bm{\tau}_1\mathbf{G}^*\vec{\gamma}^s{}^* -\mathbf{G}\bm{\tau}_1\vec{\gamma}^s{}^* = \mathbf{G}\vec{\gamma}^s = \vec{\alpha}. 
\end{align}
Now, consider the matrix $\mathbf{G}' \equiv \mathbf{G} - \vec{\alpha}\vec{\gamma}^s{}^\dag/\normnoscl{\vec{\gamma}^s}^2$. We claim that $\mathbf{G}'$ may be interpreted as a dynamical matrix of some other QBL and that $\mathbf{G}'\vec{\gamma}^s = 0$. The first claim follows $\mathbf{G}'$ obeys the only constraint set on dynamical matrices:
\begin{align*}
\bm{\tau}_1\mathbf{G}'{}^*\bm{\tau}_1 &= \bm{\tau}_1\mathbf{G}^*\bm{\tau}_1 -\bm{\tau}_1\vec{\alpha}^*\vec{\gamma}^s{}^T\bm{\tau}_1 
\\
&= -\mathbf{G} - (\bm{\tau}_1\vec{\alpha})^*(\bm{\tau}_1\vec{\gamma}^s{}^*)^\dag 
\\
&= -\mathbf{G} + \vec{\alpha}\vec{\gamma}^s{}^\dag = -\mathbf{G}',
\end{align*}
where we have used Eq.\,\eqref{alphaAH} in the third equality. The second claim is verified directly
\begin{align*}
\mathbf{G}'\vec{\gamma}^s = \mathbf{G}\vec{\gamma}^z - \vec{\alpha}\frac{\vec{\gamma}^s{}^\dag\vec{\gamma}^s}{\normnoscl{\vec{\gamma}^s}^2} = \vec{\alpha}-\vec{\alpha} = 0.
\end{align*}

Now, any QBL with dynamical matrix $\mathbf{G}'$ has $\gamma^s$ as an exact \rev{SG}. We may then directly apply Theorem 1 to find a canonically conjugate \textit{exact} \rev{ZM} $\gamma^z$. In Nambu space, this means there is a vector $\vec{\gamma}^z = -\bm{\tau}_1\vec{\gamma}^z{}^*$ such that $\widetilde{\mathbf{G}}'\vec{\gamma}^z = 0$, with
\begin{align*}
\widetilde{\mathbf{G}}' = \bm{\tau}_3\mathbf{G}'{}^\dag\bm{\tau}_3 = \widetilde{\mathbf{G}} - \bm{\tau}_3\vec{\gamma}^s\vec{\alpha}^\dag\bm{\tau}_3,
\end{align*}
and $\vec{\gamma}^s{}^\dag\bm{\tau}_3\vec{\gamma}^z = i$, so that $[\gamma^s,\gamma^z]=i$. It follows that
\begin{align*}
\normnoscl{\widetilde{\mathbf{G}}\vec{\gamma}^z} &= \normnoscl{\widetilde{\mathbf{G}}'\vec{\gamma}^z +\frac{\bm{\tau}_3\vec{\gamma}^s\vec{\alpha}^\dag\bm{\tau}_3}{\normnoscl{\vec{\gamma}^s}^2}\vec{\gamma}^z} 
\\
&= \frac{1}{\normnoscl{\vec{\gamma}^s}^2}\normnoscl{\bm{\tau}_3\vec{\gamma}^s\vec{\alpha}^\dag\bm{\tau}_3\vec{\gamma}^z} 
= \frac{|\vec{\alpha}^\dag\bm{\tau}_3\vec{\gamma}^z|}{\normnoscl{\vec{\gamma}^s}}.
\end{align*}
Using the Cauchy-Schwarz inequality, we may upper-bound the numerator of the right hand-side by $\normnoscl{\vec{\alpha}}\normnoscl{\vec{\gamma}^z} < \epsilon\normnoscl{\vec{\gamma}^z}$. Thus,
$\normnoscl{\widetilde{\mathbf{G}}\vec{\gamma}^z} < \epsilon {\normnoscl{\vec{\gamma}^z}} /{\normnoscl{\vec{\gamma}^s}} .$ 
An additional application of the Cauchy-Schwarz inequality to the identity $\vec{\gamma}^s{}^\dag\bm{\tau}_3\vec{\gamma}^z = i$ yields $1/\normnoscl{\vec{\gamma}^s}<\normnoscl{\vec{\gamma}^z}$, so that
\begin{align*}
\normnoscl{\widetilde{\mathbf{G}}\vec{\gamma}^z} < \epsilon\normnoscl{\vec{\gamma}^z}^2,
\end{align*}
as claimed. The converse holds by simply replacing $\mathbf{G}$ with $\widetilde{\mathbf{G}}$ and $\vec{\gamma}^s$ with $\vec{\gamma}^z$. 
\end{proof}

We remark that canonically conjugate modes can {\em always} be rescaled in a natural way to make the accuracies $\epsilon$ and $\epsilon'$ equal. Generally, if $[\gamma^s,\gamma^z]=i$, then $[\gamma^s{}',\gamma^z{}']=i$, with $\gamma^s{}' = \mathcal{M}\gamma^s$ and $\gamma^z{}'=\gamma^z/\mathcal{M}$ for any $\mathcal{M}>0$. Taking $M=\normnoscl{\vec{\gamma}^z}$ provides approximate SGs and ZMs of accuracy $\mathcal{M}\epsilon$.  The most interesting examples are then those in which $\epsilon \ll 1/\mathcal{M}$.

\subsection{An isospectral mapping between the purely dissipative 
and the dissipative bosonic Kitaev chains}
\label{PDCdetails}

Here, we establish an isospectral mapping between the PDMC and the DBKC. Let $\mathbf{G}^\text{DBKC}(\kappa,J,\Delta)$ and $\mathbf{G}^\text{PDC}(\mu_F,J_F,\Delta_F)$ denote the dynamical matrices of the DBKC and the PDMC  under OBCs, respectively. Note that we have distinguished the FKC hopping, pairing, and onsite potential with a subscript $F$. Define the momentum-space translation operator 
\begin{equation*}
\bm{\Lambda}(\delta k) \equiv \text{diag}(e^{-i\delta k},\ldots,e^{-iN{\color{revred}\delta k}})\otimes \mathds{1}_2.
\end{equation*}
With this, one may verify that 
\begin{equation}
\label{DBKCPDC}
\bm{\Lambda}\left(\frac{\pi}{2}\right) \mathbf{G}^\text{PDC}\left(0,-\frac{J}{2},-\frac{\Delta}{2}\right)\bm{\Lambda}^{-1}\left(\frac{\pi}{2}\right) = i \mathbf{G}^\text{DBKC}(0,J,\Delta).
\end{equation}
That is, the dynamical matrices are unitarily equivalent up to a phase when $\mu_F=\kappa=0$. In particular, the $\mu_F=0$ OBC rapidity spectrum for the PDMC is equal to $i$ times that of the DBKC, with the identifications $\kappa=0$, $J_F\leftrightarrow -J/2$, and $\Delta_F\leftrightarrow -\Delta/2$. This allows us to establish Eq.\,\eqref{PDCRap} from the known spectral properties of the DBKC \cite{Bosoranas}.

\subsection{On the numerical determination of Majorana bosons}
\label{NumMBs}

In this appendix, we present a procedure for numerically computing MBs in topologically metastable QBLs. The basic inputs are (i) a family of dynamical matrices $\mathbf{G}_N$, one for each system size $N$, corresponding to a topologically metastable QBL under OBCs; and (ii) a sufficiently small (in a sense to be explained) accuracy parameter $\epsilon$. This procedure outputs pairs of canonically conjugate $\epsilon$-pseudoeigenvectors corresponding to $\epsilon$-pseudoeigenvalue 0 for sufficiently large $N$ (to be explained). The pseudospectra considered are $2$-norm pseudospectra.

To begin, we define a positive-semidefinite matrix $\mathbf{R}_N \equiv  \mathbf{G}_N^\dag\mathbf{G}_N$. The eigenvalues of $\mathbf{R}$ are the squares of the singular values of $\mathbf{G}_N$. In the language of App.\,\ref{dblmatps}, topological metastability ensures that $\mathbf{R}_N$ possesses a finite number (say, $m$) of midgap modes, i.e., eigenvectors corresponding to eigenvalues that decay to $0$ as $N$ increases. The remaining $2N-m$ eigenvalues are bounded below by a gap parameter $\lambda_\text{gap}$. Fix $0<\epsilon<\lambda_\text{gap}^{1/2}$ and let $N_0$ be the smallest $N$, such that the $m$ midgap eigenvalues are less than $\lambda_\text{gap}$. By construction, the $m$ normalized midgap eigenvectors, which we will label $\vec{u}_j$, $j=1,\ldots,m$, will be $\epsilon$-pseudoeigenvectors of $\mathbf{G}_N$ corresponding to $\epsilon$-pseudoeigenvalue zero for all $N>N_0$. Explicitly, we have 
\begin{equation*}
\norm{\mathbf{G}_N\vec{u}}_2 = \left(\vec{u}^\dag\mathbf{R}_N\vec{u}\right)^{1/2} = s_j < \epsilon ,
\end{equation*} 
with $s_j$ the singular value corresponding to $\vec{u}_j$. Equivalently, $s_j^2=\lambda_j$ the corresponding midgap eigenvalue of $\mathbf{R}_N$.

Despite being pseudoeigenvectors of $\mathbf{G}_N$, these need not yet provide approximate SGs. In particular, they need not generate a Hermitian linear form. For this, we leverage the property $\mathcal{C}\mathbf{G}_N\mathcal{C}^{-1} = -\mathbf{G}_N$. This implies $\mathcal{C}\mathbf{R}_N\mathcal{C}^{-1} = \mathbf{R}_N$. Thus, if $\vec{u_j}$ is an eigenvector of $\mathbf{R}$, so is $\mathcal{C}\vec{u}_j$ with the same eigenvalues. In particular, the set $\{\vec{u}_j\}$ is invariant under $\mathcal{C}$. If $\mathcal{C} \vec{u}_j = e^{i\theta} \vec{u}_j$, then we can normalize to obtain a new vector $\vec{\gamma}_j^s =-e^{i\theta}\vec{u}_j$ that defines a Hermitian approximate SG $\widehat{\vec{\gamma}}^s$. If there are degenerate singular values, then the restriction of $\mathcal{C}$ to this subspace can be diagonalized to obtain a basis of vectors $\vec{\gamma}_j^s$ that each define Hermitian SGs. 

Constructing the approximate ZMs follows analogously, but with $\mathbf{G}$ replaced with $\widetilde{\mathbf{G}}$. Let $\{\vec{\gamma}_j^s{}'\}$ and $\{\vec{\gamma}_j^z{}'\}$ be the outputs of these two procedures. It follows that canonical commutation relations can  be ensured if the matrix $\mathbf{F}_{jk} = \vec{\gamma}_j^s{}'{}^\dag\bm{\tau}_3\vec{\gamma}_k^z{}'$ is diagonalizable. In all examples considered in this paper, this is the case.

\section{The time-evolution of bosonic parity in a Markovian system prepared in a cat state}
\label{CatParApp}

We present here the explicit calculation of the cat-state parity dynamics under the pure SS DBKC. As we noted in the main text, this model can be seen as a set of independent damped quantum harmonic oscillators in the normal-mode basis. This decoupling allows us to reduce the problem to that of computing the parity dynamics of a single-mode cat-state under damped harmonic motion. The multimode generalization then follows naturally.

Let $\ket{\alpha}$, $\alpha\in\mathbb{C}$, denote a single-mode coherent state and define the single-mode cat state $\ket{\mathcal{C}_\phi(\alpha)} \equiv \mathcal{N}_\phi(\alpha)\left(\ket{\alpha} + e^{i\phi} \ket{-\alpha}\right)$, with $\mathcal{N}_\phi(\alpha)$ a normalization constant. We wish to compute the expectation value of parity $P=e^{i\pi a^\dag a}$ in the time-evolved state $\rho_{\alpha,\phi}(t)$, resulting from the initial condition $\rho(0) = \ket{\mathcal{C}_\phi(\alpha)}\bra{\mathcal{C}_\phi(\alpha)}$ under the QBL
\begin{equation*}
\mathcal{L}(\rho) = -i[\omega a^\dag a,\rho] + 2\kappa\Big( a \rho a^\dag - \frac{1}{2}\{a^\dag a,\rho\}\Big) .
\end{equation*}
Here, $\omega$ is the oscillator frequency and $\kappa$ is the damping rate. We may compute $\rho_{\alpha,\phi}(t)$ exactly utilizing known results \cite{FujiiDissQHO,KorschDissQHO}. We may write 
\begin{eqnarray*}
&& 
\rho_{\alpha,\phi}(t) 
= e^{t\mathcal{L}}\left(\rho_{\alpha,\phi}(0)\right) 
\\&& 
= \mathcal{N}_\phi(\alpha)^2\left( \sigma_\alpha(t) + \sigma_{-\alpha}(t) + e^{i\phi} \chi_\alpha(t)  + e^{-i\phi} \chi_{-\alpha}(t) \right),
\end{eqnarray*}
where
\begin{equation*}
\sigma_\alpha(t) \equiv e^{t\mathcal{L}}(\ket{\alpha}\bra{\alpha}),\quad 
\chi_\alpha(t) \equiv e^{t\mathcal{L}}(\ket{\alpha}\bra{-\alpha}) .
\end{equation*}
The terms $\sigma_{\pm\alpha}(t)$ can be quoted directly as
\begin{equation*}
\sigma_{\pm\alpha}(t) = \ket{\pm \alpha(t)}\bra{\pm\alpha(t)},\quad \alpha(t) = e^{-(\kappa+i\omega) t}\alpha (0). 
\end{equation*}
The terms $\chi_{\pm\alpha}(t)$ are slightly more complicated,
\begin{equation*}
\chi_\alpha(t) = D(t) \bigg(\sum_{k=0}^\infty \frac{(2 e^{-\kappa t}\sinh(\kappa t))^k}{k!} a^k \chi_\alpha(0) a^\dag{}^k\bigg)D(t)^\dag , 
\end{equation*}
with $D(t)\equiv e^{-(\kappa+i\omega)ta^\dag a}$. Now,
\begin{equation*}
a^k \chi_\alpha(0) a^\dag{}^k = \alpha^k \chi_\alpha(0) (-\alpha^*)^k = (-|\alpha|^{2})^k\chi_{\alpha}(0), 
\end{equation*}
which leads us to
\begin{equation*}
\chi_\alpha(t) = \exp[-2 |\alpha|^2 e^{-\kappa t}\sinh(\kappa t)]\, D(t) \chi_\alpha(0)D(t)^\dag.
\end{equation*}
The remaining time-dependence may be computed as
\begin{equation*}
D(t) \chi_\alpha(0)D(t)^\dag \!=\! \exp[-2 |\alpha|^2 e^{-\kappa t}\sinh(\kappa t)] \ket{\alpha(t)}\bra{-\alpha(t)} .
\end{equation*}
Finally,  we have 
\begin{align*}
\chi_\alpha(t) &= \exp[-4 |\alpha|^2  e^{-\kappa t}\sinh(\kappa t)] 
\ket{\alpha(t)}\bra{-\alpha(t)} \\
& \equiv f_\alpha(t)\ket{\alpha(t)}\bra{-\alpha(t)},
\end{align*}
with $\chi_{-\alpha}(t)$ following accordingly. 

With the exact time dependence of $\rho(t)$ computed, we can now evaluate the expectation value of parity. A particularly useful identity is $P\ket{\alpha} = \ket{-\alpha}$. Using this, we can compute $P_{1}^{\alpha}(t) = \tr[P\sigma_\alpha(t)]$ and $P_{2}^{\alpha}(t) = \tr[P\chi_\alpha(t)]$ to obtain 
\begin{equation*}
\braket{P}\!(t) = \mathcal{N}_\phi(\alpha)^2 [P_{1}^{\alpha}(t)+P_{1}^{-\alpha}(t)+e^{i\phi}P_{2}^{\alpha}(t)+e^{-i\phi} P_{2}^{-\alpha}(t)].
\end{equation*}
Proceeding, we find 
\begin{eqnarray*}
P_{1}^{\alpha}(t) \!=\!\tr[P\ket{\alpha(t)}\bra{\alpha(t)}]
&=& \braket{\alpha(t)|-\alpha(t)}\\
&=& \exp\left(-2|\alpha|^2 e^{-2\kappa t}\right), 
\end{eqnarray*}
\begin{eqnarray*}
P_{2}^{\alpha}(t) \!= \!f_\alpha(t)\tr[P\ket{\alpha(t)}\bra{-\alpha(t)}]
&=& f_\alpha(t) \braket{-\alpha(t)|-\alpha(t)}  \\
&=& f_\alpha(t).
\end{eqnarray*}
Putting this all together yields the single-mode evolution, 
\begin{equation*}
\braket{P}\!(t) = \frac{e^{-2|\alpha|^2}+\cos(\phi) e^{-2 |\alpha|^2(1-e^{-2\kappa t})}}{1+\cos(\phi)e^{-2|\alpha|^2}}, 
\end{equation*}
from which the multimode generalization given by Eq.\,\eqref{Monster} in the main text follows.

\section{A dissipative double-winding bosonic chain}
\label{DDW}

It is natural to wonder to what degree we can control the number of ZMs and SGs on each edge of a topologically metastable chains. The properties of Toeplitz matrices ensure that each band with a positive winding number yield a left-localized ZM and a right-localized SG. Conversely, negative winding numbers yield a right-localized ZM and left-localized SGs. In all the examples discussed in the main text, the number of ZMs (SGs) on a particular edge has been either zero or one. Surprisingly, we can write down a range-one model with a single internal degree of freedom that has {\em two} ZMs (SGs) on a particular edge.
The Hamiltonian is given by
\begin{align*}
H_\text{DW} = \frac{i}{2}\sum_{j=1}^N\left(\Delta_h a_{j}^\dag a_{j+1}^\dag + \mu a_j^\dag{}^2-\text{H.c.}\right),
\end{align*}
with the usual BC conventions. Here, $\Delta_h,\mu\in\mathbb{R}$ denote the non-degenerate and degenerate parametric amplification strength, respectively. 
The dissipator consists of three distinct contributions $\mathcal{D}_\text{DW} \equiv \mathcal{D}_{-,0} + \mathcal{D}_{+,0} + \mathcal{D}_{p,1}$. Explicitly,
\begin{eqnarray*}
\mathcal{D}_{-,0} &=& 2\kappa_- \sum_{j=1}^N\mathcal{D}[a_j,a_j^\dag] , \quad 
\\
\mathcal{D}_{+,0} &= & 2\kappa_+ \sum_{j=1}^N\mathcal{D}[a_j^\dag,a_j] ,
\\
\mathcal{D}_{p,1} &=& \frac{\Delta_d}{2}\sum_{j=1}^N\mathcal{D}[a_j^\dag,a_{j+1}^\dag] - \mathcal{D}[a_{j+1}^\dag,a_{j}^\dag] - (a_j^\dag\leftrightarrow a_j) .
\end{eqnarray*}
In words, we have uniform onsite loss with strength $\kappa_-\geq 0$, uniform onsite gain with strength $\kappa_+\geq 0$, and NN incoherent pairing with strength $\Delta_d$. Positivity is the associated GKLS matrix demands that $4\kappa_+\kappa_-\geq \Delta_d^2$. For simplicity, we additionally take $\Delta_d\geq \Delta_h \geq 0$.
If $2\kappa_\pm\geq \Delta_d$, we can write a simple diagonal representation:
\begin{align*}
\mathcal{D}_\text{DW} &= (2\kappa_--\Delta_d)\sum_{j=1}^N \mathcal{D}[a_j] + (2\kappa_+-\Delta_d)\sum_{j=1}^N \mathcal{D}[a_j^\dag]
\\
&+ \frac{\Delta_d}{2}\sum_{j=1}^N\mathcal{D}[a_j-a_{j+1}^\dag] + \frac{\Delta_d}{2}\sum_{j=1}^N\mathcal{D}[a_j+a_{j+1}^\dag].
\end{align*}

The symbol of the dynamical matrix is given by $-i\mathbf{g}_\text{DW}(k) = -\kappa \mathds{1}_2 + (\mu+\Delta_h\cos(k) + i\Delta_d \sin(k))\bm{\sigma}_1$ with $\kappa \equiv \kappa_--\kappa_+$. The rapidity bands are
\begin{equation*}
\lambda_\pm(k) = -\kappa \pm(\mu+\Delta_h \cos(k) + i \Delta_d \sin(k)).
\end{equation*}
The winding numbers $\nu_\pm$ of the bands $\lambda_\pm(k)$ are given by
\begin{equation*}
\nu_\pm = \begin{cases}
1 & -1 <\tilde{\kappa}\mp \tilde{\mu}<1,
\\
0 & \text{otherwise,} 
\end{cases}
\end{equation*}
with $\tilde{\kappa}\equiv\kappa/\Delta_h$ and $\tilde{\mu}\equiv\mu/\Delta_h$. By tuning $\tilde{\kappa}$ and $\tilde{\mu}$ it is possible to obtain a total winding $\nu_++\nu_-$ equal to $0$, $1$, or $2$.

\vfill

\end{document}